\documentclass[a4paper,11pt]{article}
\usepackage{pos}
\usepackage{graphicx,url,wrapfig}
\usepackage[english]{babel}   
\usepackage[utf8]{inputenc}  
\usepackage[OT2,T1]{fontenc}
\usepackage{anyfontsize}
\usepackage{verbatim}
\usepackage{listings}
\usepackage{xcolor}
\usepackage{latexsym}
\usepackage{hyperref}
\usepackage{amsmath,mathrsfs,enumitem}
\usepackage[capitalize]{cleveref}
\usepackage{fancyvrb}
\usepackage{longtable}
\usepackage{enumitem}
\usepackage[titletoc,title]{appendix}
\usepackage{float}
\usepackage{verbatim}
\usepackage[section]{placeins}
\usepackage{bm,bbm}
\usepackage{mathtools}

\newcommand{\nonBulletListing}[1]{\vspace{1mm}
\newline
\noindent
\textbf{#1}
\newline
\noindent}

\newcommand{\au}[2]{#1.~#2}
\newcommand{\arX}[1]{\href{http://arxiv.org/abs/#1}{{\cob arXiv:#1}}}
\newcommand{\oarX}[1]{\href{http://arxiv.org/abs/#1}{{\cob arXiv:#1}}}

\newcommand{\book}[5]{\emph{#1} (#2, #3, #5)}
\newcommand{\books}[4]{\emph{#1} (#2, #3, #4)} 

\newcommand{\procm}[6]{in \emph{#1}, ed.\ by #2 (#3, #4, #6)}

\newcommand{\procsinm}[5]{in \emph{#1}, ed.\ by #2 (#3, #4, #5)}
\newcommand{\doin}[6]{\href{http://dx.doi.org/#1}{\cob #2\ #3 {\bf #4}, #5 (#6)}}

\newcommand{\doinn}[5]{\href{http://dx.doi.org/#1}{{\cob #2 {\bf #3}, #4 (#5)}}}
\newcommand{\doij}[5]{\href{http://dx.doi.org/#1}{{\cob #2 {\bf #3}, #4 (#5)}}}
\newcommand{\ndoinn}[5]{\href{#1}{{\cob #2 {\bf #3}, #4 (#5)}}}
\newcommand{\tia}[1]{{\it #1},}

\def\cob{\color{blue}}

\title{Lectures on classical and quantum cosmology}

\author*[a]{Gianluca Calcagni}
\author[b,c,d]{Maria Grazia Di Luca}
\author[e]{Tom\'a\v{s} Fodran}

\affiliation[a]{Instituto de Estructura de la Materia -- CSIC,\\
calle Serrano 121, 28006 Madrid, Spain}

\affiliation[b]{Scuola Superiore Meridionale\\
Largo S. Marcellino 10, I-80138 Napoli, Italy}

\affiliation[c]{Dipartimento di Fisica Ettore Pancini, Universit\`a di Napoli ``Federico II'',\\
Complesso Univ. Monte S. Angelo, I-80126 Napoli, Italy}

\affiliation[d]{INFN, Sezione di Napoli}

\affiliation[e]{Department of Astrophysics, IMAPP, Radboud University,\\
P.O. Box 9010, 6500 GL Nijmegen, The Netherlands}

\emailAdd{g.calcagni@csic.es}
\emailAdd{mariagrazia.diluca@unina.it}
\emailAdd{t.fodran@astro.ru.nl}

\abstract{These lecture notes introduce the reader to the hot big bang model, cosmological perturbations, gravitational waves, the cosmic microwave background, inflation, the singularity problem, the cosmological constant problem and the cosmology of quantum gravity.}

\FullConference{%
  Corfu Summer Institute 2021 ``School and Workshops on Elementary Particle Physics and Gravity''\\
  29 August - 9 October 2021\\
  Corfu, Greece
}

\tableofcontents

\begin{document}
\maketitle

\section*{Recommended reading}

These lectures notes are moderately self-contained for readers already familiar with the subject of cosmology. We suggest to consult the textbooks \cite{wei72,LiL,Muk,Calcagni:2017sdq} for further study. We base our presentation mainly on \cite{Calcagni:2017sdq}; some figures and several parameter estimates are updated with respect to their 2017 values.

\section{Hot big bang model}

We can define \emph{cosmology} as the study of Nature at very large scales, ranging from about 1 kpc (the size of a galaxy) to 1 Mpc (the order of typical inter-galactic distances in the local group of galaxies), up to thousands of Mpc (roughly corresponding to the farthest observable light source, the cosmic microwave background).

The galaxy distribution near our local group ($\sim$ tens of Mpc) looks irregular and it is characterized by regions where matter is much more clustered. Patterns of strongly over-dense and under-dense regions can also be observed at larger scales ($\sim$ hundreds of Mpc) \cite{Hoffman:2017ako}. However, in general, observations of the large-scale structure ($\gtrsim$ hundreds of Mpc) suggest that our universe is homogeneous and isotropic as a first approximation, meaning that its metric properties are the same at any point in space and in any direction. Such an assumption, known as the cosmological principle, is supported especially at larger scales ($\sim 3000$ Mpc) by the cosmic microwave background (CMB). The CMB is a radiation background of cosmic origin, dating back to more than 13 billion years ago when matter became transparent to radiation. Today, the CMB has cooled down to a temperature of $T_0=2.7255\pm 0.0006\,{\rm K}$, with very small fluctuations of order $\Delta T / T \sim 10^{-5}$; that is, the early universe is homogeneous and isotropic up to one part over a hundred thousands.

\subsection{Friedmann--Lema\^itre--Robertson--Walker (FLRW) metric}

We can mathematically formalize the cosmological principle in the framework of Einstein's theory of General Relativity (GR). The line element of a 4-dimensional isotropic and homogeneous universe is given in terms of the Friedmann--Lema\^itre--Robertson--Walker (FLRW) metric as
\begin{equation}
  \mathrm{d} s^{2}=g_{\mu \nu} \mathrm{d} x^{\mu} \mathrm{d} x^{\nu}=-\mathrm{d} t^{2}+a^{2}(t) \gamma_{\alpha \beta} \mathrm{d} x^{\alpha} \mathrm{d} x^{\beta},
\end{equation}
where $x^0=t$ is the synchronous or proper time. The spatial line element
\begin{equation}
  \gamma_{\alpha \beta} \mathrm{d} x^{\alpha} \mathrm{d} x^{\beta}=\frac{\mathrm{d} r^{2}}{1-\textsc{k} r^{2}}+r^{2} \mathrm{~d} \Omega_{2}^{2},
\end{equation}
corresponds to a maximally symmetric 3-dimensional space of constant sectional curvature $\textsc{k}$. For $\textsc{k}=-1,0,1$, the universe is open, flat, or closed, respectively. In this model of the universe, the scale factor $a(t)$ and the spatial curvature sign encode all there is to know about the geometry. It is sometimes convenient to work in terms of the conformal time $\tau$, defined as
\begin{equation}
  \mathrm{d} \tau:=\frac{\mathrm{d} t}{a(t)}.
\end{equation}
The line element then becomes
\begin{equation}
  \mathrm{d} s^{2}=a^{2}(\tau)\left(-\mathrm{d} \tau^{2}+\gamma_{\alpha \beta} \mathrm{d} x^{\alpha} \mathrm{d} x^{\beta}\right),
\end{equation}
\begin{wrapfigure}{r}{4cm}
		\vspace{-12pt}
    \includegraphics[width=4cm]{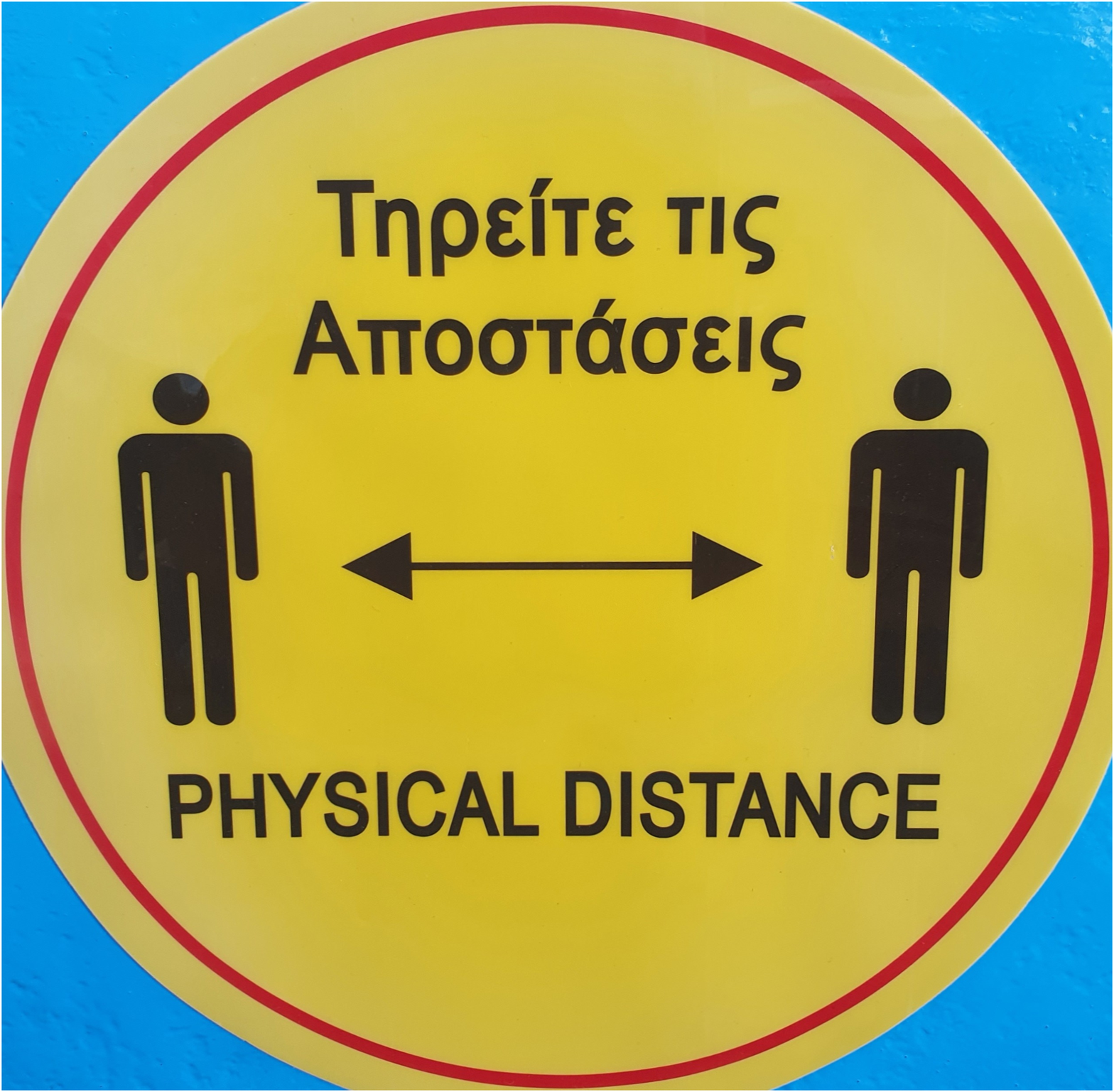}
\end{wrapfigure}
and, for $\mathrm{K}=0$, it is conformally equivalent to Minkowski. Distances computed in terms of the coordinates $x^\alpha$ are called \textit{comoving} (because comoving observers are those at rest with respect to the content of the universe). The scale factor relates the comoving and proper (physical) distances according to
\begin{equation}
    (\mathrm{proper \; distance})=a(t) \times(\mathrm{comoving \; distance}). 
\end{equation}

\subsection{Hubble parameter and \texorpdfstring{$H_0$}{H0} tension}

From the observations, we know that the universe is expanding and therefore the scale factor increases in time, $\dot{a}(t)>0$. Thus, we can introduce the redshift $z$ as a relational time:
\begin{equation}
  1+z:=\frac{a_{0}}{a},
\end{equation}
where $a_0:=a(t_0)$ is the scale factor today. Since the scale factor is not physically observable, its normalization is arbitrary and one typically chooses $a_0=1$. A local observer is at $z=0$ and distant objects are at $z>0$. We define the Hubble parameter as the relative expansion rate of the universe
\begin{equation}
  H(t):=\frac{\dot{a}(t)}{a(t)}.
\end{equation}
In the early universe, the inverse Hubble parameter gives an approximate estimate of the particle radius at time $t$, which is the maximal distance of points in causal connection. We call this distance the Hubble radius (in $c=1$ units)
\begin{equation}
  R_{H}(t):=\frac{1}{H(t)}.
\end{equation}
The value of the Hubble parameter today, the Hubble constant $H_0$, is expressed in terms of a dimensionless parameter $h$, $H_0=100 h\, \mathrm{~km} \mathrm{~s}^{-1} \mathrm{Mpc}^{-1}=h\times (9.778\,{\rm Gyr})^{-1}$ \cite{ParticleDataGroup:2020ssz}. This parameter has been estimated in several independent ways \cite{Verde:2019ivm,Perivolaropoulos:2021jda}, which can be grouped into two clusters:\footnote{CL: confidence level. $\Lambda$CDM: standard model of the universe which assumes GR, cold dark matter and a cosmological constant $\Lambda$.}
\begin{itemize}
\item Small-$z$, late-time measurements: time delays in gravitational lensing of quasars ($h=0.733^{+0.017}_{-0.018}$ \cite{Wong:2019kwg}, assuming a $\Lambda$CDM model), distance-ladder measurements using extra-galactic Cepheids ($h=0.7403\pm0.0142$ \cite{Riess:2019cxk}, model independent) or the calibration of the tip of the red giant branch ($h=0.698\pm0.008\,{\rm (stat)}\pm0.17{\rm (syst)}$ \cite{Freedman:2019jwv}, model independent), and geometric distance measurements to megamaser-hosting galaxies ($h=0.739\pm0.030$ at the 68\% CL \cite{Pesce:2020xfe}, assuming a $\Lambda$CDM model).
\item Large-$z$, early-universe measurements: baryon acoustic oscillations in large-scale structure ($h=0.6820\pm0.081$ at the 68\% CL \cite{eBOSS:2020yzd}, assuming a $\Lambda$CDM model) and CMB measurements from \textsc{Planck} ($h=0.6736\pm0.0054$ at the 68\% CL \cite{Planck:2018vyg}, assuming a $\Lambda$CDM model) and the Atacama Cosmology Telescope ($h=0.676\pm0.011$ at the 68\% CL \cite{ACT:2020gnv}, assuming a $\Lambda$CDM model).
\end{itemize}
While the measurements at high redshift all agree, there is a mismatch between these and small-redshift measurements, ranging between $4.0$ and $5.7$ standard deviations. The source of this so-called \emph{$H_0$ tension} \cite{Verde:2019ivm,Perivolaropoulos:2021jda} is not known. A conservative explanation might lie in an inadequate treatment of systematic errors but there is the possibility of a failure of the $\Lambda$CDM model and the need of a dynamical dark-energy component or, in the most extreme case, new physics beyond GR or beyond the Standard Model of particles.

The inverse of the Hubble constant provides an estimate of the age of the universe $t_{0} \simeq H_{0}^{-1}=9.778 h^{-1} \mathrm{~Gyr} \simeq 13.2-14.5 \mathrm{~Gyr}$, depending on the value of $h$. A more accurate calculation taking into account the different phases of the cosmic evolution, from radiation domination to matter domination to dark-energy domination, yields a similar range of values.

\subsection{Action and Einstein equations}

Consider the following conventions for the Levi-Civita connection, the Riemann and Ricci tensors and the Ricci scalar:
\begin{alignat}{-1}
&\Gamma_{\mu \nu}^{\rho} &&:= \frac{1}{2} g^{\rho \sigma}\left(\partial_{\mu} g_{v \sigma}+\partial_{\nu} g_{\mu \sigma}-\partial_{\sigma} g_{\mu \nu}\right), \\
&R^\rho_{~\mu\sigma\nu} &&:= \partial_{\sigma} \Gamma_{\mu \nu}^{\rho}-\partial_{\nu} \Gamma_{\mu \sigma}^{\rho}+\Gamma_{\mu \nu}^{\tau} \Gamma_{\sigma \tau}^{\rho}-\Gamma_{\mu \sigma}^{\tau} \Gamma_{\nu \tau}^{\rho}, \\
&R_{\mu \nu} &&:= R_{\ \mu \rho \nu}^{\rho}, \\
&R &&:= R_{\mu \nu} g^{\mu \nu}.
\end{alignat}
In GR, we can derive the equations of motion from the action
\begin{equation}
  S = S_g+S_{\rm m},
\end{equation}
where
\begin{equation}
  S_{g}=\frac{1}{2 \kappa^{2}} \int \mathrm{d}^{4} x \sqrt{|g|}(R-2 \Lambda),
\end{equation}
$g:=\det(g_{\mu\nu})$, $ \kappa^2=8\pi G$ and $\Lambda$ is the cosmological constant. The action $S_g$ is the Einstein--Hilbert action for the gravitational field, while $S_{\rm m}$ is the matter action. Applying the variational principle, one obtains the Einstein equations
\begin{equation}\label{eq:eineq}
  G_{\mu \nu}+g_{\mu \nu} \Lambda=\kappa^{2} T_{\mu \nu},
\end{equation}
where
\begin{equation}
  G_{\mu \nu}:=R_{\mu \nu}-\frac{1}{2} g_{\mu v} R
\end{equation}
is the Einstein tensor and
\begin{equation}
  T_{\mu \nu}:=-\frac{2}{\sqrt{|g|}} \frac{\delta S_{\mathrm{m}}}{\delta g^{\mu \nu}}=-2 \frac{\partial \mathcal{L}_{\mathrm{m}}}{\partial g^{\mu \nu}}+g_{\mu \nu} \mathcal{L}_{\mathrm{m}}
\end{equation}
is the stress-energy tensor of matter.
%
In the case of a perfect fluid,
\begin{equation}
  T_{\mu \nu}=(\rho+P) u_{\mu} u_{v}+P g_{\mu \nu},
\end{equation}
where $P$ is the isotropic pressure, $\rho$ is the energy density and $u^{\mu }$ is the velocity of the fluid constituents along their world-lines. In the comoving reference frame, the 3-velocity of such constituents is zero by definition, so that
\begin{equation}
  u^{\mu}=\left(|g_{00}|^{-1/2}, 0, 0, 0\right)^{\mu}.
\end{equation}
A real scalar field $\phi$ with potential $V(\phi)$ is a specific example of perfect fluid. The Lagrangian density reads
\begin{equation}
  \mathcal{L}_{\mathrm{m}}=\mathcal{L}_{\phi}=-\frac{1}{2} \partial_{\mu} \phi \partial^{\mu} \phi-V(\phi).
\end{equation}
%
One can easily show that the energy density and isotropic pressure on a homogeneous background are given by
\begin{alignat}{-1}
&\rho_{\phi} && =\frac{\dot{\phi}^{2}}{2}+V, \\
&P_{\phi} && =\frac{\dot{\phi}^{2}}{2}-V,
\end{alignat}
respectively, with $\dot{\phi}:=\partial_t \phi$. The equation of motion takes then the form
\begin{equation} \label{eq:dalambert}
  \Box \phi-V_{, \phi}=0,
\end{equation}
where $\Box \phi:=\frac{1}{\sqrt{|g|}} \partial_{\mu}\left(\sqrt{|g|} \partial^{\mu} \phi\right)$.
On a FLRW background, if the field only depends on the time coordinate, the equation turns into
\begin{equation} \label{eq:harOsc}
  \ddot{\phi}+3 H \dot{\phi}+V_{, \phi}=0.
\end{equation}
When $V=m^2\phi^2/2$, (\ref{eq:harOsc}) is the equation for a harmonic oscillator with a background-dependent friction term $3H\dot{\phi}$.

\subsection{Friedmann equations}

The Einstein equations on a homogeneous, isotropic setting reduce to the Friedmann equations. Given the FLRW metric, the only non-vanishing components of the Levi-Civita connection are
\begin{equation}
  \Gamma_{\alpha \beta}^{0}=H g_{\alpha \beta}, \quad \Gamma_{\alpha 0}^{\beta}=H \delta_{\alpha}^{\beta}, \quad \Gamma_{\alpha \beta}^{\lambda}=\Gamma_{\alpha \beta}^{\lambda}\left[\gamma_{\alpha \beta}\right],
\end{equation}
where $\alpha,\beta=1,2,3$ and $\Gamma_{\alpha \beta}^{\lambda}\left[\gamma_{\alpha \beta}\right]$ is the Levi-Civita connection of the three-dimensional, maximally symmetric manifold $\Sigma$ with metric $\gamma_{\alpha \beta}$. The non-vanishing components of the Ricci tensor are
\begin{alignat}{-1}
    R_{00} &=-3\left(H^{2}+\dot{H}\right)=-3 \frac{\ddot{a}}{a}, \\
R_{\alpha \beta} &=\left(\frac{2\textsc{k}}{a^{2}}+3 H^{2}+\dot{H}\right) g_{\alpha \beta}.
\end{alignat}
We can express the Ricci scalar in terms of $ { }^{3} R$, the three-dimensional Ricci scalar of the manifold $\Sigma$, as
\begin{equation}
R ={ }^{3} R+3\left( 4 H^{2}+2 \dot{H}\right)\,.
\end{equation}
Recalling the form of $T^{\mu\nu}$ for a perfect fluid, we can now write the Friedmann equations as
\begin{alignat}{-1}
&H^2 && =\frac{\kappa ^2}{3}  \rho+\frac{\Lambda}{ 3}-\frac{\textsc{k}}{a^2}, \label{eq:1stFE}\\
&\frac{\ddot{a}}{a} &&=-\frac{\kappa^2}{6}(\rho+3P)+\frac{\Lambda}{3}. \label{eq:2ndFE}
\end{alignat}
The first Friedmann equation (\ref{eq:1stFE}), which is the 00 component of Einstein equations (\ref{eq:eineq}), contains a term proportional to the energy density, a term proportional to the cosmological constant and a pure GR term (which we cannot obtain in a Newtonian derivation) proportional to spatial curvature. Equation (\ref{eq:2ndFE}), given by the combination of the 00 component and the trace of Einstein equations, is about the acceleration in the expansion of the universe. We see that the cosmological constant term gives a positive contribution to $\ddot{a}$. To solve these equations, we need to specify the equation of state $P=P(\rho)$.

Let us only consider the equation of state of a barotropic fluid, whose pressure is proportional to the energy density (such a simplification, as we will see, is suitable to describe the content of our universe):
\begin{equation}
  P=w\rho,
\end{equation}
where $w$ is a constant referred to as the barotropic index. We can find exact solutions to the Friedmann equations in the power-law form
\begin{equation}
  a= t^{p},
\end{equation}
where $p={2}/[3(1+w)]$; the Hubble parameter scales as the inverse of proper time, $H={p}/{t}$. Notice that, for such solutions, when $t\rightarrow 0$, $a\rightarrow 0$, the metric becomes degenerate and $H\rightarrow \infty$. Since also $R\rightarrow\infty$, we have a physical divergence. This singularity is known as the big bang and its existence in GR, argued in several focussing and singularity theorems, is called the big bang problem.

\subsection{Content of the universe}

We define the critical energy density $\rho_{\text{crit}}$ as the one we would have in a spatially flat universe,
\begin{equation}
  \rho_{\text {crit }}:=\frac{3 H^{2}}{\kappa^{2}}.
\end{equation}
Let us introduce the density parameter $\Omega$, defined as the ratio between the total energy density and the critical energy density:
\begin{equation}
  \Omega:=\frac{\rho}{\rho_{\text{crit}}}.
\end{equation}
Since the energy density is an additive quantity, we can write the density parameter as
\begin{equation}
  \Omega=\Omega_{\mathrm{r}}+\Omega_{\mathrm{m}}+\Omega_{\Lambda}+\Omega_{\textsc{k}},
\end{equation}
where $\Omega_\textsc{k}:=-{\textsc{k}}/({a^2 H^2})$ is the deviation from flatness and $\Omega_{\Lambda}:={\Lambda}/({3H^2})$ is the contribution from the cosmological constant. The $\Omega_{\rm r}$ and $\Omega_{\rm m}$ are the density parameters associated with radiation and pressure-less matter (dust), respectively.

These components are described by the following barotropic indices and the evolution of their energy densities:
\begin{itemize}
\item Cosmological constant:
\begin{equation}
  \\ w=-1, \qquad \rho_\Lambda=\frac{\Lambda}{\kappa^2}={\rm const}\,.
\end{equation}
\item Radiation (ultra-relativistic particles):
\begin{equation}
  \\ w=\frac13, \qquad \rho_r(z)=\rho_{{\rm r},0}(1+z)^4.
\end{equation}
\item Pressure-less matter (non-relativistic particles):
\begin{equation}
    w=0, \qquad \rho_{\mathrm{m}}(z)=\rho_{\mathrm{m},0}(1+z)^{3}.
\end{equation}
\end{itemize}
Radiation density decreases faster than matter density during the expansion ($z\rightarrow 0$). Therefore, the radiation component dominates over matter at early times, but matter eventually takes over. The moment $z_{\text{eq}}$ at which the two densities coincide is called equality and by requiring $\rho_{\rm r}(z_{\text{eq}})=\rho_{\mathrm{m}}(z_{\text{eq}})$ we find $z_{\text{eq}}\approx 3400$.

Lastly, we also notice that the first Friedmann equation can be recast as the constraint
\begin{equation}
  \Omega_{\mathrm{r}}+\Omega_{\mathrm{m}}+\Omega_{\Lambda}+\Omega_{\textsc{k}}=1.
\end{equation}
What are the values of such densities today? For the radiation component, the only contribution comes from photons (given the universe temperature, there is no other ultra-relativistic particle) and $\Omega_{{\rm r},0}$ is negligible. For the matter component, we have three contributions:
\begin{equation}
 \Omega_{\mathrm{m},0}=\Omega_{\mathrm{b},0}+\Omega_{\nu,0}+\Omega_{\mathrm{nb},0},
\end{equation}
respectively baryons, neutrinos and non-baryonic matter. In particular, the latter is also known as dark matter, since it does not interact with photons and thus we can observe it only indirectly. Evidence favors dark matter being made of particles moving at non-relativistic speed and lying outside the Standard Model of particle physics. The neutrino contribution is negligible today, while $\Omega_{\mathrm{b},0}\sim 0.05$ and $\Omega_{\mathrm{nb},0}\sim 0.25$.

From observations, we know that our universe is going through a phase of accelerated expansion. Consider the second Friedmann equation
\begin{equation}
  \frac{\ddot{a}}{a}=-\frac{1}{2} H^{2}(1+3 w) \Omega=-\frac{1}{2} H^{2}(1+3 w).
\end{equation}
In order to have $\ddot{a}>0$, the expansion must be driven by an energy component with $w<-1/3$. This component is called dark energy and, from observational constraints, it accounts for $\sim 70\%$ of the total energy budget of our universe. These results allow one to conclude that $\Omega_{\textsc{k}}\approx 0$.

The standard cosmological model with a cosmological constant, cold dark matter and zero curvature is called flat $\Lambda$CDM and is the reference model for many parameter estimations.

\subsection{Acceleration}

In order to describe cosmic acceleration, it is convenient to introduce the first slow-roll parameter
\begin{equation}
  \epsilon:=-\frac{{\rm d}\ln H}{{\rm d}\ln a}=-\frac{\dot{H}}{H^2}=1-\frac{\ddot{a}}{aH^2}=1+\frac{1}{2}(1+3w)\Omega=\frac{3}{2}(1+w).
\end{equation}
To have acceleration, we must require $\epsilon<1$ (which in turn implies $w<-1/3$). According to the value of $\epsilon$, we can distinguish three different acceleration regimes
\begin{alignat}{-1}
    &\text{super-acceleration:} \quad && &\epsilon<0 && &\quad\Longleftrightarrow\quad && &w<-1, \\
    &\text{de Sitter acceleration:} \quad && &\epsilon=0 && &\quad\Longleftrightarrow\quad && &w=-1, \\
    &\text{sub-acceleration:} \quad && &0<\epsilon<1 && &\quad\Longleftrightarrow\quad && -1< &w<-\frac{1}{3}.
\end{alignat}
The above expressions hold for a universe filled by a single perfect fluid with barotropic index $w$. The dark-energy component dominates at late times, but only by two thirds of the total density. Assuming that $w=w_0$ is redshift-independent, the \textsc{Planck}$+$lensing$+$SNe$+$BAO estimate for the equation of state of dark energy is \cite{Planck:2018vyg}
\begin{equation}\label{w0est} 
w_0=-1.028\pm 0.031\qquad \textrm{(68\% CL)}\,,
\end{equation}
compatible with a cosmological-constant-dominated universe and the $\Lambda$CDM model. Recall, however, that a pure cosmological constant leads to tensions in the estimate of $H_0$.

\subsection{Hot big bang model}
The CMB has a Planckian, quasi-isotropic distribution. This observation indicates that the early universe was in thermal equilibrium and that the CMB is the remnant of an initial very hot phase. For this reason, we can describe the expansion of our universe using its temperature $T$ as a relational time. Thus, we can express all the relevant physical quantities in terms of $T$. In particular, the generalized Stefan--Boltzmann law for the radiation-dominated era reads
\begin{equation}
  \rho_r=\frac{\pi^2}{30}g_*(T)\,T^4,
\end{equation}
where $g_*(T)$ accounts for the different species of particles that are relativistic at temperature $T$. This relation allows us to express $T$ in terms of the redshift:
\begin{equation}
  T(z) \simeq \frac{3.24}{g_{*}^{1/4}}(1+z)\,.
\end{equation}
In Tab.~\ref{tab:uniTimeline}, we report the main events of the universe thermal history. In the first seconds after the big bang, the temperature was around $10^{16}$ K and there were no atomic structures in the universe. After some minutes, the first nuclei started to form, a process known as big bang nucleosynthesis (BBN). The BBN plays a crucial role in putting several constraints on the cosmological parameters using the present abundances of atomic nuclei. When the universe was about 70,000 years old, radiation-matter equality was established and, from that moment on, matter became dominant. The CMB was formed at $t\approx 3.8\times10^5$ years, at a temperature of about $3,000$ K, i.e., one thousand times hotter than today. First stars formed when the universe was $10^8$ years old and the corresponding size of causal correlation was around 110 Mpc, which is of the same order of magnitude of the typical length scale of galaxy clusters.

\begin{table}[H] 
    \centering
    \resizebox{\columnwidth}{!}{%
    \begin{tabular}{|l|l|l|l|l|l|}
    \hline$T(\mathrm{~K})$ & $T(\mathrm{eV})$ & $t$ & $z$ & $R_{\mathrm{p}}$ & Event \\
    \hline $10^{16}$ & $10^{12}$ & $10^{-12} \mathrm{~s}$ & $10^{16}$ & $0.2 \mathrm{~mm}$ & Highest energy probed in laboratory (LHC). \\
    \hline $10^{10}$ & $10^{5}$ & $2 \mathrm{~s}$ & $10^{10}$ & $10^{5} \mathrm{~km}$ & Formation and destruction of nuclei begins. \\
    \hline $10^{9}$ & $10^{4}$ & $200 \mathrm{~s}$ & $10^{9}$ & $10^{7} \mathrm{~km}$ & Nucleosynthesis of light ions. \\
    \hline $10^{8}$ & $3 \times 10^{3}$ & $20 \mathrm{~min}$ & $10^{8}$ & $10^{9} \mathrm{~km}$ & BBN ends. \\
    \hline $10^{4}$ & 1 & $7.0 \times 10^{4} \mathrm{yr}$ & 3400 & $39 \mathrm{kpc}$ & Radiation-matter equality. Matter domination begins. \\
    \hline 3000 & $0.3$ & $3.8 \times 10^{5} \mathrm{yr}$ & 1090 & $289 \mathrm{kpc}$ & Decoupling of matter and radiation. CMB forms. \\
    \hline 100 & $10^{-2}$ & $10^{8} \mathrm{yr}$ & 25 & $110 \mathrm{Mpc}$ & First stars. \\
    \hline 3 & $3 \times 10^{-4}$ & $14 \times 10^{9} \mathrm{yr}$ & 0 & $14.2 \mathrm{Gpc}$ & Today. \\
    \hline
    \end{tabular} 
    }
    \caption{Timeline of the universe thermal history. The table reports the temperature in $\mathrm{~K}$ and $\mathrm{eV}$, the corresponding cosmological time and redshift for each event. The particle horizon $R_{\rm p}$ is computed too.}
    \label{tab:uniTimeline}
\end{table}


\section{Cosmological perturbations and gravitational waves}

The cosmological principle allows us to describe the properties of our universe when averaging over $O({\rm Mpc})$ distances. As we have already mentioned, the universe looks strongly inhomogeneous at smaller scales, with matter concentrated in gravitationally bound structures. Inhomogeneous settings may require a more realistic description than the FLRW background. A first step in that direction, which works extraordinarily well in all situations where deviations from the cosmological principle are small, is to allow for tiny perturbations of the FLRW background. The metric takes the form
\begin{equation}
  g_{\mu v}=\tilde{g}_{\mu \nu}+h_{\mu \nu} ,
\end{equation}
where $\tilde{g}_{\mu\nu}$ is the background and $h_{\mu\nu}$ is the cosmological perturbation.
We can classify perturbations into three sectors according to their properties under spatial coordinate transformations: the scalar, vector and tensor perturbations. We will not focus on the vector sector because the vector perturbations are typically damped away in the early universe.

\subsection{Linear tensor perturbations (gravitational waves)}

We first consider tensor perturbations in the linear regime. This approach means that we neglect any term of order $O(h^2)$ in our equations. In order to simplify our description, we can get rid of the non-physical degrees of freedom, making use of invariance under spacetime diffeomorphisms. In particular, we ask the perturbation tensor not to have mixed time-space components, to have zero divergence and zero trace:
\begin{equation}
\quad h_{\mu 0}=0, \qquad \partial^{\nu} h_{\mu \nu}=0, \qquad h=h_\mu^{\ \mu}=0.
\end{equation}
This is called the transverse-traceless gauge. The perturbed metric takes form
\begin{equation}
 \mathrm{d} s^{2}=a^{2}(\tau)\left[-\mathrm{d} \tau^{2}+\left(\delta_{\alpha \beta}+h_{\alpha \beta}\right) \mathrm{d} x^{\alpha} \mathrm{d} x^{\beta}\right],
\end{equation}
where $\alpha,\beta=1,2,3$ are spatial indices. We are working with conformal time and considering a spatially flat background. Since we are in a transverse gauge, the gravitational wave $h_{\alpha\beta}$ reduces to a $2\times 2$ traceless symmetric matrix. Thus we are only left with two degrees of freedom. Assuming, without loss of generality, that $x^3$ is the propagation direction, $h_{\alpha\beta}$ can be decomposed in the $(x^1,x^2)$ plane into two independent polarization scalar modes $h_+(x)$ and $h_{\times}(x)$
\begin{equation}
  h_{\alpha \beta}(x)=h_{+}(x) e_{\alpha \beta}^{+}+h_{\times}(x) e_{\alpha \beta}^{\times}\,,
\end{equation}
where the two polarization matrices are given as
\begin{equation}
 e^{+}=\left(\begin{array}{cc}
1 & 0 \\
0 & -1
\end{array}\right),
\qquad
e^{\times}=\left(\begin{array}{ll}0 & 1 \\ 1 & 0\end{array}\right).
\end{equation}
One can show that, in the transverse-traceless gauge, the linear perturbations to the Ricci tensor and scalar are given as 
\begin{equation} \label{eq:TTgaugeLinPert}
  \delta R_{\mu \nu}=-\frac{1}{2} \tilde{\Box} h_{\mu \nu}, \qquad \delta R=0,
\end{equation}
respectively. In order to obtain first-order equations of motion, we can perturb the action at second order:
\begin{equation}
  \delta^{(2)} S_{h}=\frac{1}{4 \kappa^{2}} \sum_{\lambda=+, \times} \int \mathrm{d}^{4} x \sqrt{|\tilde{g}|}\left(h_{\lambda} \tilde{\Box} h_{\lambda}-2 \Lambda h_{\lambda}^{2}\right) .
\end{equation}
We now consider, separately, the two polarization modes. Calling $h_{\lambda}=\varphi$ and $m^{2}=2 \Lambda$, in a flat FLRW background we have to solve the following equation in momentum space:
\begin{equation}\label{eq:ms}
  \varphi_{\mathbf{k}}^{\prime \prime}+2 \mathcal{H} \varphi_{\mathbf{k}}^{\prime}+\left(k^{2}+m^{2} a^{2}\right) \varphi_{\mathbf{k}}=0,
\end{equation}
where the prime stands for the derivative with respect to the conformal time $\tau$ and $\mathcal{H}:=\partial_{\tau}a/a$. We first consider the asymptotic solutions in the two limits for large and small values of the product $k|\tau|$, where $k=|\mathbf{k}|$ is the comoving wavenumber. At very small scales ($k|\tau| \gg 1$), spacetime curvature is negligible and one can ignore the Hubble friction term $2 \mathcal{H} \varphi_{\mathbf{k}}^{\prime}$. The perturbation is well inside the horizon and it obeys the harmonic oscillator equation
\begin{equation}
  \varphi_{\mathbf{k}}^{\prime \prime}+k^{2}\varphi_{\mathbf{k}}\simeq0.
\end{equation}
Incoming and outgoing plane waves give the independent solutions
\begin{equation}
  \varphi_{\mathbf{k}} \simeq A_{k} \mathrm{e}^{\pm \mathrm{i} k \tau}.
\end{equation}
On the other hand, outside the horizon ($k|\tau| \ll 1$), the effective mass term can be ignored and, assuming $\Lambda=0$, the perturbation is approximately constant:
\begin{equation}
  \varphi_{\mathbf{k}} \simeq C_{k}.
\end{equation}
The perturbation freezes when it crosses the horizon and then it keeps the constant value $C_k$ determined by the continuity condition at horizon crossing. In order to obtain the full analytic solutions, consider the rescaling of the field $w_{k}:=a \varphi_{\mathbf{k}}$. Equation (\ref{eq:ms}) becomes
\begin{equation}\label{eq:musa}
  w_{k}^{\prime \prime}+\left(k^{2}-M^{2}\right) w_{k}=0\,,
\end{equation}
with $M^{2}=\left[2- \frac{m^{2}}{H^{2}}\right] \frac{1}{ \tau^{2}}$. This is called the Mukhanov--Sasaki equation and $w_k$ is called the Mukhanov--Sasaki variable. On a de Sitter background, the general solution is a superposition of Bessel functions
\begin{equation}
  w_k=C_1\sqrt{k|\tau|}J_{\nu}(k|\tau|)+C_2\sqrt{k|\tau|}Y_{\nu}(k|\tau|),
\end{equation}
which reproduces the previous asymptotic behaviors.

\subsection{Linear scalar perturbations} \label{sec:scalarpert}

Let us now focus on the scalar sector. We can write the most general perturbed metric as
\begin{equation}
 \mathrm{d} s^{2}=a^{2}(\tau)\left\{-(1+2 \Phi) \mathrm{d} \tau^{2}+2\left(\partial_{\alpha} B\right) \mathrm{d} x^{\alpha} \mathrm{d} \tau\right. \left.+\left[(1-2 \Psi) \delta_{\alpha \beta}+2 \partial_{\alpha} \partial_{\beta} E\right] \mathrm{d} x^{\alpha} \mathrm{d} x^{\beta}\right\},
\end{equation}
where $\Phi$, $\Psi$, $B$ and $E$ are scalar perturbations of the metric components. These perturbations are generated by fluctuations of the matter content of the universe. Therefore, we need to account for the correspondent perturbations in the stress-energy tensor
\begin{equation}
\rho=\tilde{\rho}+\delta \rho, \quad P=\tilde{P}+\delta P.
\end{equation}
In this case, it is convenient to work with a gauge-invariant quantity, called curvature perturbation on uniform density hypersurfaces
\begin{equation}
  \zeta=-\Psi-H \frac{\delta \rho}{\dot{\rho}}.
\end{equation}
With the appropriate background-dependent rescaling, following the same steps as in the previous case, one ends up again with a Mukhanov--Sasaki equation. In theories beyond GR, tensor and scalar perturbations may still obey a Mukhanov--Sasaki equation but with different variable, mass term and propagation speed.

\subsection{Astrophysical gravitational waves}

The formalism we are using to deal with cosmological perturbations is very general and we can apply it to different backgrounds to treat other kinds of perturbations. In GR, we describe weak gravitational signals propagating as transverse waves at the speed of light, called gravitational waves (GWs). Astrophysical GWs were detected for the first time in September 2015 by the Laser Interferometer Gravitational-wave Observatory (LIGO) and Virgo \cite{Abbott:2016blz,TheLIGOScientific:2016src} and, since then, several GW events from the merger of compact objects have been observed. 

We consider the case of tensor perturbations again, but this time we linearize the Einstein equations (\ref{eq:eineq}) in the presence of matter; for a Minkowski background with $\Lambda=0$, we find
\begin{equation}
  \Box_{\eta} h_{\mu \nu}=-2 \kappa^{2} S_{\mu \nu},
\end{equation}
where $S_{\mu\nu}=T_{\mu\nu}-(1/2)\eta_{\mu\nu}T^{\ \rho}_{\rho}$. The solution is the convolution of the source term with the retarded Green's function:
\begin{alignat}{-1}
& h_{\mu \nu}(x) =-2 \kappa^{2} \int d^{4} x^{\prime} S_{\mu \nu}\left(x^{\prime}\right) {\cal G}^{\mathrm{ret}}\left(x-x^{\prime}\right),\label{hS} \\
& \Box_{\eta} {\cal G}^{\mathrm{ret}}\left(x-x^{\prime}\right) =\delta^{D}\left(x-x^{\prime}\right), \\
&\left.{\cal G}^{\mathrm{ret}}\right|_{t<t^{\prime}} =0.
\end{alignat}
We can solve the problem in the local wave-zone, that is, at distances much larger than the characteristic wave-length $r \gg \lambda$ (so that we are quite far from the source) but much shorter than cosmological distances (so that we do not need to account for cosmological effects in the propagation). The retarder Green's function then takes the form
\begin{equation}\label{hG}
 {\cal G}^{\mathrm{ret}}(t,r)\simeq C\frac{\delta(t-r)}{r},
\end{equation}
where $C$ is a constant. The GW amplitude decreases as the inverse distance from the source.

\subsection{Luminosity distance}\label{ludis}

When dealing with cosmological scales, the expansion of the universe affects signal propagation. As a consequence, in cosmology, we have different definitions of distances that are not equivalent. We now introduce the concept of luminosity distance, which is an actual observable. Consider a source of electromagnetic waves with intrinsic luminosity ${\rm L}$ and let ${\rm F}$ be the flux measured at Earth. The luminosity distance $d_{L}^\textsc{em}$ is then defined by inverse-square-distance law
\begin{equation}
{\rm F}=: \frac{{\rm L}}{4 \pi\left(d_{L}^\textsc{em}\right)^2}\,.
\end{equation}
We call standard candles sources of known intrinsic luminosity whose properties do not change with the redshift. Standard candles allow us to measure the luminosity distance of close-by objects and host galaxies. One can show that the luminosity distance is a rescaling of the comoving distance $r$ \cite{CoLu}:
\begin{equation}\label{dem}
  d_{L}^\textsc{em}=\frac{a_{0}^{2}}{a} r=(1+z) \int_{t(z)}^{t_{0}} \frac{d t}{a}=(1+z) \int_{a(z)}^{1} \frac{d a}{H a^{2}}=(1+z) \int_{0}^{z} \frac{d z}{H},
\end{equation}
where we expressed the ratio $a_0/a$ in terms of the redshift and we used the definition of comoving distance.
%

On the other hand, we can consider sources of gravitational waves and give an analogous definition of luminosity distance $d_{L}^\textsc{gw}$, referring this time to the GW signal. Equivalently, the inverse of the amplitude $h_{\mu\nu}$ is defined to be proportional to the GW luminosity distance. Schematically
\begin{equation}\label{dgwdef}
h=: \frac{1}{d_{L}^\textsc{gw}}\,.
\end{equation}
Sources of both electromagnetic and gravitational waves are called standard sirens.

For standard sirens, we can measure the ratio
\begin{equation}\label{ratiod}
  \frac{d_{L}^\textsc{gw}(z)}{d_{L}^\textsc{em}(z)},
\end{equation}
which can give us important information about the cosmic expansion and possible deviations from GR or the standard model of cosmology \cite{Belgacem:2019pkk,Calcagni:2019ngc,Belgacem:2020pdz}. In fact, one can show that (again, schematically)
\begin{equation}\label{dhem}
h= \frac{1}{d_{L}^\textsc{em}}\qquad \textrm{in GR},
\end{equation}
which can be guessed by combining (\ref{hS}), (\ref{hG}) and (\ref{dem}). Therefore, in GR the ratio (\ref{ratiod}) is equal to 1. Notice that (\ref{dhem}) is the result of a calculation \cite{CDL}, while (\ref{dgwdef}) is a definition.

\subsection{Primordial stochastic GW background (SGWB)}

So far, interferometers have detected only GWs coming from individual sources. However, primordial perturbations in the tensor sector should give rise to a stochastic background of superposing GWs, filling our universe together with the CMB.
 The energy density associated to such a background is given by the average of the kinetic term of tensor perturbations:
\begin{equation}
  \rho_{\textsc{gw}}:=\frac{M_{\mathrm{Pl}}^{2}}{8 a^{2}}\left\langle\left(\partial_{\tau} h_{\alpha\beta}\right)^{2}+\left(\nabla h_{\alpha\beta}\right)^{2}\right\rangle=\frac{M_{\mathrm{Pl}}^{2}}{4} \int {\rm d} \ln k\left(\frac{k}{a}\right)^{2} \frac{k^{3}}{\pi^{2}} \sum_{\lambda}\left|h_{k}^{\lambda}\right|^{2},
\end{equation}
where $M_{\mathrm{Pl}}=\kappa^{-1}$ is the reduced Planck mass. During the radiation- and matter-dominated eras, the primordial tensor spectrum $\mathcal{P}_{\mathrm{t}}(k)$, which is the two-point correlation function of the tensor perturbation in comoving momentum space, evolves into a stochastic background $\Omega_{\mathrm{GW}}(k, \tau)$. This process can be encoded into a transfer function $\mathcal{T}(k, \tau_{0})$, which describes the deformation of the primordial spectrum by the evolution of the tensor modes after horizon crossing:
\begin{equation}
  \Omega_{\textsc{gw}}(k, \tau_0):=\frac{1}{\rho_{\text {crit }}} \frac{{\rm d} \rho_{\textsc{gw}}}{{\rm d} \ln k}=\frac{k^{2}}{12 a_{0}^{2} H_{0}^{2}} \mathcal{P}_{\mathrm{t}}(k) \mathcal{T}^{2}(k, \tau_{0}).
\end{equation}
As we will see, primordial perturbations naturally arise in the context of inflation and, for the tensor sector, one predicts a very small amplitude. This smallness can explain why the current generation of interferometers has not detected any SGWB yet.


\section{Cosmic microwave background}\label{sec:3}

One second after the big bang, the universe was a hot plasma of matter and radiation in thermal equilibrium. Since there were no atomic structures, free electrons and baryons could interact with photons and among themselves. As the universe expanded, the plasma started to cool down and the recombination process began. After the atomic structures formed, the density of diffusion centers decreased and the universe became transparent to radiation. The CMB is a snapshot of our universe at this matter-radiation decoupling. The surface from which we receive the CMB photons is called the last-scattering sphere and it represents the horizon for electromagnetic observations of our universe. No photons from earlier epochs can come through to us. In contrast GWs can filter through the CMB and become a valuable probe of the very early universe.

\begin{figure}[!htb]
    \centering
    \includegraphics[width=0.9\linewidth]{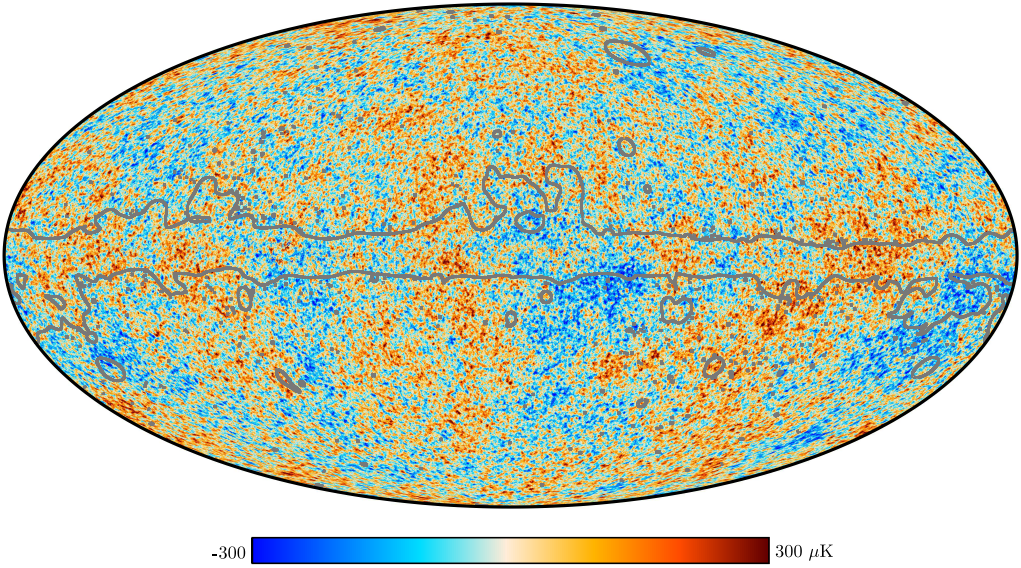}
    \caption{The 2018 \textsc{Planck} map of the temperature anisotropies of the CMB. The gray outline shows the extent of the confidence mask (\copyright ESA and the \textsc{Planck} Collaboration \cite{Planck:2018nkj}).
}
\label{fig:CMBmap}
\end{figure}

The CMB was discovered by chance in 1964. The signal was found consistent with a black-body spectrum at a temperature of about 3~K. Further investigations inaugurated by COBE \cite{Fix96} confirmed the black-body spectrum and found tiny anisotropies in an otherwise highly isotropic background. We entered the so-called era of precision cosmology thanks to the space observatories WMAP \cite{WMAP,Ben12,Hin12} and \textsc{Planck} \cite{Planck:2018nkj}, the latter having reached the highest sensitivity as far as temperature mapping is concerned (Fig.~\ref{fig:CMBmap}).

\subsection{Temperature spectrum}

In order to describe the temperature fluctuations of the CMB, it is convenient to introduce polar coordinates on the last-scattering surface $S^2$. We can, then, decompose the fluctuations in spherical harmonics $Y_{\ell m}(\vartheta, \varphi)$ defined as
\begin{equation}
    Y_{\ell m}(\vartheta, \varphi) =(-1)^{\frac{m+|m|}{2}} \sqrt{\frac{2 \ell+1}{4 \pi} \frac{(\ell-|m|) !}{(\ell+|m|) !}} P_{\ell|m|}(\cos \vartheta)\, \mathrm{e}^{\mathrm{i} m \varphi},
\end{equation}
where the $P_{\ell|m|}(y)$ are the associated Legendre polynomials of degree $\ell$. In particular, $\ell$ is called multipole moment of the expansion. The temperature fluctuation (with respect to the mean value $T_0$) observed in the direction of the unit vector $\mathbf{e}$ whose tip lies on the sphere $S^2$ of comoving radius $\chi$ centered at the observer is given by
\begin{equation}
    \frac{\delta T(\mathbf{e} \chi)}{T_{0}} =\sum_{\ell=0}^{+\infty} \sum_{m=-\ell}^{\ell} a_{\ell m} Y_{\ell m}(\vartheta, \varphi),
\end{equation}
where $a_{\ell m}$ are the coefficients of the expansion. Since ${\delta T(\mathbf{e} \chi)}/{T_{0}}$ is a real quantity, using the properties of spherical harmonics, one can easily prove that $a_{\ell 0}$ is real. Let us assume that the anisotropies come from a random process with a Gaussian distribution (the inflationary paradigm predicts this). Under these assumptions, the real and imaginary part of the $a_{\ell m}$ coefficients are independent. Now consider for each $\ell$ the $(2\ell+1)$-dimensional set
\begin{equation}
    \left\{\alpha_{\ell, \bar{m}} \mid \bar{m}=1,2, \ldots, 2 \ell+1\right\}:=\left\{a_{\ell 0}, \sqrt{2} \operatorname{Re}\left(a_{\ell m}\right), \sqrt{2} \operatorname{Im}\left(a_{\ell m}\right) \mid 1 \leqslant m \leqslant \ell\right\}.
\end{equation}
We can then define for each $\ell$ the temperature angular spectrum $C_{\ell}^{T T}$ as the square of the half width of the Gaussian distribution
\begin{equation}\label{eq:gauss}
    f_{\ell}\left[\alpha_{\ell, \bar{m}}\right]=\frac{1}{\sqrt{2 \pi \sigma_{\ell}^{2}}} \mathrm{e}^{-\frac{\alpha_{\ell, \bar{m}}^{2}}{2 \sigma_{\ell}^{2}}},
\end{equation}
where $\sigma_{\ell}^2$ is the variance of the distribution $f_{\ell}$ defined as
\begin{equation}
    \sigma_{\ell}^{2}:=C_{\ell}^{T T}=\frac{1}{(2 \ell+1)}\left\langle a_{\ell}^{2}\right\rangle , \qquad   
    a_{\ell}^{2}:=\sum_{m=-\ell}^{\ell}\left|a_{\ell m}\right|^{2},
\end{equation}
and $\left\langle a_{\ell}^{2}\right\rangle$ is the average over an ensemble of skies, i.e., a collection of skies observed at different points in the universe for a given $\ell$. The temperature angular spectrum is related to the primordial spectrum $\mathcal{P}_{\delta T}(k)$ (the two-point correlation function of temperature fluctuations in comoving momentum space) by the following integration:
\begin{equation}
    C_{\ell}^{T T}=4 \pi \int_{0}^{+\infty} \frac{\mathrm{d} k}{k} \mathcal{P}_{\delta T}(k) j_{\ell}^{2}\left(k \tau_{0}\right),
\end{equation}
where $j_{\ell}$ is the spherical Bessel function of degree $\ell$. Since, for the time being, we can only observe the universe from one point, we cannot average over all skies. Let us assume the ergodic hypothesis, which implies that the average over the ensemble of all possible skies is equivalent to spatial averages over one sky
\begin{equation}
    \left\langle {\cal O}\right\rangle=\left\langle {\cal O}\right\rangle_{\rm sky}:=\frac{1}{4\pi}\int_{S^2} {\rm d}\Omega _2\, {\cal O}.
\end{equation}
The ergodic hypothesis only works for random fields with a continuous spectrum, while the angular spectrum $C_{\ell}^{T T}$ is labeled on a discrete set. For this reason, our assumption implies a theoretical error that can be estimated. For small angular scales, we can average over a large number of pairs of independent directions separated by the same angle. In fact, for $\ell\gg 1$, we have many modes labeled by $m$, so that the theoretical error will be smaller. On the contrary, at large scales, we have a large intrinsic error. This effect is called cosmic variance and it prevents us from performing complete measurements of our theoretical quantities. The cosmic variance can be computed as
\begin{equation}\label{eq:cosmicVariance}
    \sigma ^2 _{C_{\ell}}:=\left\langle (C_{\ell}^{\rm obs}-C_{\ell})^2\right\rangle=\left\langle C_{\ell}^{{\rm obs}\,2}\right\rangle-C_{\ell}^2=\frac{2}{2\ell+1}C_{\ell}^2.
\end{equation}
From equation (\ref{eq:cosmicVariance}), we see that the higher the multipole moment $\ell$, the lower the cosmic variance.
%
%
This means that, if we wanted to use CMB data to test a cosmological model alternative to the standard one and this model predicted deviations at low multipoles, such deviations should be larger than cosmic variance.

From the CMB data, we can extract important information to estimate the cosmological parameters of models of the early universe. We briefly illustrate the necessary steps. First of all, from our instruments, one gets a set of time-ordered data that can be compressed into sky maps at different frequencies to minimize the noise. Then, one extracts a single map, from which one reconstructs the angular power spectrum. Lastly, one can constrain the cosmological parameters from the power spectrum features.

\begin{figure}[!htb]
    \centering
    \includegraphics[width=0.9\linewidth]{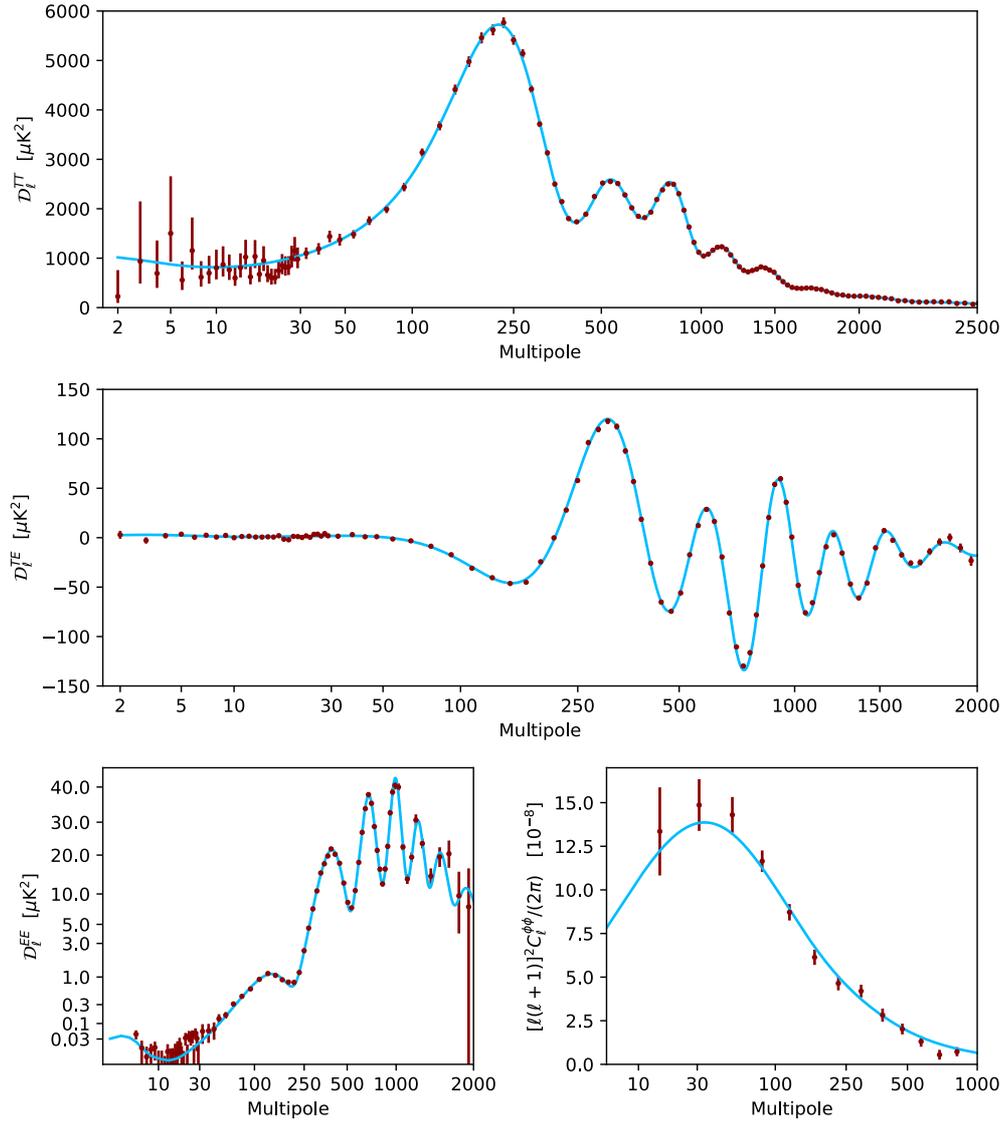}
    \caption{The 2018 \textsc{Planck} CMB power spectra ${\cal D}_\ell=\ell(\ell+1)C_\ell/(2\pi)$ for temperature ($TT$, top), the temperature-polarization cross-spectrum ($TE$, middle), the E mode of polarization ($EE$, bottom left) and the lensing potential ($\phi\phi$ bottom right). Within $\Lambda$CDM these spectra contain the majority of the cosmological information available from \textsc{Planck} and the blue lines show the best-fit model. The uncertainties of the $TT$ spectrum are dominated by sampling variance at all scales below about $\ell=1800$, a scale at which the CMB information is essentially exhausted within the framework of the $\Lambda$CDM model. The $TE$ spectrum is about as constraining as the $TT$ one, while the $EE$ spectrum still has a sizeable contribution from noise. The anisotropy power spectra are plotted here with a multipole axis that goes smoothly from logarithmic at low $\ell$ to linear at high $\ell$ (\copyright ESA and the \textsc{Planck} Collaboration \cite{Planck:2018nkj}).
}
\label{fig:CMBSpectrum}
\end{figure}

Let us roughly sketch the main features of the observed spectrum reported in the top panel of Fig.~\ref{fig:CMBSpectrum}.
At low multipoles ($\ell \ll 100$), the spectrum shows a plateau behavior. The plateau is caused by primordial scalar perturbations of the metric affecting gravitational potential experienced by photons traveling from the last-scattering surface to the observer. This is called the Sachs--Wolfe effect and it dominates the CMB anisotropies at low multipoles. It is possible to show that the asymptotic power spectrum in this regime takes the power-law form
\begin{equation}
    \mathcal{P}_{\delta T}(k)\simeq A(\tau_0k)^{n} .
\end{equation}
If the primordial spectrum is scale invariant (Harrison--Zel'dovich spectrum) or almost scale invariant, which means $n=0$ or $n\approx 0$, then we can explain the plateau. However, why should we have an almost scale-invariant primordial spectrum? This condition poses a fine-tuning problem which is solved in the context of inflation.

Other features to consider are those related to acoustic peaks. Because of density fluctuations, the baryonic matter distribution presents potential wells. Photons coming from the last-scattering surface go through these wells and the interaction can be described via a harmonic-oscillator model. This process gives rise to acoustic peaks in the power spectrum. 

\begin{figure}[!htb]
    \centering
    \includegraphics[width=0.9\linewidth]{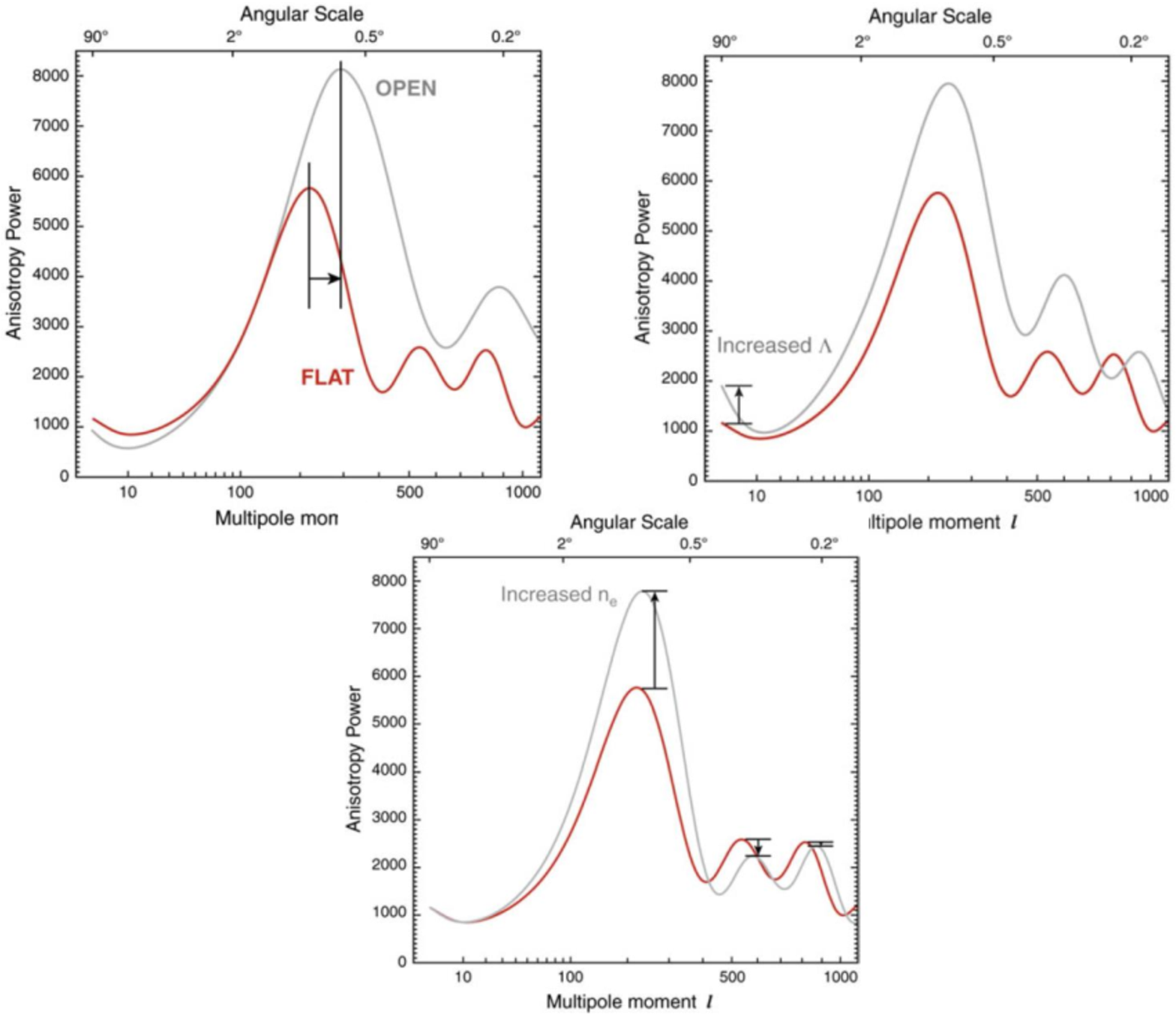}
    \caption{Dependence of the temperature power spectrum on different cosmological parameters. Setting $\textsc{k}=-1$, one gets higher amplitudes of the acoustic peaks and the whole spectrum is shifted to the right. Similarly, if we increase $\Lambda$ the first two peaks gets higher. Lastly, also the value of the baryon density $n_{\rm e}$ is strongly constrained by the first peak amplitude (Credit: NASA/WMAP Science Team \cite{WMAPc4}).
}
    \label{fig:changedCMBSpectrum}
\end{figure}

Amplitude, number and position of the peaks are strongly dependent on the cosmological parameters (Fig.~\ref{fig:changedCMBSpectrum}).
For instance, if we move from the flat spatial model to the open-universe case, we get a higher amplitude of the first peak and a spectrum shift to the right. Moreover, the pattern and size of hot and cold spots in the CMB map also depends on spatial curvature. For a closed universe, we would see a thick-grain structure. These considerations allow us to conclude that observations are consistent with a spatially flat universe. From these examples, the reader can get an intuitive notion of how much CMB data can constrain alternative cosmological models. 

\begin{figure}
\centering
\includegraphics[width=10cm]{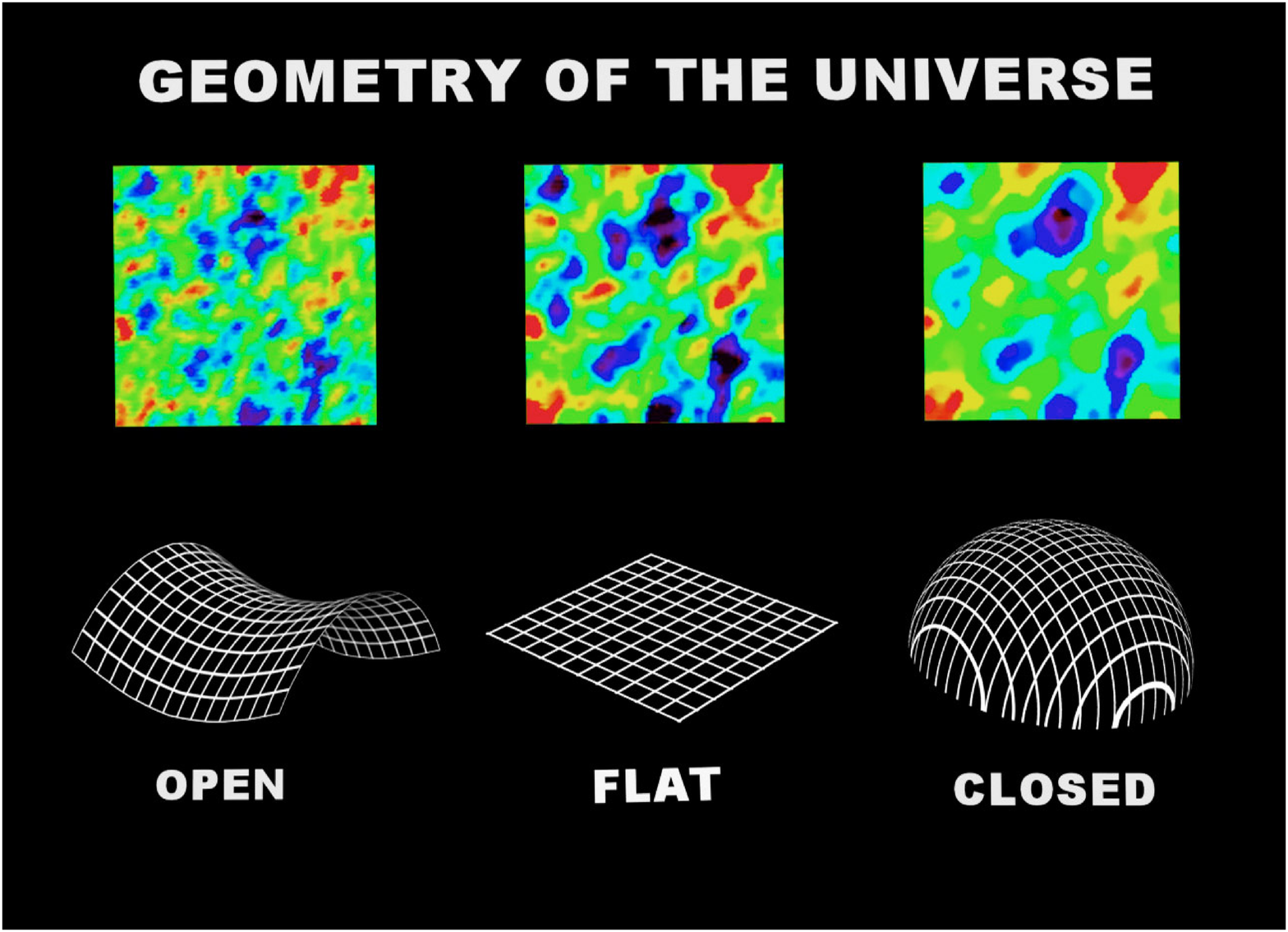}
\caption{\label{fig:curva} Anisotropy distortion from geometry. Simulated CMB features are characterized by, respectively, larger/smaller angular scales in a closed/open universe (Credit: NASA/WMAP Science Team \cite{WMAPc4}).}
\end{figure}

The effect of the curvature of the universe is shown in simulated patches of the CMB map in Fig.~\ref{fig:curva}. A non-zero curvature would bend the CMB photon geodesics and make the observer see a more coarse-grained or fine-grained pattern of hot and cold temperature spots. Observations favour the flat universe.


\subsection{Scalar and tensor spectra}

We saw that we need an almost scale-invariant primordial spectrum of scalar perturbations to explain the low-multipole plateau in the CMB power spectrum. To be more quantitative, we now introduce some parameters to describe both primordial scalar and tensor spectra. We recall that the scalar spectrum is the two-point function of the curvature perturbation on uniform density hypersurfaces $\zeta$,
\begin{equation}
    \mathcal{P}_{\mathrm{s}}(k):=\mathcal{P}_{\zeta}(k).
\end{equation}
The scalar spectral index is defined as the logarithmic variation of $\mathcal{P}_{\mathrm{s}}$ with respect to the comoving wavenumber $k$,
\begin{equation}
    n_{\mathrm{s}}-1:=\frac{\mathrm{d} \ln \mathcal{P}_{\mathrm{s}}}{\mathrm{d} \ln k}.
\end{equation}
For a power-law spectrum $n_{\mathrm{s}}$ is constant and, in particular, $n_{\mathrm{s}}=1$ if the spectrum is scale invariant. However, if we want to account for deviations from the power-law behaviour we can introduce the running of the spectral index and the parametrization
\begin{equation}
    \alpha_{\mathrm{s}}:=\frac{\mathrm{d} n_{\mathrm{s}}}{\mathrm{d} \ln k}\qquad \Longrightarrow\qquad \mathcal{P}_{\mathrm{s}}(k)=A_{\mathrm{s}}\left(\frac{k}{k_{0}}\right)^{n_{\mathrm{s}}\left(k_{0}\right)-1+\frac{1}{2} \ln \left(\frac{k}{k_{0}}\right) \alpha_{\mathrm{s}}\left(k_{0}\right)},
\end{equation}
where $A_{\mathrm{s}}$ is the amplitude parameter and $k_0$ is the pivot scale, which depends on the features of the experiment ($k_0=0.05\,{\rm Mpc}^{-1}$ for the \textsc{Planck} experiment). In the case of the tensor spectrum, we can proceed analogously. Recall that we now have two independent polarization modes,
\begin{equation}
    \mathcal{P}_{\mathrm{t}}(k):=\sum_{\lambda=+, \times} \mathcal{P}_{h_{\lambda}}(k)=2 \mathcal{P}_{h_{\lambda}}(k).
\end{equation}
In the tensor case, the spectral index is defined as
\begin{equation}
    n_{\mathrm{t}}:=\frac{\mathrm{d} \ln \mathcal{P}_{\mathrm{t}}}{\mathrm{d} \ln k}.
\end{equation}
The spectrum is scale invariant when $n_\mathrm{t}=0$. In principle, we could also define a tensor running index, but we can neglect it according to observations, if we limit the discussion to CMB scales. We then parametrize the spectrum in the following way:
\begin{equation}
    \mathcal{P}_{\mathrm{t}}(k)=A_{\mathrm{t}}\left(\frac{k}{k_{0}}\right)^{n_{\mathrm{t}}\left(k_{0}\right)}.
\end{equation}
However, instead of using the amplitude $A_{\mathrm{t}}$ as a parameter, one typically defines the tensor-to-scalar ratio
\begin{equation}
    r:=\frac{A_{\mathrm{t}}}{A_{\mathrm{s}}}.
\end{equation}
From observations, we know that the scalar spectrum is deviating by $8\sigma$ from the Harrison–Zel'dovich spectrum \cite{Planck:2018vyg}:
\begin{equation}\label{nsobs}
    n_{\mathrm{s}}=0.9649 \pm 0.0042\qquad {\rm (68\%\,CL)}\,,
\end{equation}
assuming $\alpha_{\rm s}=0$. Although we have not observed yet the primordial tensor spectrum, there is an upper bound on the tensor-to-scalar ratio at $k_0=0.05\,{\rm Mpc}^{-1}$ \cite{BICEP:2021xfz}:
\begin{equation}\label{eq:rup}
    r<0.036\qquad {\rm (95\%\,CL)}\,.
\end{equation}
This small value of $r$ has already ruled out many inflationary models.
%
%
%
\section{Inflation}
\subsection{Problems of hot big bang}

The hot big-bang model can explain experimental data using a reduced number of parameters, about $O(10)$. However, this approach presents the following key issues:
\begin{itemize}
    \item \textit{Flatness problem.} 
    Why is the universe so close to flatness? We can rephrase the flatness problem as an initial condition problem. It is simple to prove that, in a decelerating universe originally filled with radiation and matter, $\Omega_\textsc{k}(z)$ decreases with the redshift, so that it becomes large at early times. In contrast, according to the current constraints on $\Omega_{\textsc{k},0}$, the density parameter associated to the curvature should have been of order of $10^{-64}$ when the universe had the Planck energy density. This small value implies extreme fine-tuning.

    \item \textit{Horizon problem.} 
    Why do we have such a strong correlation between distant regions of the last-scattering surface?
    Viewed from Earth, the horizon at last scattering subtends angles ${\sim} 1.3^{\circ}$. Thus, if the universe had been decelerating from the beginning, two regions of the sky separated by larger angular distances had never been in causal contact when matter and radiation decoupled. Nevertheless, as we know, temperature fluctuations of the CMB are tiny everywhere in the sky vault, ${\delta T}/{T}_0 \sim 10^{-5}$. This observation would require many disconnected regions to have the same initial conditions. Once again, this discrepancy translates into a fine-tuning problem.

    \item \textit{Primordial seeds problem.} What originated the anisotropies?
    We saw in section \ref{sec:3} how we could explain the CMB spectrum in terms of a primordial scalar spectrum that is almost scale-invariant. Nevertheless, we have simply assumed that primordial scalar and tensor perturbations exist. We have offered no mechanism to explain where they originated from or why scalar perturbations are dominant compared to the tensor ones.
\end{itemize}
\noindent
Inflation can solve all these problems.

\subsection{Inflation: kinematics}

We can define inflation as any phase of accelerated expansion:
\begin{equation}
    \ddot{a} > 0\,.
\end{equation}
This condition is purely kinematic, since it does not say anything about the origin of acceleration. As we already know, we can give an equivalent description in terms of the first slow-roll parameter $\epsilon$ or of the cosmic fluid equation of state. In GR and in the presence of only one fluid,
\begin{equation}
 \ddot{a} > 0\qquad \Longleftrightarrow\qquad \epsilon < 1\qquad \Longleftrightarrow\qquad  P < - \frac{1}{3} \rho\, .
\end{equation}
A prototypical background for inflation is de Sitter spacetime. Since in this universe the Hubble parameter is constant, the scale factor grows exponentially and the first slow-roll parameter is equal to zero at any time,
\begin{equation}
    \quad H(t) = H\qquad \Longrightarrow\qquad   a(t) = {\rm e}^{Ht}\qquad \Longrightarrow   \qquad \epsilon(t) = 0\, .
\end{equation}
A quasi-de Sitter acceleration is driven by a cosmological-constant-like fluid with $P\simeq-\rho$.
\subsection{Solution to the flatness problem}

\begin{figure}[!htbp]
    \centering
    \includegraphics[width=0.75\linewidth]{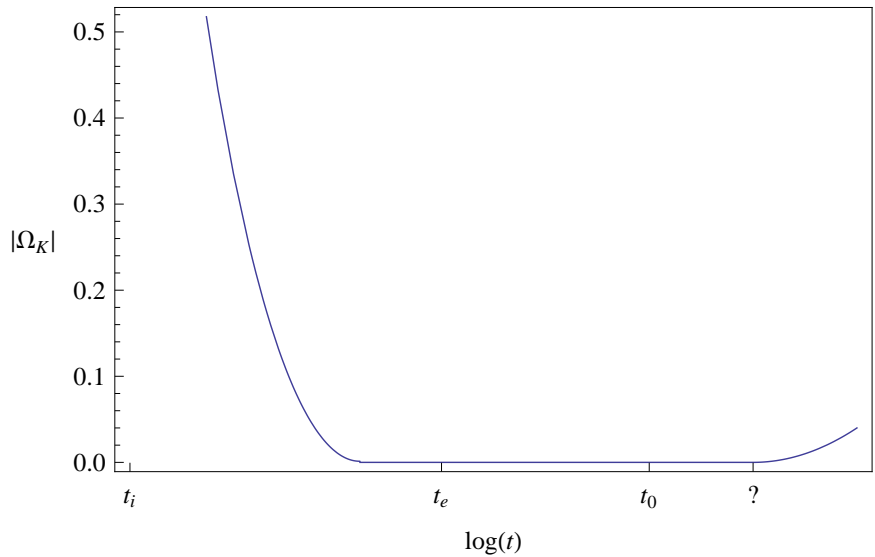}
    \caption{Time evolution of the curvature density parameter. $|\Omega_\textsc{k}|$ decreases almost exponentially until the end of inflation at $t=t_{\rm e}$, then it starts to grow slowly. At the present time $t_0$, we still have $|\Omega_\textsc{k}|\approx 0$. Reprinted with permission from \cite{Calcagni:2017sdq}, \copyright 2017 Springer International Publishing Switzerland.}
    \label{fig:curvatureEvolution}
\end{figure}

We want to prove that an accelerated expansion phase in the very early universe can solve the flatness and horizon problem. Recall the definition of $\Omega_\textsc{k}$:
\begin{equation}
    \left| \Omega_\textsc{k} \right|=\frac{\left| \textsc{k} \right|}{(aH)^2}= \frac{\left| \textsc{k} \right|R_H^2}{a^2} =\left| \textsc{k} \right| r^2_H  ,
\end{equation}
where we introduce the comoving Hubble radius $r_H=R_H/a$. For an accelerating universe, the Hubble radius $R_H$ is a reasonable estimate of the particle horizon $R_{\rm p}$. We can describe the behaviour in time of the comoving Hubble radius in terms of the first slow-roll parameter,
\begin{equation}
   \dot{r}_H = - \ddot{a}r^2_H = - \frac{1}{a}(1- \epsilon )\qquad \Longrightarrow\qquad \dot{r}_H< 0\qquad \Longleftrightarrow\qquad \epsilon<1 .
\end{equation}
The time evolution of the curvature density parameters is shown on Fig.~\ref{fig:curvatureEvolution}. During matter and radiation domination, $r_H$ grows in time and so does $|\Omega_\textsc{k}|$. However, if we have acceleration ($\epsilon<1$), $|\Omega_\textsc{k}|$ decreases and, if inflation lasts long enough, it will take a very long time for $|\Omega_\textsc{k}|$ to grow again. As a consequence, we could start with an arbitrarily large curvature component, the initial value would decrease (almost) exponentially and today we would still observe a very small curvature term.

\subsection{Solution to the horizon problem}

\begin{figure}[H] 
\centering
\includegraphics[height=5cm]{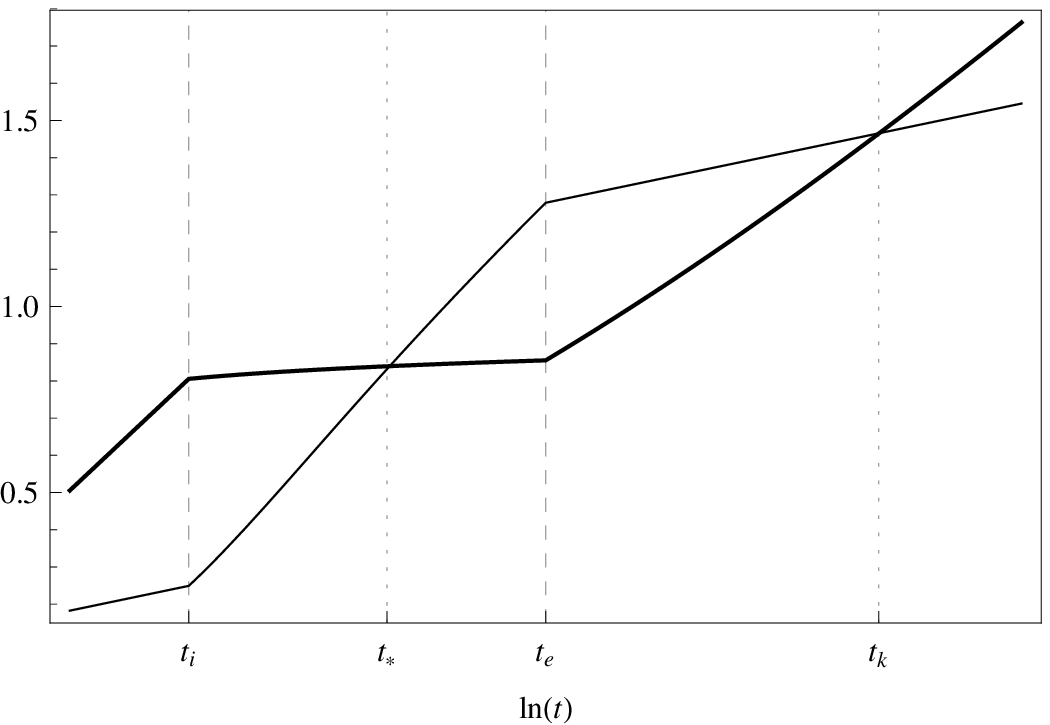}\hspace{.6cm}\includegraphics[height=5cm]{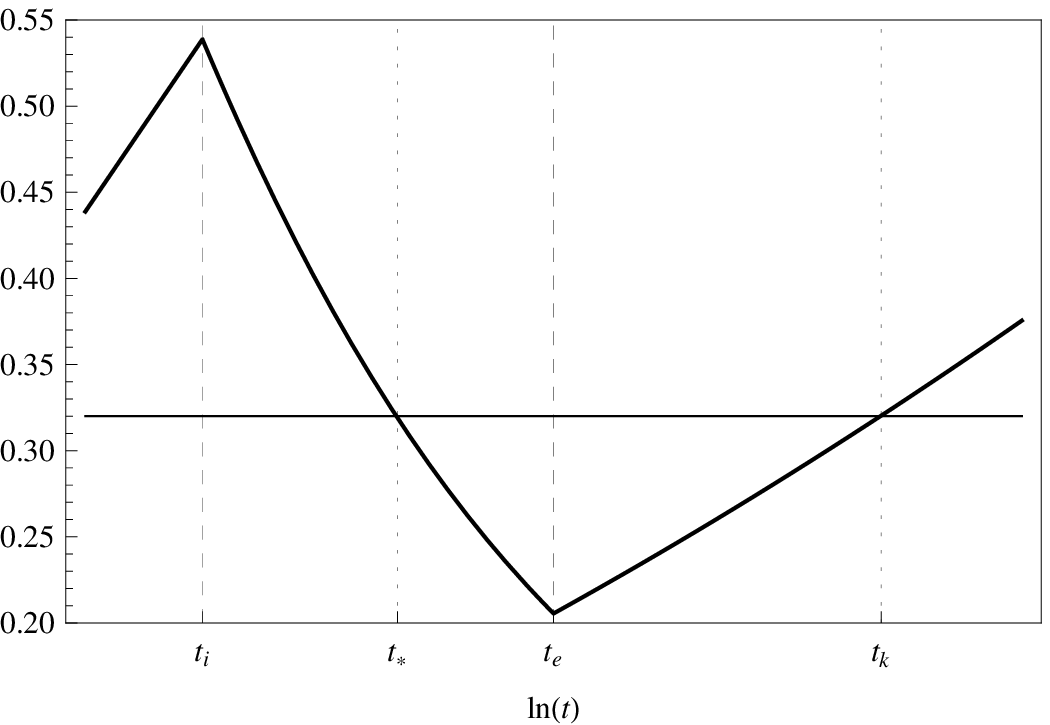}
\caption{Left: the thin line is the proper wave-length $\lambda = a(t)\,\lambda_{\rm com}$ of scalar perturbations. It grows proportionally to the scale factor. The thick line is the Hubble radius $R_H$. During inflation, the Hubble radius is almost constant and the mode exits the horizon at $t=t_*$, to reenter it a a later time $t=t_k$. Right: the thin line is the comoving wave-length $\lambda_{\rm com}$ of scalar perturbations, which is constant in time. The thick line is the comoving Hubble radius $r_H=R_H/a$. Introducing an early inflationary phase, the comoving Hubble radius is shrunk below the comoving wave-length, while it increases during the post-inflationary decelerated expansion until encompassing the comoving wave-length again. Reprinted with permission from \cite{Calcagni:2017sdq}, \copyright 2017 Springer International Publishing Switzerland.}\label{fig:comoving}
\end{figure}

Consider the proper wave-lengths of scalar perturbations. Since these wave-lengths are proportional to the scale factor, they always grow in time with a time-dependent rate. During radiation and matter domination, as we saw, the Hubble radius grows too, but at a larger rate. A mode crossing the horizon at the present time is never below the horizon during radiation and matter domination. However, the Hubble radius is (almost) constant during inflation, while the accelerated expansion exponentially stretches the proper wavelengths. Again, requiring that inflation lasts for long enough, we reach back in time a point at which the mode exits the horizon (see Fig.~\ref{fig:comoving}a). Perturbations are pushed outside the horizon and, when inflation ends, the Hubble radius starts to grow again until catching up. Perturbations entering the horizon today were already inside the casual patch during inflation before horizon exit.

Analogously, we can look at the comoving picture that is shown in Fig.~\ref{fig:comoving}b. Comoving wave-lengths are constant, while the comoving Hubble radius increases during the decelerated expansion and decreases during the accelerated expansion. Inflation shrinks the comoving Hubble radius below a given wave-length. Then, during radiation and matter domination, the comoving Hubble radius grows and includes the wave-length again.

The causal patch in the CMB can be seen at larger separations not because we are violating causality but because those regions were in causal contact at very early times.

\subsection{Inflation: dynamics}

In order to face the primordial seeds problem, we need to approach inflation from the dynamical viewpoint and answer the question of what drives inflation. The simplest candidate is a real scalar field $\phi$ slowly rolling over its potential $V(\phi)$. We saw that a scalar field behaves like a perfect fluid of barotropic index
\begin{equation}
    w_{\phi}=\frac{P_{\phi}}{\rho_{\phi}}=\frac{\frac{1}{2} \dot{\phi}^{2}-V(\phi)}{\frac{1}{2} \dot{\phi}^{2}+V(\phi)}.
\end{equation}
The extreme slow-roll condition reads
\begin{equation}
    \dot{\phi}^{2} \ll V,
\end{equation}
so that the kinetic term is subdominant with respect to the potential and with $w_{\phi}\approx-1$ we reproduce the equation of state for the cosmological constant, which determines a quasi-de Sitter expansion. The field $\phi$ is called the inflaton. 

According to this cold big bang model, at its beginning the universe is only filled with the inflaton field and undergoes an acceleration phase. After reaching a potential well at some point, the field starts to roll quickly and it decays in ultra-relativistic particles giving rise to a very hot plasma, thus recovering the evolution of the hot big bang model. This last phase is called reheating. From observational constraints, we know that it takes place at temperatures of order $10^{16}$ GeV (Grand Unification scales)
\begin{equation}
    T_{\mathrm{reh}} \simeq\left(\frac{30}{\pi^{2}} \frac{\rho_{\mathrm{reh}}}{106.75}\right)^{\frac{1}{4}}<\left(\frac{30}{\pi^{2}} \frac{3 H_{*}^{2}}{106.75 \kappa^{2}}\right)^{\frac{1}{4}}<7.6 \times 10^{15} \mathrm{GeV},
\end{equation}
where $\rho_\mathrm{reh}$ is the energy density at reheating and $H_*$ is the Hubble parameter at horizon exit, i.e., before reheating. 

The duration of inflation is a crucial quantity if we want to solve the horizon and the flatness problem. We can quantify it in terms of the number of e-foldings. There are two possible definitions of this number. We can define the total time measured from the beginning until the end of inflation or from a point when a perturbation mode crosses (exits) the horizon until inflation ends. In general, it is convenient to use the second definition, which involves the values of the co-moving Hubble radius at horizon crossing and at the end of inflation:
\begin{equation}
    \mathcal{N}_{k}:=\mathcal{N}_{\mathrm{e}}-\mathcal{N}_{*}:=\ln \frac{\tau_{*}}{\tau_{\mathrm{e}}} \simeq 52-\ln \left(k \tau_{0}\right)-\frac{2}{3} \ln \frac{10^{15} \mathrm{GeV}}{\rho_{\mathrm{e}}^{1 / 4}}-\frac{1}{3} \ln \frac{10^{10} \mathrm{GeV}}{T_{\text {reh }}}.
\end{equation}
This expression depends on the reheating temperature. Typically, the duration is assumed to be at least 50 e-foldings, which is the minimum value required to solve the flatness problem. 

We can now consider the condensed timeline of the universe thermal history reported in Tab.~\ref{tab:uniTimeline2}, including this very early phase, where we use as a relational time the relative dimensions of the universe with respect to present time. 
\begin{table}[H]
    \centering
    \begin{tabular}{|l|l|l|l|l|}
    \hline$\rho^{1 / 4}~(\mathrm{GeV})$ & $T~(\mathrm{GeV})$ & $t~(\mathrm{s})$ & $r_{H}\left(t_{0}\right) / r_{H}(t)$ & Event \\
    \hline$\sim 10^{19}$ & $\sim 0$ & $10^{-44}$ & $>10^{52}$ & Cold big bang? Inflation begins at Planck scale? \\
    \hline$\sim 10^{15}$ & $\sim 0$ & $10^{-36}$ & $10^{24}$ & Inflation ends at GUT scale? Reheating begins? \\
    \hline & $\sim 10^{10}$ & $10^{-7}$ & $10^{22}$ & Reheating ends? \\
    & $10^{-2}$ & $10^{-2}$ & $10^{10}$ & Latest end of cold big bang model \\
    \hline & $10^{-5}$ & 200 & $10^{8}$ & BBN \\
    & $10^{-9}$ & $10^{12}$ & 100 & Radiation-matter equality  \\
    & $10^{-13}$ & $10^{17}$ & 1 & Today \\
    \hline
    \end{tabular}
    \caption{Condensed timeline of the universe thermal history including inflation. The fourth column reports the relative comoving size of the universe with respect to present time, which can be used as a relational time. Events associated with inflation come with question marks. Specifically, the typical energy scales are strongly model-dependent.}
    \label{tab:uniTimeline2}
\end{table}
The radiation-matter equality occurred when the universe was about 100 times smaller than today; the BBN, when it was about $10^8$ times smaller. We know from the BBN constraints that the latest time when inflation could end is about $10^{-2}$ seconds. The very early phase is strongly model-dependent. In particular, we expect the universe to start with zero temperature and very high energy, although the scale of inflation itself depends on the particle physics model we use.

Despite all these question marks, inflation is essential to explain what we observe today, so that what we might call the standard cosmological model has to include inflation. For the moment, there are no alternative models capable of explaining as many data as inflation.

\subsection{Slow-roll parameters}

We saw that we can describe acceleration using the first slow-roll parameter $\epsilon$. Analogously, the second slow-roll parameter $\eta$ is useful when we want to account for the duration of the accelerated expansion phase. In the context of inflation, we can introduce the slow-roll parameters in two ways. One is in terms of the Hubble parameter and the time derivatives of the field,
\begin{alignat}{-1} \label{eq.slow-roll}
\begin{split}
    \epsilon &:= \frac{3\dot{\phi}^2}{\dot{\phi}^2+2V} ,
    \\ 
    \eta &:=-\frac{\mathrm{d} \ln \dot{\phi}}{\mathrm{d} \ln a} =-\frac{\ddot{\phi}}{H \dot{\phi}} .
\end{split}
\end{alignat}
The second way, more convenient when we know the details of the inflaton dynamics, is in terms of the potential,
\begin{alignat}{-1} \label{eq.slow-roll_V}
\begin{split}
    \epsilon_V &:=\frac{M_{\mathrm{Pl}}^2}{2}\left(\frac{V_{, \phi}}{V}\right)^{2} , \\ 
    \eta_V &:=M_{\mathrm{Pl}}^2 \frac{V_{, \phi \phi}}{V} .
\end{split}
\end{alignat}
During inflation, the slow-roll parameters take very small values and are almost constant, since their time derivatives are of second order in the parameters themselves:
\begin{equation}
    \epsilon \ll 1, \qquad |\eta| \ll 1, \qquad
    \dot{\epsilon}=2 H \epsilon(\epsilon-\eta), \qquad \dot{\eta}=H\left(\epsilon \eta-\epsilon^{2}\right).
\end{equation}
The slow-roll parameters define a so-called ``tower'' because we have leading order terms $\epsilon$ and $\eta$ and, then, we can define higher-order parameters that are progressively smaller. In this approximation, we can relate the definitions in equations (\ref{eq.slow-roll}) and (\ref{eq.slow-roll_V}) as
\begin{equation}
    \epsilon \simeq \epsilon_{V} \quad  \mathrm{and}  \quad \eta \simeq \eta_{V}-\epsilon_{V}.
\end{equation}
Lastly, the number of e-folds is given by the integral
\begin{equation}
    \mathcal{N}_{k} \simeq \operatorname{sgn}(\dot{\phi}) \frac{\sqrt{2}}{M_{\mathrm{Pl}}} \int_{\phi_{*}}^{\phi_{\mathrm{e}}} \frac{\mathrm{d} \phi}{\sqrt{\epsilon(\phi)}}\,,
\end{equation}
where $\phi_{*}$ is the value of the inflaton at the time the perturbation with comoving wavenumber $k$ exits the horizon. Once a model is chosen, one can compute all these quantities.

\subsubsection{Example 1: monomial potential}

As a first example, consider a monomial potential, a simple power law in the form
\begin{equation}
    V(\phi)=\frac{\sigma_{n}}{n} \phi^{n}; \quad \sigma_{n}>0,
\end{equation}
with $n=2,3,4,\dots$. The first two slow-roll parameters are
\begin{alignat}{-1}
    \begin{split}
    \epsilon &= \frac{n^2}{2}\frac{M_{\mathrm{Pl}}^2}{\phi^2}, \\
    \eta &= \frac{n(n-2)}{2}\frac{M_{\mathrm{Pl}}^2}{\phi^2}.
    \end{split}
\end{alignat}
Further, we can calculate the value of the field at the beginning and the end of inflation as
\begin{alignat}{-1}
    \phi_{\mathrm{i}}^{2} &= \left[\frac{n(8\pi)^2 M_{\mathrm{Pl}}^{4}}{\sigma_{n}}\right]^{\frac{2}{n}},
\intertext{and}
    \phi_{\mathrm{e}}^{2} &= \frac{n^{2}}{2} M_{\mathrm{Pl}}^{2} \ll \phi_{\mathrm{i}}^{2},
\end{alignat}
respectively. Lastly, we can integrate to get $\phi_*$ as a function of the number of e-folds,
\begin{equation}
    \phi_*^2=\frac{n(4\mathcal{N}_{k}+n)}{2}M_{\rm Pl}^2.
\end{equation}
Therefore, we can fix the number of e-folds to some value (typical or minimum), then use it to calculate the slow-roll parameters. In this way, we get a testable prediction since, as we will see, the slow-roll parameters enter the primordial power spectra.

\subsubsection{Example 2: natural inflation}

Natural inflation gives us another interesting example, which is motivated by a symmetry-breaking mechanism. We have a periodic potential in the form
\begin{equation}
    V(\phi)=\frac{M^{4}}{2}\left(1+\cos \frac{\phi}{f}\right),
\end{equation}
where $f$ and $M$ are mass scales, and again we can calculate the slow parameters
\begin{alignat}{-1}
    \epsilon &\simeq \frac{1}{8} \frac{M_{\mathrm{Pl}}^{2}}{f^{2}}\left(\frac{\phi}{f}\right)^{2}, \\
    \eta &\simeq-\frac{1}{2} \frac{M_{\mathrm{Pl}}^{2}}{f^{2}}.
\end{alignat}
We notice that $|\eta| \gg \epsilon$ is the main contribution to the power spectra. The values of the field at the beginning and the end of inflation are given by
\begin{alignat}{-1}
    \phi_{\mathrm{i}} &\simeq\left(1-64 \pi^{2} \frac{M_{\mathrm{Pl}}^{4}}{M^{4}}\right)^{\frac{1}{2}} 2 f, \\
    \phi_{\mathrm{e}} &\simeq \sqrt{8} \frac{f^{2}}{M_{\mathrm{Pl}}},
\end{alignat}
respectively. Then,
\begin{equation}
    \phi_{*} \simeq \sqrt{8} \frac{f^{2}}{M_{\mathrm{Pl}}} \exp \left(-\frac{M_{\mathrm{Pl}}^{2}}{f^{2}} \frac{\mathcal{N}_{k}}{2}\right).
\end{equation}


\subsection{Solution of the primordial seeds problem}

We can now discuss the origin of the temperature fluctuations in the CMB. Cosmology is the area where particle physics and gravity meet. In particular, when studying inflation, we are doing quantum cosmology because a deeper understanding of the inflaton fluctuations requires quantum mechanics. By a unique \emph{coincidentia oppositorum}, a coincidence of opposites, the macrocosm turns out to be a window on the microcosm of quantum fields. In fact, it is a quantum mechanism to solve the primordial seeds problem. The inflaton is a scalar field. Therefore, it behaves as a quantum object and undergoes quantum fluctuations around its vacuum expectation value (VEV). These fluctuations generate scalar, vector and tensor fluctuations in the background metric and, if they are small enough, they are compatible with the expected primordial scalar and tensor spectra. In particular, the classical field $\phi(\tau)$ coincides with the vacuum expectation value of the quantum field $\hat{\phi}(\tau, \mathbf{x})$
\begin{equation}
    \langle\hat{\phi}(\tau, \mathbf{x})\rangle:=\langle 0|\hat{\phi}(\tau, \mathbf{x})| 0\rangle=: \phi(\tau) ,
\end{equation}
where we assume the vacuum state $|0\rangle$ to be spatially homogeneous and isotropic so that the VEV only depends on conformal time $\tau$.
We can then decompose the quantum field into a classical part (the VEV) plus a small quantum fluctuation,
\begin{equation}
    \hat{\phi}(\tau, \mathbf{x})=\phi(\tau)+\delta \hat{\phi}(\tau, \mathbf{x}) .
\end{equation}
For convenience, we introduce a rescaled field $\hat{u}(\tau, \mathbf{x})=f(\tau) \delta \hat{\phi}(\tau, \mathbf{x})$, as we did for the metric perturbations. We can then decompose the new quantum field in terms of creation and annihilation operators,
\begin{equation}
    \hat{u}_{\mathbf{k}}(\tau)=u_{k}(\tau) a_{k}+u_{k}^{*}(\tau) a_{-k}^{\dagger}.
\end{equation}
Since our field lives on a curved background, the vacuum is not unique and neither are the creation and annihilation operators. Therefore, we need to choose our vacuum according to some criterion. In particular, it is possible to choose a vacuum state which reduces to Minkowski spacetime asymptotically (Bunch--Davies vacuum). We then need to find the expression for the modes $u_{k}(\tau)$ and $u_{k}^{*}(\tau)$. With a proper choice of the function $f(\tau)$, our modes are solutions of the Mukhanov--Sasaki equation (\ref{eq:musa}) for a certain background-dependent mass $M(\tau)$. For small values of $k\tau$, the modes freeze when leaving the horizon. The primordial power spectrum is nothing but the two-point correlation function evaluated at horizon crossing,
\begin{alignat}{-1}
    \left\langle\left|\hat{u}_{\mathbf{k}}(\tau)\right|^{2}\right\rangle &:=\langle 0|\hat{u}_{\mathbf{k}} \hat{u}_{\mathbf{k}}^{\dagger}| 0\rangle=\left|u_{k}(\tau)\right|^{2} , \\
\hat{\mathcal{P}}_{u}(k) &:=\frac{k^{3}}{2 \pi^{2}}\left|u_{k \tau \ll 1}\left(\tau_{*}\right)\right|^{2}\big|_{k=aH}.
\end{alignat}
According to the inflationary paradigm, we can explain the CMB anisotropies as originated by quantum fluctuations of the inflaton field. This is consistent with the smallness of temperature fluctuations, which amount to one part over $10^5$. 

\subsection{Inflationary spectra}

For the scalar field, the two-point correlation function in the de Sitter background is constant and it is proportional to the square of $H$,
\begin{equation}\label{eq:Pdep}
    \mathcal{P}_{\delta \phi}=\left(\frac{H}{2\pi}\right)^{2}.
\end{equation}
However, we recall from section \ref{sec:scalarpert} that the observed spectrum of scalar fluctuations is not just the spectrum of the field fluctuations but of the curvature perturbation on uniform density hypersurfaces $\zeta$ (which is a gauge-invariant combination of energy density fluctuations and scalar metric perturbations). It turns out that the scalar spectrum is equal to (\ref{eq:Pdep}) divided by the first slow-roll parameter
\begin{equation}
    \mathcal{P}_{\mathrm{s}}=\frac{1}{2M_{\rm Pl}^2} \frac{1}{\epsilon}\left(\frac{H}{2 \pi}\right)^{2}.
\end{equation}
Since in our model $\epsilon$ is very small, this quantity is quite large when compared to the tensor spectrum, which is simply given by the scalar-field spectrum normalized by the Planck mass,
\begin{equation}
    \mathcal{P}_{\mathrm{t}}=\frac{8}{M_{\rm Pl}^2}\left(\frac{H}{2 \pi}\right)^{2}.
\end{equation}
We recall that all fluctuations, when conveniently rescaled, satisfy the same Mukhanov--Sasaki equation in GR. That is why we have the same spectra apart for the normalization. Further, we can compute the scalar-to-tensor ratio, which is linear in the parameter $\epsilon$ and, therefore, very small:
\begin{equation}
    r \simeq 16 \epsilon.
\end{equation}
We have already seen that this prediction is consistent with our data. We now have a natural mechanism to explain why scalar perturbations are dominant with respect to the tensor ones. Moreover, we find that both the scalar and the tensor spectral indices are linearly proportional to the first-order slow-roll parameters:
\begin{equation}
    n_{\mathrm{s}}-1 \simeq 2 \eta-4 \epsilon, \quad n_{\mathrm{t}} \simeq-2 \epsilon.
\end{equation}
Therefore, inflation predicts an almost (but, crucially, not exactly) scale-invariant spectrum that is consistent with observations.

Furthermore, inflation also justifies the assumption we made when describing the CMB spectrum. We assumed that the distribution of the multipole coefficients $a_{\ell m}$ is Gaussian, equation (\ref{eq:gauss}). Inflation naturally predicts this property because the smallness of the perturbations allows one to perturb the Einstein equations (\ref{eq:eineq}) at the first order and stop at the linear approximation. In this case, inflationary fluctuations are Gaussian because the real and imaginary parts of the Mukhanov--Sasaki variable $u$ are linearly independent. This property is called Gaussianity. Thus, in the linear approximation, inflationary fluctuations have a Gaussian probability distribution and we can completely describe them using only the power spectrum in momentum space. Higher-order $n$-point correlation functions vanish for odd $n$, while for even $n$ they are functions of the power spectrum only. All these statistical properties are evaluated in the ensemble of spatial points in the sky vault. However, this is not the end of the story, since the Gaussianity assumption only works at the linear level. At second order in cosmological perturbations, due to self-interactions inflation also predicts a small non-Gaussian component, which has not been detected yet.

\begin{figure}[!htbp]
    \centering
    \includegraphics[width=0.95\linewidth]{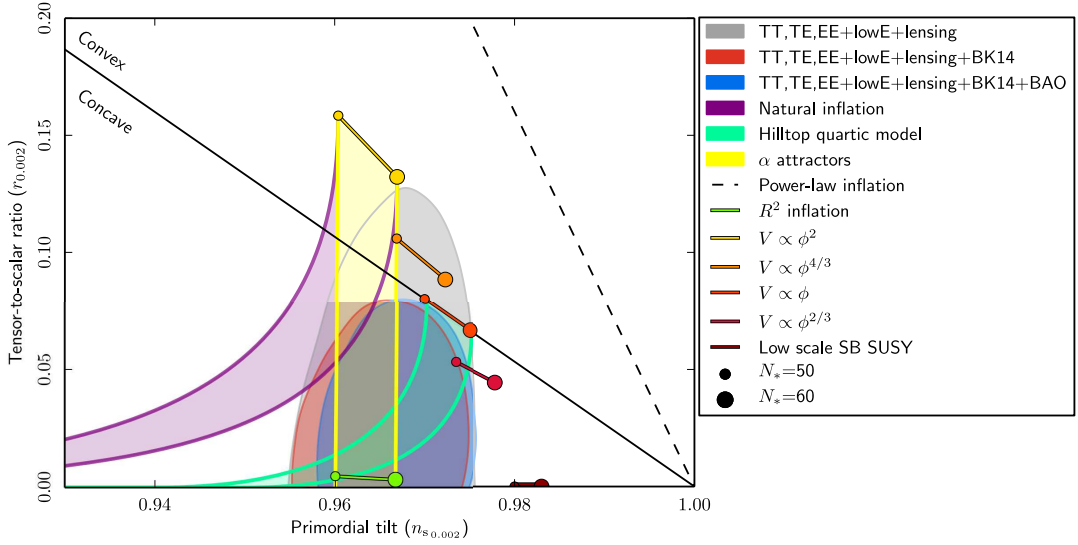}
    \caption{Marginalized 68\% and 95\% CL likelihood contours in the plane $r$ (tensor-to-scalar ratio, evaluated at the pivot scale $k_0=0.002\,{\rm Mpc}^{-1}$) versus $n_{\mathrm{s}}$ (scalar spectral index). Regions of different colors correspond to the joint marginalized constraints from 2018 \textsc{Planck} alone and in combination with other data sets. Starobinsky inflation is denoted as ``$R^2$ inflation'' (\copyright ESA and the \textsc{Planck} Collaboration \cite{Planck:2018nkj}).
}    \label{fig:Planck2018}
\end{figure}

In Fig.~\ref{fig:Planck2018}, we can see which inflationary models are viable according to observational constraints. Any model falling within the confidence regions for a number of e-folds between 50 and 60 is a reasonable candidate. In particular, Starobinsky inflation (see below) falls right in the middle of the likelihood region. 

Although there are still several models compatible with such constraints, there have been remarkable steps forward in the last twenty years. For example, the quadratic potential was still acceptable according to the first-year WMAP data in 2003 \cite{WMAP:2003syu}. Ten years later, with the \textsc{Planck} 2013 data, this model was almost excluded \cite{Planck:2013jfk} and was later ruled out. The same happened in many other inflationary models. Similarly, the tensor-to-scalar ratio was bounded as $r<0.9$ in 2003 \cite{WMAP:2003syu}, while now this upper limit is more than twenty times smaller, as shown in (\ref{eq:rup}). We entered the era of precision cosmology.

\subsection{Starobinsky inflation}

Starobinsky inflation is based on a quadratic modification of the Einstein--Hilbert action so that the action has the form \cite{Starobinsky:1980te}
\begin{equation}\label{starog}
    S=\frac{1}{2 \kappa^{2}} \int {\rm d}^{4} x \sqrt{|g|}\left[R+ \frac{R^2}{6m^2}\right] ,
\end{equation}
where $m$ is a mass scale. Just like any $f(R)$ model, this is not a theory of quantum gravity (QG) because, in general, terms that depend on the Ricci tensor are always generated at loop level if we quantize gravity as a perturbative quantum field theory (QFT). Nevertheless, this is one of the best inflationary models phenomenologically speaking.

If we make a conformal transformation of the metric to the Einstein frame, it can be recast as standard GR plus a scalar field,
\begin{equation}
    S = \int {\rm d}^{4} x \sqrt{|g|}\left[ \frac{\hat{R}}{2 \kappa^{2}} - \hat{\partial}_\mu\phi \hat{\partial}^\mu\phi- V(\phi) \right],
\end{equation}
with the potential
\begin{equation}
    V(\phi) = \frac{3m^2}{4 \kappa^{2}} \left(1 - {\rm e}^{-\sqrt{\frac{2}{3}} \kappa \phi} \right)^2 .
\end{equation}
Starobinsky inflation is in excellent agreement with the CMB observations because it predicts a scalar spectral index $n_{\rm s}=0.967$ and a tensor-to-scalar ratio $r=O(10^{-3})$, which is well within the likelihood region of Fig.~\ref{fig:Planck2018}.

\subsection{Inflation from particle physics}

We conclude with the following list of considerations on how to implement the inflaton field within particle physics.
\vspace{1.5mm}
\nonBulletListing{Minimally coupled Higgs field}
The first idea is to identify the inflaton with the Higgs boson since it is the only fundamental  scalar available in the Standard Model of particle physics. However, if we assume minimal coupling to the geometry, this approach does not work because the Higgs mass (125 GeV) is much smaller than the mass we would require to sustain inflation in the early stages of acceleration ($m \sim 10^{13}\, \textrm{GeV}$).
\vspace{1.5mm}
\nonBulletListing{Non-minimally coupled Higgs field}
If we couple the Higgs non-minimally to gravity, i.e., by considering a term $|\phi|^2R$, we obtain a model that can sustain inflation. However, this approach has unitarity issues and is not satisfactory \cite{BLT1,BaEs}. One can rescue Higgs inflation by adding new physics and field couplings \cite{BGS2,BMSS}. Although this model works, it is not as appealing as the idea of just identifying the inflaton with the Higgs boson \cite{AtC2}.
\vspace{1.5mm}
\nonBulletListing{Supergravity inflation}
In supergravity (SUGRA) models, there are plenty of boson fields that are reasonable candidates for the inflaton, but they are affected by the so-called $\eta$-problem: when computing the second slow-roll parameter $\eta$, one typically finds a value close to one. Thus one of the hypotheses behind inflation is completely spoiled. Several models characterized by supersymmetry breaking on the de Sitter vacuum state solve this issue in SUGRA inflation, but they also require a number of non-perturbative ingredients. Reviews on SUGRA inflation with references can be found in \cite{Calcagni:2017sdq,Yamaguchi:2011kg}.

\section{Big bang (singularity) problem}\label{bbp}

The standard cosmological model predicts the big bang. The model, in fact, assumes that the universe expands from a configuration where the scale factor is zero and, consequently, the metric is degenerate and the energy density diverges. The universe temperature at the big bang can be zero or infinite depending on whether one has inflation or not. This pathological behaviour implies that our picture of the universe starts from a point where the laws of GR are no longer valid; this is called the big bang problem.

\subsection{Origin and avoidance}

In GR, the solutions of Einstein equations (\ref{eq:eineq}) are plagued by singularities, as for instance in black holes. Since the theory is not predictive at singularities, this has been viewed as a major theoretical problem and a motivation for quantum gravity. In fact, even if we find one regular solution at $t=0$, it does not mean that the big bang problem is solved. One can make this argument more rigorous through a series of focusing theorems started by Hawking and Penrose in the 1960s \cite{Pen65,Haw1,Haw2,Haw3,Ger66,Haw4,HaP}. These theorems make various assumptions about the universe matter content or the type of dynamics. However, in 2003 Borde, Guth and Vilenkin (BGV) proved a very interesting theorem \cite{BGV} not based on any specific gravitational theory. Also, it does not assume anything about the matter content or equation of state. The theorem implies that, if we change dynamics from Einstein gravity to a more general one (for instance, $R\rightarrow f(R)$), we do not solve the singularity problem and can still find singularities.

\subsection{BGV theorem}

We can formulate the BGV theorem as follows. Let $(\mathcal{M}, g)$ be a spacetime with a congruence $u^{\mu}$ continuously defined along any time-like or null geodesic $v^{\mu}(\sigma)$ (the observer). Further, we define the following quantities
\begin{equation}
    \begin{split}
        \gamma &:=-u_{\mu}v^{\mu}\geq 0,\qquad u^2=-1\,, \\
        w^{\mu} &:= \frac{v^{\mu}-\gamma u^{\mu}}{\gamma^2+v^2}\qquad \Longrightarrow\qquad w^2=1 , \\
        \mathscr{H} &:= w_{\mu}w^{\nu}\nabla_{\nu}u^{\mu} ,\\
        \mathscr{H}_{\mathrm{av}} &:= \frac{1}{\sigma _e -\sigma _i} \int _{\sigma_{i}} ^{\sigma _e}{\rm d}\sigma  \mathscr{H}(\sigma) ,
    \end{split}
\end{equation}
where $\sigma$ is a proper-time or an affine parameter along the observer's geodesic, running from some initial value $\sigma_i$ to some end value $\sigma_e$. For instance, an observer in the congruence defined by the vector $u^{\mu}$ is like traveling through a flux of particles. We introduced a local notion of volume expansion $\mathscr{H}$, which is a generalization of the Hubble parameter. We then defined an average expansion $\mathscr{H}_{\mathrm{av}}$ over the observer's world-line.

Let the averaged expansion condition $\mathscr{H}_{\mathrm{av}} > 0$ hold for almost any $v^{\mu}$. Then, $(\mathcal{M}, g)$ is geodesically past-incomplete: geodesics have a finite proper or affine length. The proof is straightforward. Noting that $\mathscr{H}=\dot F=: v^\mu\nabla_\mu F$, where $F=\gamma^{-1}>0$ for null geodesics ($v^2=0$) and $F={\rm arctanh}(\gamma^{-1})>0$ for time-like geodesics ($v^2=-1$), one notes that the average expansion is just a positive finite constant $C$ divided by the length of our geodesic, which is then finite: 
\begin{equation}
    0<\mathscr{H}_{\mathrm{av}}=\frac{F(\gamma _e)-F(\gamma _i)}{\sigma _e -\sigma _i}=\frac{C}{\sigma _e -\sigma _i}.
\end{equation}

To get this result, we did not need to assume inflation, only that the universe expanded on average. We did not even need homogeneity or isotropy. This theorem is very powerful and one might think it seems to put the final word on the singularity problem. However, this is not accurate. Having past-incomplete world-lines is not enough to prove the existence of a global singularity. For that, we would need to show that past-incompleteness happens for all observers at the same time.

\subsection{Jumping into the big bang: BKL conjecture}

Let us assume we cannot get rid of the big bang. What would we see if we went towards the singularity? Surprisingly, the following result holds in several models, not only in GR but also if we modify gravity or change the number of spacetime dimensions. 

\begin{figure}[!htbp]
    \centering
    \includegraphics[width=0.95\linewidth]{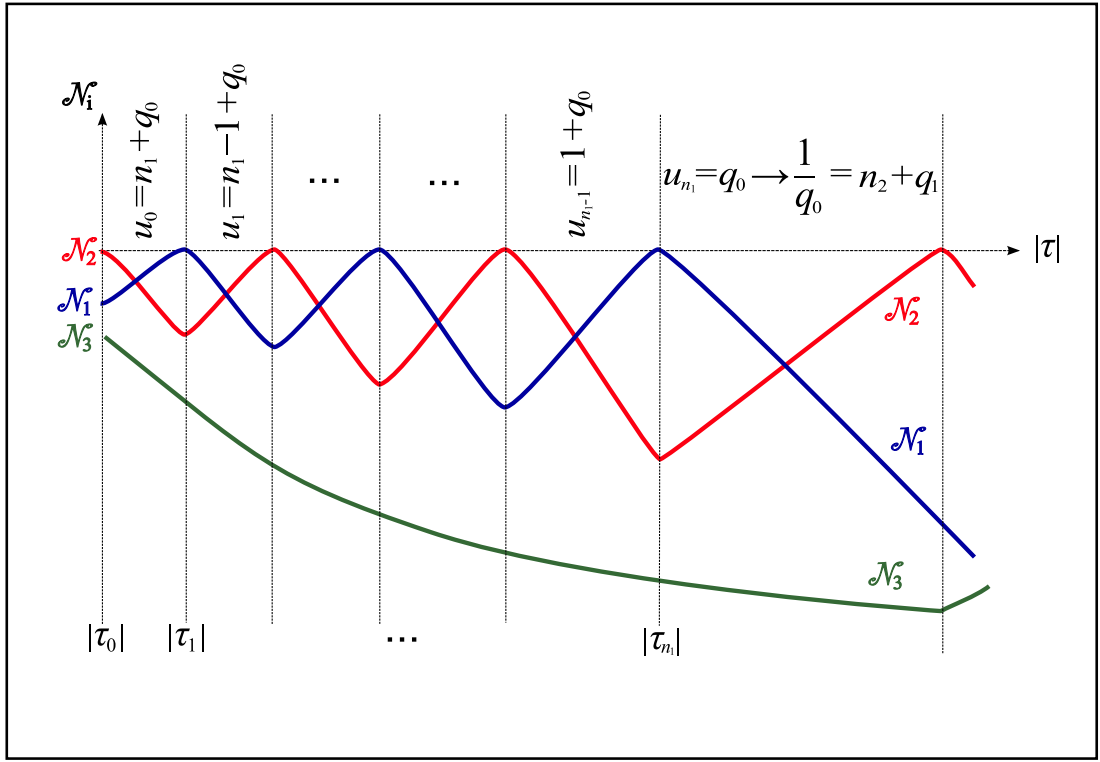}
    \caption{BKL singularity. Time flows backwards towards the singularity to the right, where $\tau:=\int {\rm d} t/(a_1a_2a_3)$ and $u_n$ parametrizes the power laws $a_i^{(n)}\sim t^{p_i(u_n)}$ within each epoch $n$. Reprinted with permission from \cite{Calcagni:2017sdq}, \copyright 2017 Springer International Publishing Switzerland.}\label{fig:BKL}
\end{figure}

Near the singularity, spatial points decouple and spacetime becomes homogeneous. In particular, time derivatives dominate over spatial derivatives. Belinsky, Khalatnikov, and Lifshitz (BKL) proposed this conjecture in the late 60s and early 70s, based on converging results in different settings \cite{Mis69,BKL70,BKL82}. The approach towards the singularity turns out to be either monotonic or oscillatory in the scale factors. Spacetime becomes homogeneous but not isotropic, meaning that different directions can behave differently. For instance, if we have three independent scale factors $a_1(t)$, $a_2(t)$ and $a_3(t)$, which are all functions of time, we are in the so-called Bianchi~IX model. There are many possible parametrizations of the scale factors, such as
\begin{equation}
    a_{1,2}=: \mathrm{e}^{-\Omega+\beta_{+} \pm \sqrt{3} \beta_{-}}, \qquad a_{3}=: \mathrm{e}^{-\Omega-2 \beta_{+}},\qquad \Omega:=-\frac{1}{3} \ln \left(a_{1} a_{2} a_{3}\right).
\end{equation}
The coefficients $\beta_\pm$ of this parametrization enter the potential associated with the Bianchi IX cosmology. During the evolution towards singularity, the universe bounces back and forth along the walls of this potential. This process is the reason why such a configuration is also called cosmological billiard: the scale factors chaotically change their behaviour during the evolution backwards in time. One can understand the BKL singularity also by parametrizing the scale factors in terms of the number of e-folds ${\cal N}_i=\ln a_i$. Towards the singularity, one scale factor always decreases, but the other two oscillate in opposite directions, so that at any time there is always one scale factor increasing backwards in time. This behaviour is shown on Fig.~\ref{fig:BKL}. This process happens infinitely many times before reaching the singularity.

Although here we have focused on the BKL big bang, there are other models of singularity (e.g., coming from SUGRA), some of which even reach the point of describing the shape of the singularity itself \cite{Dad}. 

\section{Cosmological constant problem}

Compared to the singularity problem, the cosmological constant one is of broader interest because it is an observational problem. While the big bang and black-hole singularities undeniably constitute an issue for GR as a theoretical framework, we do not observe them. In contrast, we do have direct access to the effects of the cosmological constant. Also, there are many model-dependent solutions to the big bang problem, but there are many more for the cosmological constant problem. This, and the fact that there is no common agreement on which solution is the best, are telltales indicating the scope of the task ahead.

We know that dark energy accounts for about 3/4 of the total cosmic energy budget but, nevertheless, it is not dense. Its energy density is about 120 orders of magnitude smaller than the reduced Planck energy density
\begin{equation}\label{rholam}
    \rho_{\Lambda} \sim 10^{-48} \mathrm{GeV}^{4} \sim 10^{-120} M_{\mathrm{Pl}}^{4}.
\end{equation}
Why is there such a hierarchy of extremely small numbers? 

We can state the cosmological constant problem in different ways. Here we recall four of them: the old problem, the new problem, the broken-symmetry problem and the $4\pi$ puzzle.

\subsection{Old problem}\label{oldpro}

One of the first formulations is called the old problem, reviewed in \cite{Martin:2012bt}. According to equation (\ref{w0est}), the dark-energy component looks like a cosmological constant because its equation of state is compatible with a barotropic index equal to $-1$. For this reason, we can naturally frame the problem in the formalism of QFT. Since dark energy looks like a pure cosmological constant $\Lambda={\rm const}$, we may try to interpret it as the zero-point energy of matter fields. Unfortunately, QFT cannot explain the magnitude of $\Lambda$ in terms of vacuum energy density. First of all, we recall that the vacuum energy density (or zero-point energy) is the eigenvalue of the system Hamiltonian when acting upon the physical vacuum state. In QFT, it is equivalent to the contribution $Z[0]$ of all bubble diagrams, i.e., Feynman diagrams with no external legs.

If we compute this energy using the electroweak sector of the Standard Model of particle physics, at tree level we obtain
\begin{equation}
    \left|\rho_{\text {vac }}^{(0)}\right| \sim \frac{m_{h}^{2}}{8 \sqrt{2} G_{\mathrm{F}}} \sim 1.2 \times 10^{8}\, \mathrm{GeV}^{4} \sim 10^{-65} M_{\mathrm{Pl}}^{4} \sim 10^{56} \rho_{\Lambda},
\end{equation}
where $m_h$ is the Higgs mass and $G_{\mathrm{F}}$ is the Fermi coupling. This value is 56 orders of magnitude bigger than what we observe.

Going to one-loop level does not alter this estimate. One must compute the following integral:
\begin{equation}
    \rho_{\text {vac }}^{(1)}=\sum_{i} \frac{N_{i}}{2} \int_{-\infty}^{+\infty} \frac{d^{3} \mathbf{k}}{(2 \pi)^{3}} \sqrt{|\mathbf{k}|^{2}+m_{i}^{2}},
\end{equation}
where $N_i$ is the number of one-particle states of the species $i$. Using a cut-off regularization scheme, one ends up with the famous estimate $\rho_{\mathrm{vac}, i}^{(1)}\sim E_{\max}^{4}$, where $E_{\max}$ is an energy ultraviolet (UV) cut-off \cite{Zel68}. Thus if we take the reduced Planck mass as a cut-off, $E_{\max}=M_{\rm Pl}$, we get $\rho_{\mathrm{vac}, i}^{(1)}\sim 10^{120} \rho_{\Lambda}$.

However, such a regularization breaks Lorentz invariance and gives a wrong result. Using instead the dimensional regularization scheme, one gets \cite{Akhmedov:2002ts,KoP}
\begin{equation}
\rho_{\mathrm{vac}}^{(1)}=\sum_{i} \frac{N_{i} m_{i}^{4}}{(8 \pi)^{2}} \ln \left(\frac{m_{i}^{2}}{M^{2}}\right) \sim 10^{-65} M_{\mathrm{Pl}}^{4} \sim 10^{56} \rho_{\Lambda}\,,
\end{equation}
where $M$ is the renormalization scale chosen as the average between $H_0$ and the energy of photons coming from supernov\ae\ \cite{KoP}. Thus, we end up with the same order of magnitude as for the tree level. It is not as big as the often-quoted $10^{120}$, but it is still huge. One can show that going to the higher loops does not improve the discrepancy.  

\subsection{Problem of small/large numbers}

We stress that the old cosmological constant problem is not just an artifact stemming from applying QFT for the sake of it. The problem is neither the smallness itself nor the fact that ordinary QFT does not work, but that we cannot explain the value of $\rho_\Lambda$ with known, well-established physics. For instance, at matter-radiation equality, the energy density is about 110 orders of magnitude smaller than the Planck density,
\begin{equation}
    \rho_{\mathrm{eq}} \approx 1.5 \times 10^{-110} M_{\mathrm{P}}^{4}.
\end{equation}
Such an abysmal gap is comparable with the one in equation (\ref{rholam}), being $\rho_{\mathrm{eq}}$ ``only'' ten orders of magnitude larger than $\rho_\Lambda$. However, this is not a hierarchy problem because we can predict the value of $\rho_{\mathrm{eq}}$ using the particle physics of the Standard Model.

In order to further understand the problem of small or large numbers in relation with the cosmological constant, we propose an analogy taken from our own build-up. Suppose we want to estimate the probability that one person has a specific face. In the language of genetics, this would amount to ask what the probability is to have a certain phenotype. In turn, this is an expression of the individual's genotype, i.e., the information encoded in the genes. Genes are made of DNA rolled into chromosomes and the composition of the latter is determined at conception by the combination of two\begin{wrapfigure}{r}{4cm}
    \includegraphics[width=4cm]{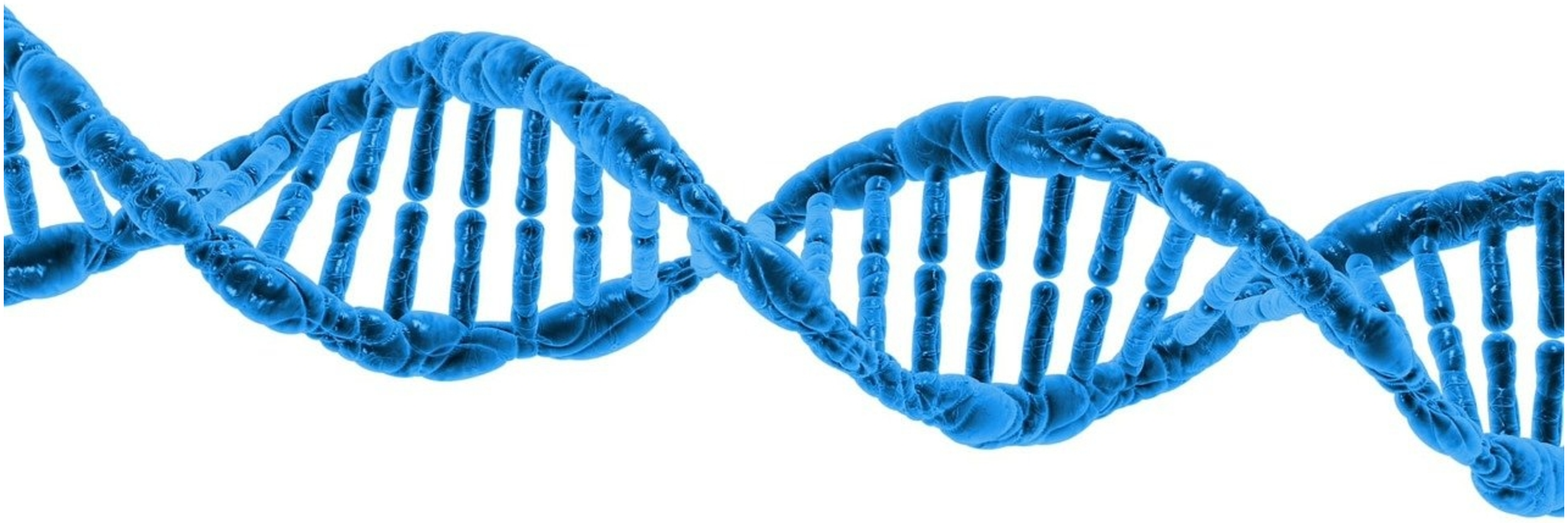}
\end{wrapfigure} gametes, the haploid cells carrying the father's and the mother's genetic information. Now, within the DNA, a gene locus is said to be in heterozygosis when it admits two different alleles of the gene. Therefore, the variety of faces may be linked to genetic recombination and to how many loci are in heterozygosis. 

Out of a total of about $20,000$ coding genes in the complete haploid set of 23 chromosomes \cite{ensembl,Ezk14}, about 16-20\% are polymorphic \cite{Lew67,HaHo,Wil13}, meaning that, in average, each individual's genome has about $3200-4000$ loci in heterozygosis. This means that genetic recombination can produce 
\begin{equation}\label{gamete}
2^{3200}-2^{4000}\sim 10^{960}-10^{1200}
\end{equation}
different gametes, a tremendous number by any account. This number is much superior even to the famous $10^{500}$ vacua in string theory (section \ref{sec:land}), not to mention the puny $10^{120}$ of the hierarchy problem for the cosmological constant. However, its explanation is straightforward once we know genetics and the workings of the DNA. Coming back to physics, the cosmological constant problem might exist simply because we still do not know what a more complete cosmological theory (``genetics'') looks like and how it would predict the content (the ``genome'') of the universe.


\subsection{New and coincidence problem}

The so-called new cosmological constant problem focuses on the fact that the cosmological constant energy density today is of the same order as the matter-energy density,
\begin{equation}
    \rho_{\Lambda}=O\left(\rho_{\mathrm{m}}\right)\,.
\end{equation}
Why is that so? Another facet of the same observation is the coincidence problem. 

The cosmological constant becomes dominant at very small redshifts: Why so recently? However, let us note that in terms of cosmic time $\Lambda$ domination begun when the universe was about half its present age. Therefore, this type of fine tuning depends on the chosen time variable.

\subsection{Broken symmetry problem}

There is another way to state the cosmological constant problem, proposed by Weinberg \cite{Wei89}. When we are in Minkowski spacetime, we can shift the matter Lagrangian by a constant:
\begin{equation}
\mathcal{L}_{\mathrm{m}} \rightarrow \mathcal{L}_{\mathrm{m}}+\rho_{0} \qquad \Longrightarrow \qquad T_{\mu}^{\nu} \rightarrow T_{\mu}^{\nu}+\rho_{0} \delta_{\mu}^{\nu} 
\end{equation}
This shift does not change the equations of motion $\nabla_{\nu} T_{\mu}^{\ \nu}=0$. However, as soon as we switch on gravity, we no longer have this shift symmetry because the Einstein equations (\ref{eq:eineq}) are not invariant due to the cosmological constant:
\begin{equation}
    \Lambda \rightarrow \Lambda-\kappa^{2} \rho_{0} .
\end{equation}
This sudden breaking of the shift symmetry is peculiar from the point of view of field theory. In field theory, vacuum solutions are usually symmetric with respect to some subgroup of the full symmetry group. In the case of classical gravity, when $\Lambda=0$ a vacuum solution is the Minkowski metric $g_{\mu \nu}=\eta_{\mu \nu}$, which is invariant under the Poincar\'e group. However, when $\Lambda\neq 0$ this symmetry is broken and solutions preserving space-time translations require a fine tuning on $\Lambda$. This way is another way to state that we cannot apply all the field-theory concepts to GR straightforwardly when the cosmological constant is different from zero.

\subsection{\texorpdfstring{$4\pi$}{4pi} puzzle}

We conclude by recasting the problem in terms of the $4\pi$ puzzle, which is a numerical coincidence based on classical cosmology \cite{Pad14}. For the observed value of $\rho_{\Lambda}$, the duration of the matter-radiation era from the end of inflation at time $t_{\rm e}$ (reheating) until the onset of late-time acceleration at time $t_\Lambda$ can be computed in terms of the number of e-foldings and it is equal to $4 \pi \pm 10^{-3}$:
\begin{equation}
    \mathcal{N}_{\mathrm{e}\Lambda}=\int_{k_{\rm e}}^{k_\Lambda} \frac{\mathrm{d}^{3} k}{(2 \pi)^3} \mathcal{V}_\textrm{com}=\frac{2}{3 \pi} \ln \frac{a_{\rm e} H_{\rm e}}{a_\Lambda H_\Lambda} \simeq \frac{1}{6 \pi} \ln \left[10^{-3} \frac{\rho_{\mathrm{reh}}}{\rho_{\Lambda}}\right] \approx 4 \pi ,
\end{equation}
where $\mathcal{V}_{\text {com }}\propto (aH)^{-3}$ is the comoving Hubble volume. This value is based on the history of the universe and, specifically, on the onset of the dark energy era. It also depends on the reheating because it depends on when the radiation era starts. However, these uncertainties do not alter the above estimate.

\subsection{Attempts to solving the \texorpdfstring{$\Lambda$}{Lambda} problem}

Let us consider some possible solutions to the cosmological constant problem not invoking quantum gravity. Simple ideas that, however, do not work or are still under development are the following.
\nonBulletListing{Quintessence: $\rho_{\Lambda}=\rho_{\phi}$}
As we saw, we can explain the acceleration in the early universe by introducing a dynamical scalar field $\phi$. It is reasonable to ask if this approach also works for the late-time universe. This minimally coupled scalar field is called quintessence \cite{Tsu13}. The problem is that one has to choose an \emph{ad hoc} potential to fit observations and, more importantly, one has to fine-tune the initial conditions. Assuming the quintessence field to be non-minimally coupled to matter (extended quintessence) \cite{Wetterich:1994bg,Amendola:1999er}, one can get acceleration with $w_{0} \approx-1$ and no fine-tuning \cite{Peri05} but the $\Lambda$CDM model is still statistically favoured \cite{Davari:2019tni}. Also, both minimally and non-minimally coupled models of quintessence tend to favour lower local values of $H_0$ compared with $\Lambda$CDM, thus exacerbating the $H_0$ tension and facing an observational problem \cite{Banerjee:2020xcn,Lee:2022cyh}.
\nonBulletListing{Modified gravity: $f(R)$}
Another explanation may come from modified-gravity models such as $f(R)$, where late-time acceleration is sustained by geometry itself \cite{Tsu10,ClFPS}. However, we do not know how to theoretically justify observationally viable models because, simply put, we do not know where these Lagrangians come from. Attention to phenomenology should go hand by hand with care for a top-down motivation, something about which $f(R)$ and similar \emph{ad hoc} proposals tend to fail. Lastly, many of these models require moderate fine-tuning.
\nonBulletListing{Void models}
The cosmological principle works on average, but we know that large void regions surround the clusters of galaxies. A possibility is that acceleration is a spurious effect of assuming a meaningful average energy density $\langle\rho\rangle$ and the FLRW background, where one is not taking into account the fact that signals propagate through voids where $\rho_{\rm void}\ll\langle\rho\rangle$. These models (see \cite[section 7.6.2]{Calcagni:2017sdq} for references) are attractive because they do not add new physics to GR, only a more sophisticated background and matter distribution. However, most of them are in tension with data, entail some fine tuning or remove the fine tuning at the price of simplicity.


\section{Cosmology of quantum gravity}

We now present some cosmological models emerging from QG. It is legitimate to ask what quantum gravity predicts about cosmological observations. In this section, we partially try to describe the state of the art in this field.
%
\subsection{Ideal QG proposal}\label{ideal}

An ideal QG proposal should satisfy some minimal requirements which, for generality, we state for a theory of everything (i.e., including all sectors of gravity and matter):
\begin{enumerate}
    \item \textit{Theoretically rigorous, viable and inclusive.} 
    First of all, a theory of QG should be theoretically rigorous, with good properties such as no (or a finite number of) infinities in the UV, no or harmless violations of unitarity, and so on. In particular, on the side of theoretical consistency, it should solve the singularity problem. On the side of inclusiveness, it should reproduce portions of well-established theories already tested by experiments, in particular, a good low-energy or low-curvature limit recovering GR, an early-universe inflationary mechanism, the Standard Model of particles, and so on.
    \item \textit{Retrodictive.} A good theory should give retrodictions, i.e., observational predictions that have already been tested, but stemming from a new theoretical set-up. This is a step further on the side of inclusiveness. For instance, its inflationary spectra should be computable and consistent with observations, it should solve the cosmological constant problem, the $H_0$ tension or explain some loose strand of the Standard Model, such as the origin of neutrino masses.
    \item \textit{Predictive and testable.} It should also give unique predictions of yet-unobserved phenomena that could be checked against near-future observations in the laboratory, particle physics, astrophysics or cosmology.
\end{enumerate}
We are currently far from this ideal framework but there are many proposals with some, if not most, of the above properties. Nowadays, the question is not whether one can quantize gravity (we did it already!), but which of the existing theories is closer to fulfill this program. Since it is possible to advance our knowledge of QG phenomenology in fundamental theories or, at least, through models strongly tied to fundamental theories, we should make a warning remark on back-of-the-envelope arguments discarding the possibility of QG effects on the ground that they are too small or confined to Planck scales. In general, we have two possibilities to approach QG:
\begin{itemize}[leftmargin=*]
    \item \textit{Perturbative QG.} One can realize QG as a perturbative QFT when taking a modification of GR as the underlying classical theory (GR cannot be perturbatively quantized consistently \cite{GoSa2}). For instance, one can add higher-order curvature operators to the action. However, in this case we expect QG effects to become relevant at high curvature or short distances. From the point of view of homogeneous cosmology, we only have two scales available, the Planck length $\ell_{\mathrm{Pl}}$ and the Hubble parameter $H$, so that corrections would be expressed as powers of the product $\ell_{\mathrm{Pl}} H$. Using the value of the Hubble parameter today, $H_0$, we get a highly-suppressed value of $10^{-60}$, possibly exponentiated to some yet smaller quantity.
    \item \textit{Non-perturbative QG effects.} Things change if we consider non-perturbative quantizations of QG or, in general, non-perturbative effects regardless of the type of quantization. As a simple example, in some settings the fundamental building blocks may dynamically evolve to a semi-classical spacetime limit where an intermediate length scale $L\gg \ell_{\mathrm{Pl}}$ emerges. This scale could be the characteristic length of some quantum states of geometry \cite{Bojowald:2011hd}. Then, one can construct dimensionless quantities such as $ \ell_{\mathrm{Pl}}^{a} H^{b} L^{c}$ if the powers sum up to zero, $a-b+c=0$; some quantum corrections would be proportional to this quantity and not all of them would be small. String theory and string cosmology are another clear demonstration of how non-perturbative physics can make the microscopic visible.
\end{itemize}
The type of back-of-the-envelope, curvature-corrections-based reasoning described in the first bullet is often invoked to dismiss QG as a relevant actor in the cosmology theater. These arguments have also been refuted time and again by the fact, among several others, that quantum theories beyond Einstein gravity can significantly modify the standard inflationary spectrum, thanks also to the non-perturbative effects mentioned in the second bullet.

\subsection{Inflation in QG}

There are many models of QG or inspired by QG which embed inflation. Here we give just three examples. The first is non-local quantum gravity, the second is a class of models coming from loop quantum gravity and the third is a class of models from string theory.
\nonBulletListing{Non-local Starobinsky inflation}
Non-local quantum gravity is a perturbative QFT approach where the dynamics is characterized, at the fundamental level, by a certain class of non-local operators \cite{Modesto:2017sdr,Calcagni:2021hve}. This theory is unitary and renormalizable or even finite in some realizations. In the so-called Weyl basis and in four dimensions, the generic action of non-local quantum gravity is
\begin{equation}
    S=\frac{1}{2 \kappa^{2}} \int {\rm d}^{4} x \sqrt{|g|}\left[R+R \gamma_{\mathrm{S}}(\Box) R + C_{\mu \nu \rho \sigma} \gamma_{\mathrm{C}}(\Box)C^{\mu \nu \rho \sigma}+{\cal V}({\cal R})\right]\,.
\end{equation}
The action is of higher order in curvature terms in order to make the theory renormalizable. The non-local form factors $\gamma_{\mathrm{S}}(\Box),\gamma_{\mathrm{C}}(\Box)$ are introduced to preserve unitarity (ghost freedom) and improve renormalizability and depend on a fundamental scale $M_*$ possibly close or equal to $M_{\rm Pl}$. Different form factors are allowed by these self-consistency conditions imposed on the theory but they all share the property of diverging in the UV, so that their inverse (the graviton propagator) vanishes in the same limit. Last, ${\cal V}$ is a collections of local terms in curvature tensors ${\cal R}=R,R_{\mu\nu},R_{\mu\nu\rho\sigma}$ which is necessary only if we want to make the theory not only renormalizable, but also UV finite. Note that the Weyl tensor $C_{\mu \nu \rho \sigma}$ vanishes on a FLRW background.

Starobinsky inflation can be embedded in this theory. Calling $z:={\Box}/{M_*^2}$ and parametrizing the form factor in the Weyl term as
\begin{equation}
\gamma_{\rm C}(z) = \frac{1}{2M_*^2}\frac{{\rm e}^{{\rm H}_2(z)}-1}{z}\,,
\end{equation}
with a similar parametrization for $\gamma_{\mathrm{S}}$ with a different function ${\rm H}_0$, one can choose the form factors ${\rm H}_0$ and ${\rm H}_2$, which depend on two mass scales $M_*$ and $m$, such that the Starobinsky action (\ref{starog}) is recovered in the local limit. Regarding inflation, let us concentrate on the GW spectrum, which will also be examined in section \ref{sgwb}. The primordial tensor spectrum is \cite{Koshelev:2016xqb,Koshelev:2017tvv}
\begin{equation}\label{starnlP}
   \mathcal{P}_{\mathrm{t}} \simeq \frac{m^2}{2 \pi^2 M_\mathrm{Pl}^2} (1-3 \epsilon) {\rm e}^{-{\mathrm{H}}_{2}\left[-R/(2M_*^2)\right]},
\end{equation}
where the Ricci scalar $R$ is evaluated at horizon crossing. The other inflationary observables follow through.
\nonBulletListing{Loop quantum cosmology}
Loop quantum cosmology (LQC) is the cosmological realization of the Hamiltonian quantization scheme of loop quantum gravity. The latter does not aim to integrate the Standard Model, although there are proposals in that direction. There are at least three different versions of LQC, depending on how we treat cosmological perturbations and the background: the anomaly cancellation approach, the hybrid quantization approach and the dressed metric approach \cite{Calcagni:2017sdq,Barrau:2014maa,Bojowald:2015iga,ElizagaNavascues:2020uyf,Li:2021mop}. Just like in the standard model of cosmology, the inflaton is a scalar field introduced by hand, its origin is unexplained and its potential is simply assumed. Still, the QG ingredients of LQC open a window on new physics. In fact, even if the curvature scale of inflation is typically much smaller than the Planck energy, we can still get macroscopic effects in QG modifications of the CMB spectrum. 

So far, LQC models have proved themselves both retrodictive and predictive. However, we recall that deviations from GR at low multipoles should be larger than cosmic variance to be testable as far as the CMB is concerned, which happens only in a very limited number of cases. 
\nonBulletListing{String cosmology}
There are very different realizations of string inflation, depending on how we compactify the extra spatial directions if these are small, or on how the visible four-dimensional universe would be embedded in large extra dimensions. In general, in this framework, the inflaton naturally arises as a field among the many geometry-originated scalars (called moduli) of the theory. In this sense, the string approach is more satisfactory than LQC because we do not need to artificially add extra degrees of freedom.

However, there are many ways to have an early universe stage of acceleration (Tab.~\ref{tab:stringTab}) \cite{Calcagni:2017sdq,BaMcA}: moduli inflation, axion inflation, monodromy inflation, warped D-brane inflation, UV DBI inflation, IR DBI inflation, string-gas cosmology, certain versions of the ekpyrotic universe, and so on. These models are all in agreement with observations of the primordial scalar spectrum but they can give different predictions about the tensor sector and the level of non-Gaussianity. In some of these models (moduli inflation, warped D-brane inflation, IR DBI inflation and viable versions of the ekpyrotic universe), the tensor-to-scalar ratio $r$ is too small to be observable. Other models can produce larger values of $r$ closer to the current upper bound (\ref{eq:rup}). Lastly, the fine-tuning of string-cosmology models is generally acceptable and, in some cases, it is very mild. 

In the classification of section \ref{ideal}, all these models are retrodictive because the theoretical value of the scalar spectral index $n_{\rm s}$ lies in the allowed observational range coming from (\ref{nsobs}) and the tensor-to-scalar ratio $r$ respects the upper bound (\ref{eq:rup}); and they are also predictive because it is possible to calculate the theoretical value of $r$ when we still have not observed the tensor sector. But only the second group of models is testable as far as the tensor sector is concerned, since the predicted value of $r$ is large enough to be within observational reach in the near future.
\begin{table}[H]
    \centering
    \resizebox{\columnwidth}{!}{%
    \begin{tabular}{|l|l|l|c|l|l|l|l|l|}
    \hline Model & $\epsilon_{*}$ & $\left|\eta_{*}\right|$ & $0.95<n_{\mathrm{s}}<0.98$ & $n_{\mathrm{t}}$ & $r$ & $f_{\mathrm{NL}}$ & Fine tuning \\
    \hline Moduli inflation & $\sim 0$ & $\sim 10^{-2}$ & $\checkmark$ & $\lesssim 0$ & $\sim 0$ & Small & $10^{0}-10^{-3}$ \\
    \hline Axion inflation & $\sim 10^{-2}$ & $\gtrsim 10^{-2}$ & $\checkmark$ & $\sim-10^{-2}$ & $\lesssim 0.001$ & Small? & $10^{-2}-10^{-4}$ \\
    \hline Monodromy inflation & $\sim 10^{-2}$ & $\gtrsim 10^{-2}$ & $\checkmark$ & $\sim-10^{-2}$ & $0.01-0.1$ & Small & $10^{-2}$ \\
    \hline Warped D-brane inflation & $\sim 0$ & $\sim 10^{-2}$ & $\checkmark$ & $\lesssim 0$ & $\sim 0$ & Small & $10^{-2}-10^{-3}$ \\
    \hline UV DBI inflation & $\sim 10^{-1}$ & $\sim 10^{-1}$ & $\checkmark$ & $\sim-10^{-1}$ & $\gtrsim 0.01$ & Large & $10^{-2}$ \\
    \hline IR DBI inflation & $\sim 0$ & $\sim 10^{-2}$ & $\checkmark$ & $\lesssim 0$ & $\sim 0$ & Large & $10^{-1}-10^{-2}$ \\
    \hline String-gas cosmology & $-$ & $-$ & $\checkmark$ & $\sim+10^{-2}$ & $\lesssim 0.1$ & Small & $10^{-2}$ \\
    \hline Ekpyrotic universe & $\sim 10^{2}$ & $\sim 10^{2}$ & $\checkmark$ & $\gtrsim 2$ & $\sim 0$ & Large & $10^{-1}-10^{-2}$ \\
    \hline
    \end{tabular}
    }
    \caption{Main features of different models of string cosmology. The scalar spectral index is inside the $3\sigma$ experimental bound determined by (\ref{nsobs}). $f_{\mathrm{NL}}$ is a parameter estimating the level of non-Gaussianity of scalar perturbations. All the models predict small values of the slow-roll parameters evaluated at horizon crossing, except for the ekpyrotic universe. String-gas cosmology and the ekpyrotic universe are alternatives to inflation.}
    \label{tab:stringTab}
\end{table}

\subsection{Big bang problem in QG}

In the following list, we briefly summarize how different approaches to QG fare concerning the big bang problem. More references prior to 2017 can be found in \cite{Calcagni:2017sdq}.
\begin{itemize}
    \item \textit{String theory}. Singularities are resolved by T-duality \cite{GaVe4} or by other means in different corners of the parameter space of the theory \cite{LMS3,CoCo2,Cra06,BeRe}. A particularly original scenario is the $E_{10}/K(E_{10})$ coset model \cite{DaNi3}. This model arises in $D=11$ SUGRA, which can be considered as the low-energy limit of M-theory, which, in turn, is regarded as the 11-dimensional extension of string theory. This setting is a quantum version of the cosmological billiard. If one tries to approach the cosmic singularity, one finds that spatial points decouple and that spacetime becomes homogeneous, in agreement with the BKL conjecture for classical billiards. However, close to the singularity something unexpected happens. In the strong-coupling limit where the 11-dimensional Newton's constant goes to infinity, $G\to+\infty$, all the $D=11$ SUGRA dynamical matter fields and the metric reduce to just one degree of freedom in one dimension, a homogeneous dynamical variable $\nu(t)$ which takes values in the Lie group ${\cal E}_{10}$ and follows a null geodesic in the quotient space ${\cal E}_{10}/K({\cal E}_{10})$, where $K({\cal E}_{10})$ is the maximal compact sub-group of ${\cal E}_{10}$. In this regime, the big bang is never reached because spacetime dissolves into a purely algebraic structure: the Universe becomes just symmetry.		
    \item \textit{Non-local quantum gravity}. In this theory, the big bang singularity is smeared into a bounce already at the classical level, thanks to the presence of the non-local operators \cite{Biswas:2005qr,Koshelev:2013lfm,Calcagni:2013vra}. In the case of black holes, singularities can be resolved through non-locality in certain versions of the theory, while in others the help of conformal symmetry is needed.	
    \item \textit{Canonical quantum cosmology}, where the universe is described by a wave function satisfying the Wheeler--DeWitt equation. We have a probabilistic interpretation of the big bang. The wave-function solving the Hamiltonian constraint of the universe vanishes at the big bang, so that the probability to hit the initial singularity is zero \cite{De671}. In general, non-singular solutions exist \cite{Kim5} but it is not clear how typical they are, since we do not have quantum versions of the singularity theorems mentioned in section \ref{bbp}.
    \item \textit{Loop quantum gravity}. While in Wheeler--DeWitt quantum cosmology the big bang happens but with zero probability, in the case of loop quantum cosmology we actually skip the big bang \cite{Calcagni:2017sdq,Bojowald:2015iga,ElizagaNavascues:2020uyf,Li:2021mop}. In fact, as a result of the quantization of geometry, the evolution of the universe goes through discrete steps which do not include the big bang at $t=0$. In the limit of a continuous evolution, the effective Friedman equations display a finite quantum bounce where the energy density is about half the Planck density \cite{Calcagni:2017sdq,Bojowald:2015iga,ElizagaNavascues:2020uyf,Li:2021mop}. This result is fairly generic but relies on assumptions that have been criticized \cite{Bojowald:2020wuc}.
    \item \textit{Spin foams}, a covariant path-integral quantization based on the states of loop quantum gravity \cite{Per13}. The bounce predicted in LQC is confirmed \cite{Ren13}.
    \item \textit{Group field theory}, a perturbative QFT where fields live on a group manifold and in which both loop quantum gravity and spin foams are included \cite{Fousp,BaO11}. LQC dynamics, hence the bounce, are recovered solving the theory on condensate states \cite{GiSi}.	
		 \item \textit{Multifractional spacetimes}, where fields live in a spacetime with scale-dependent Hausdorff and/or spectral dimension \cite{revmu,Calcagni:2021ipd}. In the so-called stochastic interpretation of the theory, spacetime becomes fuzzy, there is a minimal uncertainty on scales and it no longer makes sense to talk about singularities at all. However, these scenarios are not yet embedded in a full theory of QG.
	  \item \textit{Asymptotic safety}, a non-perturbative realization of quantum gravity based on the functional renormalization approach \cite{NiR,Eichhorn:2018yfc,Bonanno:2020bil}.
Bouncing solutions have been found \cite{Kofinas:2016lcz} but their generality remains unclear.	Therefore, the big bang problem is not yet solved.		
		\item \textit{Unimodular gravity} is a class of theories where some components of the metric are non-dynamical \cite{Calcagni:2017sdq,BOTu}. Since it is equivalent to GR at the classical level, the cosmic singularity is still in its place classically and any news about it should come from the quantum theory, which can be defined via the path integral. However, attention has been mainly drawn to the cosmological constant problem rather than to the big bang. Therefore, the big bang problem is not yet solved.	
    \item \textit{Causal sets}, where the fundamental setting is not continuous spacetime but a discretum of points with an ordering relation \cite{Dow13,Surya:2019ndm}. The big bang problem is not yet solved.
		  \item \textit{Causal dynamical triangulation}, a path-integral quantization regularized by approximating smooth manifolds with triangulations \cite{AGJL4,Loll:2019rdj}. The big bang problem is not yet solved. 	
\end{itemize}

\subsection{Cosmological constant problem in QG}

A fundamental question is whether the solution of the cosmological constant problem is rooted in the UV or in the infrared. We may think about it as an infrared problem because it has to do with the large-scale properties of the universe, but its origin could be in short-scale physics, as in the failed QFT resolution of the old problem (section \ref{oldpro}). 

Let us now see how QG models deal with the cosmological constant problem. We further discuss the theories marked with an asterisk * in the following sub-sections.
\begin{itemize}
    \item \textit{String theory.}* We have a landscape of vacua of which a sizable portion have a small $\Lambda$.
		\item \textit{Non-local quantum gravity.} The $\Lambda$ problem is not yet addressed.
    \item \textit{Canonical quantum cosmology.}* There is a probabilistic interpretation of the value of $\Lambda$.
    \item \textit{Loop quantum gravity.}* There are very few proposals to address the issue.
    \item \textit{Spin foams.} The $\Lambda$ problem is not yet addressed.
    \item \textit{Group field theory.}* There is a probabilistic interpretation of the value of $\Lambda$.
    \item \textit{Multi-fractional spacetimes.} The $\Lambda$ problem is reformulated but not yet addressed. One multi-fractional theory can explain dark energy just as a geometrical effect \cite{Calcagni:2020ads}.		
    \item \textit{Asymptotic safety.}* There is a clear-cut prediction for $\Lambda$ which, however, relies on important assumptions.
    \item \textit{Unimodular gravity.} There is an attempt in emergent unimodula gravity to solve the $4\pi$ puzzle to explain why $\Lambda$ is small, although it does not predict the observed value \cite{PaPa2}.
    \item \textit{Causal sets.}* There is a sketchy prediction for $\Lambda$.
    \item \textit{Casual dynamical triangulations.} A de Sitter universe emerges from full QG, but the $\Lambda$ problem is not yet addressed.
\end{itemize}

\subsubsection{String theory: a landscape for \texorpdfstring{$\Lambda$}{Lambda}}\label{sec:land}

String theory offers an original probabilistic resolution of the cosmological constant problem. The theory could have as many as $10^{10}-10^{500}$ distinct vacua in the low-energy limit. However, how many of them could produce the universe that we live in? 
 It turns out that we might have enough vacua satisfying the observational constraints. We can count them with either statistical selection \cite{Dou03,Do04b,DDKa} or anthropic selection \cite{BoPo,Sus03}. Statistical selection means that we count the vacua according to joint probabilities (for instance, of having a small $\Lambda$ and a low-scale supersymmetry breaking). As for the anthropic selection, based on vacua compatible with life, there are many versions of the anthropic principle but even Weinberg's most conservative formulation \cite{Wei87} can lead to a relative estimate of the number of viable vacua.

An open question is whether stable de Sitter backgrounds are really supported by string theory. Despite many explicit examples, doubts have been cast about the actual number of viable vacua produced by the theory, being most the landscape a ``swampland'' of unphysical configurations \cite{Palti:2019pca}.

\subsubsection{Canonical quantum cosmology: a probability for \texorpdfstring{$\Lambda$}{Lambda}}

In Wheeler--DeWitt canonical quantum cosmology, we can compute the nucleation probability of the initial state of the Universe \cite{Bau83,Ha84b,Wu07}, that is, the probability for the Universe to start in a de Sitter state with a given value of $\Lambda$. Depending on whether we take the Vilenkin \cite{Vil82} or the Hartle--Hawking \cite{HaH83} wave-function $\Psi$, we get different probabilities $P\sim |\Psi|^2$, both exponential but with a crucial sign difference:
\begin{equation}\label{wdwren}
    P_{\rm V}(\Lambda) = \exp \left(-\frac{12M_{\rm Pl}^2}{\Lambda}\right), \qquad 
    P_{\rm HH}(\Lambda) = \exp \left(\frac{12M_{\rm Pl}^2}{\Lambda}\right).
\end{equation}
In the case of the Hartle--Hawking wave-function, universes with small $\Lambda$ are favored. However, we can renormalize the wave-function with a $\Lambda$-dependent constant, which may erase the effect. Therefore, there is a certain degree of arbitrariness in these results.

\subsubsection{Loop quantum gravity: a condensate for \texorpdfstring{$\Lambda$}{Lambda}}

In the case of loop quantum gravity, there are some attempts to explain $\Lambda$ from $SU(2)_q$ ``fat'' spin networks. Assuming geometries where the metric is degenerate, one can encode spacetime dynamics into a Dirac equation for spinorial degrees of freedom \cite{Alexander:2008yg}. If these fermions $\psi$ condense, then the expectation value of the condensate gives an exponentially suppressed cosmological constant which only requires a mild fine-tuning of one part over one hundred:
\begin{equation}
    \Lambda=\Lambda_{0} \exp \left(-\left\langle\bar{\psi} \gamma^{5} \gamma^{z} \psi\right\rangle\right) \ll 1,
\end{equation}
where $\gamma^5$ and $\gamma^z$ are Dirac gamma matrices. However, the number of assumptions leading to this result forbid to consider it as robust before studying the model in greater detail.

\subsubsection{Group field theory: another probability for \texorpdfstring{$\Lambda$}{Lambda}}

Group field theory (GFT) is a theory of fields living on a group manifold, i.e., a manifold made of group elements. In the case of gravity, the group is the local gauge group of gravity. This theory is argued to reproduce the states of geometry of spin foams and loop quantum gravity. Furthermore, contrary to the case of LQC in relation with loop quantum gravity, cosmology can be extracted directly from the full theory and it requires much less effort. 

If $\varphi(g)$ is a scalar field, where $g\in \mathbb{G}$ is a group element, we can write the action of a group scalar field theory as
\begin{equation}
    S_{\mathrm{GFT}}=\int_{\mathbb{G}} \mathrm{d}^{4} g\left[\int_{\mathbb{G}} \mathrm{d}^{4} g^{\prime} \varphi^{*}(g) \mathcal{K}\left(g, g^{\prime}\right) \varphi\left(g^{\prime}\right)+V(\varphi,\varphi^{*})\right],
\end{equation}
where $\mathcal{K}\left(g, g^{\prime}\right)$ is a non-local operator and the potential $V$ is a non-linear interaction of the fields. Then we can perform the usual Fock quantization with commutation relations
\begin{equation}
    \left[\hat{\varphi}(g), \hat{\varphi}^{\dagger}\left(g^{\prime}\right)\right]=\mathbbm{1}_{\mathbb{G}}\left(g, g^{\prime}\right),
\end{equation}
and, given the vacuum state $|\varnothing\rangle$, we can construct one-particle states as
\begin{equation}
    |g\rangle:=\hat{\varphi}^{\dagger}(g)|\varnothing\rangle.
\end{equation}
We can play with these quanta to get a condensate, which is a coherent state such that
\begin{equation}
    |\xi\rangle:=A e^{\hat{\xi}}|\varnothing\rangle, \quad \hat{\xi}:=\int \mathrm{d}^{4} g \xi(g) \hat{\varphi}^{\dagger}(g), \quad \hat{\varphi}|\xi\rangle=\xi|\xi\rangle .
\end{equation}
Then, if we have infinitely many particles in the same state, we get a continuous homogeneous and isotropic spacetime.

Furthermore, we can get a probabilistic prediction for the cosmological constant because the wave-function of the Universe in canonical quantum cosmology is given by the expectation value of the group field. The quantum dynamics is given by the Gross--Pitaevskii equation
\begin{equation}
    0=\langle\xi|\mathbbm{1} \hat{\mathcal{C}}| \xi\rangle=\int \mathrm{d}^{4} g^{\prime} \mathcal{K}\left(g, g^{\prime}\right) \xi\left(g^{\prime}\right)+\left.\frac{\delta V}{\delta \varphi^{*}(g)}\right|_{\varphi=\xi} ,
\end{equation}
where $\hat{\mathcal{C}}$ is the quantum version of the equations of motion. The expectation value goes as 
\begin{equation}
    \xi(\Lambda) \sim \frac{1}{\sqrt{\Lambda}} .
\end{equation}
Just like for canonical quantum cosmology, universes with small values of $\Lambda$ are favored, although here the probability distribution is less peaked at $\Lambda=0$. However, here we do not have the normalization problem we encountered in the Wheeler--DeWitt result (\ref{wdwren}), since this is a Lagrangian formulation without the ambiguities typical of the Hamiltonian formalism.

\subsubsection{Asymptotic safety: a bullseye prediction for \texorpdfstring{$\Lambda$}{Lambda}}

In asymptotic safety, we quantize our theory using the functional renormalization approach and postulate that the theory is asymptotically safe. The latter means the following. Let $\lambda_i$ be the set of dimensionless couplings of the theory and let $k$ be the energy scale. Then, asymptotic safety states that the UV limit of the renormalized couplings is non-vanishing,
\begin{equation}
	\lim _{k \rightarrow \infty} \bar{\lambda}_{i}(k)=\bar{\lambda}_{i}^{*} \neq 0 \,.
\end{equation}
If we just quantize GR in vacuum, there are only two scale-dependent couplings, Newton's constant $G_{k}=k^{-2} G_{k_{0}}$ and the cosmological constant $\Lambda_{k}=k^{2} \Lambda_{k_{0}}$. The effective action and equations of motions are given by
\begin{equation}
	\Gamma_{k}=\frac{1}{16 \pi G_{k}} \int \mathrm{d}^{D} x \sqrt{|g|}\left(R-2 \Lambda_{k}\right), \qquad \frac{\delta \Gamma_{k}}{\delta g_{\mu \nu}}\left[\left\langle g_{\mu \nu}\right\rangle_{k}\right]=0\,,
\end{equation}
where $\langle g_{\mu \nu}\rangle_{k}$ is the average of the metric over Euclideanized spacetime volumes of linear size $\sim k^{-1}$.

We can reformulate the old cosmological problem by asking why the perturbative semi-classical regime of the renormalization-group trajectory should be so much fine-tuned. From the perspective of the functional renormalization-group approach, the problem is not quite solved. In fact, on one hand, there do exist trajectories passing very close to the $\Lambda=0$ axis in the $(\Lambda_k,G_k)$ plane \cite{Fal14}, so that it is possible to realize small values of $\Lambda$. On the other hand, the physical explanation of why this happens is still unclear.

From the perspective of effective dynamics where $k$ is identified with a cosmological scale, asymptotic safety can give rise to a local-void model with dynamical $\Lambda_k$ and $G_k$ fitting all observations and being statistically equivalent to $\Lambda$CDM, but without the fine tuning implicit in a constant $\Lambda$ \cite{Anagnostopoulos:2018jdq,Anagnostopoulos:2019mrc}. This model is interesting but it is based on the phenomenological assumption that the replacement $(\Lambda,G)\to(\Lambda_k,G_k)$ in the classical equations of motions is a faithful description of the semi-classical limit of the theory.

An alternative is to formulate asymptotically free QG as a perturbative resummed QFT, which is a QFT where all the tree level propagators in Feynman diagrams are replaced by resummed propagators. Once one does that, one can calculate the vacuum expectation value of the fields of the theory. The result is equated to the energy density of dark energy today $\rho_\Lambda(t_{0})$, where $\rho_\Lambda(t)$ is defined such that $\rho_\Lambda+P_\Lambda=0$ (in agreement with (\ref{w0est})) and radiation, matter and dark energy are conserved as a whole in the continuity equation
\begin{equation}
\dot\rho+\dot\rho_\Lambda+3H[(\rho+\rho_\Lambda)+(P+P_\Lambda)]=0\,.
\end{equation}
An effective dynamics from a running $\Lambda$ is also assumed as in the previous approach, but with different scale identification, so that $\rho_\Lambda(t)\propto t^{-2}$ in terms of cosmic time. Then, one gets \cite{War13,War14,War15}
\begin{equation}
\rho_\Lambda(t_{0})=\rho_{\rm vac}\approx 9.5 \times 10^{-121} M_{\mathrm{Pl}}^{4}\,,
\end{equation}
astonishingly close to the observed value (\ref{rholam}). The number of underlying assumptions and loose threads (for instance: (\ref{rholam}) is found assuming the dynamics of GR) suggest to take this result with caution. Although it may be too good to be true, more study will be required before discarding it as sheer coincidence.

\subsubsection{Causal sets: another bullseye prediction for \texorpdfstring{$\Lambda$}{Lambda}}

In this framework, we do not have continuous spacetime but, instead, a discrete structure of partially ordered points $(x \preceq y)$. These points are randomly generated on a Lorentzian manifold in such a way (called sprinkling) to reproduce the causal structure of continuous Lorentzian geometries. Volume measurements are replaced by point counting and are subject to an intrinsic uncertainty since the number of points $N$ can be determined only probabilistically, through a Poisson distribution with standard deviation $\Delta N=\sqrt{N}$. Denoting as $\alpha^2$ the average number of points occupying a Planckian volume, in four dimensions the volume uncertainty is then $\Delta \mathcal{V}=\sqrt{\mathcal{V}}/(\alpha M_{\mathrm{Pl}}^2)$. Regular lattices or graphs do not satisfy this property. The dynamics of these models is currently under construction through different approaches. 

In this setting, the cosmological constant arises as the canonical conjugate of the volume $\mathcal{V}$, so that $\Lambda$ is the action of $\alpha^2$ fundamental elements and
\begin{equation}
    \Delta \mathcal{V} \Delta \Lambda \geq M_\mathrm{Pl}^{-2}.
\end{equation}
In terms of cosmic time, $\Lambda(t)$ is a stochastic variable with zero average, $\langle\Lambda\rangle=0$, while the corresponding energy density is
\begin{equation}
    \rho_{\Lambda} =M_\mathrm{Pl}^{2}\Delta\Lambda \sim \frac{1}{\Delta \mathcal{V}}=\frac{\alpha M_{\mathrm{Pl}}^{2}}{\sqrt{\mathcal{V}}} \sim \alpha M_{\mathrm{Pl}}^{2} H^{2},
\end{equation}
which reproduces the observed value (\ref{rholam}) provided \cite{Sor90,ADGS,AhSo}
\begin{equation}\label{alpharange}
\alpha=O(10^{-2}) - O(1)\,.
\end{equation}

It has been argued that this range of $\alpha$ clashes with the prediction for CMB fluctuations \cite{Bar06,Zun07}. Heuristically, since $10^{-5}\sim \Delta T / T \sim \Delta \Phi / \Phi \sim \delta \rho_{\Lambda} / \rho_{\mathrm{tot}} \sim \alpha$. The model from which this estimate comes from relies on the assumption that at each spacetime point one can associate a value $\Lambda(x)$ determined only by the Poisson fluctuations within the past light cone of $x$. However, such model is based on stochastic fluctuations which are local and classical rather than quantum and it has not been embedded in a self-consistent covariant causal-set theory. Although we do not have a complete cosmology of causal sets yet, a more refined application of constraints from CMB data seems to be compatible with the range (\ref{alpharange}) \cite{Zwane:2017xbg}.

\section{Can GWs probe quantum gravity?}

In recent years, there has been a great effort to check whether and how GWs can probe theories, not necessarily of QG, beyond GR. One possibility is that such theories involve modifications of the mechanisms of production of GWs, while another possibility is to have modifications in the propagation of GWs. More literature can be found in the references below and in \cite{Calcagni:2020ume}. \medskip

\noindent \textbf{Production of GWs.} Modifications to the production mechanisms of GWs can appear in:
\begin{itemize}
    \item \textit{Modified black-hole metrics}, that could give rise to modified waveforms \cite{Canizares:2012is,Yunes:2016jcc,Barausse:2020rsu}. We know that GWs are either produced by astrophysical systems or from the early universe. For instance, in the first case, we could have modifications of gravity that imply different black hole solutions. This modification reflects into the GW waveform and we can test it with interferometers.
    \item \textit{Primordial blue-tilted tensor spectra}, that could give rise to an observable SGWB \cite{Barausse:2020rsu,Calcagni:2019kzo}. In the case of primordial GWs, some theories can predict a blue tilted tensor spectrum, which, in turn, would source an exotic SGWB.
\end{itemize}
\noindent \textbf{Propagation of GWs.} Modifications to the propagation of GWs can arise from:
\begin{itemize}
    \item \textit{Modified dispersion relations}, which can affect the GW propagation speed \cite{Yunes:2016jcc,EMNan,ArCa2} and the waveform phase \cite{Yunes:2016jcc,Mirshekari:2011yq}. After the production, the GW propagates through space and time until it reaches the observer. Since a constant non-luminal group velocity $c_{\rm t}\neq 1$ is tightly constraint, the simplest thing one can consider to encode a different propagation is a modified dispersion relation, which has repercussions in the GW waveform.
    \item \textit{Modified luminosity distance}, which can be tested through the observation of standard sirens \cite{Belgacem:2019pkk,Calcagni:2019ngc}. This is an example of test of new physics in the burgeoning field of multi-messenger astronomy \cite{Addazi:2021xuf}.
\end{itemize}

\subsection{Modified dispersion relations}\label{mdr}

In this sub-section, we consider a crude way in which GW detectors can constrain the propagation of GW. In models breaking Lorentz invariance, one can consider modified dispersion relations with an extra higher-order power of the spatial momentum. The energy $\omega=k^0$ and the momentum $k=|\mathbf{k}|$ are then not proportional to each other,
\begin{equation}\label{eq.MDR}
    \omega^{2}=k^{2}\left(1 \pm \frac{k^{n}}{M^{n}}\right)+O(k^{n+3}),
\end{equation}
where $M$ is a fundamental mass scale introduced for dimensional consistency. From this phenomenological dispersion relation, one can calculate the group velocity of the wave and the difference between this group velocity and the speed of light:
\begin{equation}
    v_g=\frac{\mathrm{d} \omega}{\mathrm{d} k}\,,\qquad \Delta v:=|v_g-1|\,.
\end{equation}
From this, one gets the expression of the fundamental mass 
\begin{equation}
    M \simeq \frac{\omega}{(\Delta v)^{\frac{1}{n}}},
\end{equation}
which depends on the GW frequency and an inverse power of the velocity difference $\Delta v$. If we consider the first gravitational wave detected by the LIGO-Virgo collaboration, GW150914, its frequency is $\omega \approx 630 \mathrm{~Hz} \approx 10^{-13} \mathrm{eV}$, corresponding to very low energies and a very tight constraint on the velocity difference \cite{Abbott:2016blz,TheLIGOScientific:2016src}:
\begin{equation}
    |\Delta v|<4 \times 10^{-20}.
\end{equation}
If we take a simple polynomial correction with $n=1$ or $n=2$, which may arise in certain scenarios, one obtains very weak bounds on the mass $M$, too weak to be of any use: $M>10^{-2}-10^7 \mathrm{~eV}$. Conversely, given a minimal mass $M>10$ TeV of order of the LHC center of mass energy, the values of $n$ compatible with the LIGO-Virgo bound lie in the range \cite{ArCa2}
\begin{equation}\label{n076}
    0<n<0.76\,.
\end{equation}
Such values have a geometrical interpretation because they are typical of field theories on multi-scale spacetimes, where $n$ is related to the dimension of spacetime \cite{Calcagni:2019ngc,revmu}. This example may be interesting because when we recover the notion of spacetime in the continuum limit, in general in QG it shows a multi-scale structure and the properties of geometry change with the scale. 

These arguments are valid only for models that break Lorentz invariance.

\subsection{Luminosity distance}

As we mentioned in section \ref{ludis}, in GR the amplitude of astrophysical gravitational waves is inversely proportional to the luminosity distance of their optical counterpart, so that $d_{L}^\textsc{gw}/d_{L}^\textsc{em}=1$. Deviations of GR arise, for instance, if we modify the dynamics of GWs through the action measure and kinetic term. Then, we end up with a phenomenological expression for the ratio $d_{L}^\textsc{gw}/d_{L}^\textsc{em}$, where we have a correction depending on a scale $\ell_*$ \cite{Calcagni:2019ngc},
\begin{equation} \label{eq.lumdist}
    \frac{d_{L}^\textsc{gw}}{d_{L}^\textsc{em}}=1 \pm|\Gamma-1|\left(\frac{d_{L}^\textsc{em}}{\ell_{*}}\right)^{\Gamma-1},
\end{equation}
where $\Gamma$ is a real parameter. Even when $\ell_{*} = O(\ell_{\rm Pl})$, there is a chance to have a detectable QG effect if $\Gamma \gtrsim 1$. To formally derive this result, we assume a measure ${\rm d}\varrho(x)$ whose scaling is the Hausdorff dimension $d_\textsc{h}$ and we write the linearized action for the graviton with a modified kinetic term as
\begin{equation}
    S = \frac{1}{2} \int {\rm d}\varrho(x)\,, h_{ij} \mathcal{K} (\Box) h^{ij} ,
\end{equation}
where $i,j$ are spatial indices. Then, the scaling of our GW, which we call $\Gamma$, depends on the Hausdorff dimension of spacetime, which in turn depends on the measure, and also on how the kinetic term scales with momentum,
\begin{equation}
   [h_{ij}] = \frac{d_\textsc{h} - [\mathcal{K}]}{2} =: \Gamma\,,
\end{equation}
where $[\mathcal{K}]$ is the dimensionality of the momentum-dependent part of the kinetic term. Assuming a constant $\Gamma$, in the local wave zone the GW amplitude then goes as
\begin{equation}\label{hij}
    h_{ij} = \frac{\kappa \mathcal{F}_{ij} (t-r)}{(r^2)^{\frac{\Gamma}{2} }} \sim \frac{1}{r^\Gamma} \rightarrow \frac{1}{(d_L^\textsc{em})^\Gamma}\,,
\end{equation}
where $\mathcal{F}_{ij}$ is a tensor retarded function of time and the local radial distance $r$ from the source, while the last step generalizes the result in the local wave zone to cosmological distances. When $\Gamma$ depends on the scale, (\ref{hij}) holds only asymptotically and one can consider the more general expression
(\ref{eq.lumdist}), where the factor $|\Gamma-1|$ has been inserted to avoid an incorrect normalization when $\Gamma=1$ (GR case). The scale $\ell_*$ is hidden in the modified kinetic term $\mathcal{K}(\Box)$, whose Fourier transform contains higher-order powers of the momentum and thus a scale appears in front of the leading-order term.

\begin{figure}[!htbp]
    \centering
    \includegraphics[width=0.75\linewidth]{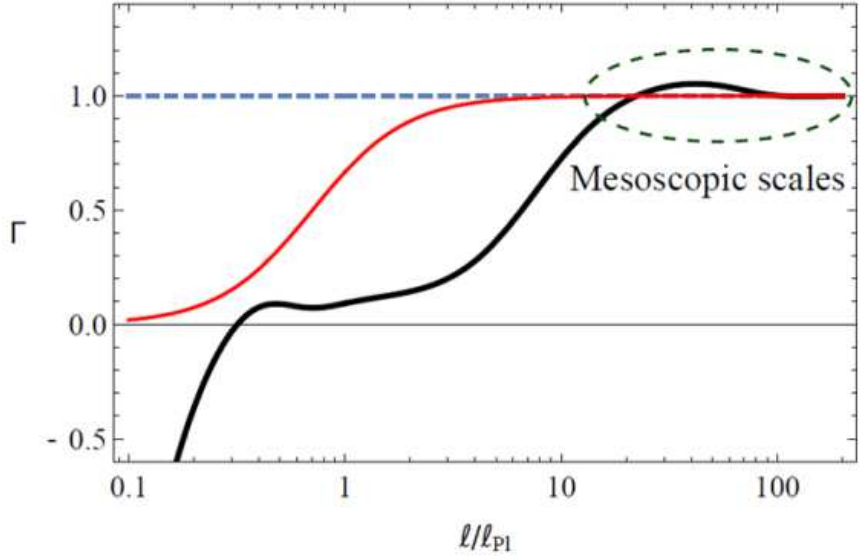}
    \caption{Dimensionality parameter $\Gamma$ of GWs as a function of the scale. The red curve corresponds to a monotonic evolution, for instance when the dimension of spacetime changes only in the deep UV, but there is no observable effect in the luminosity distance because $\Gamma$ is always smaller than one. The black curve corresponds to a generic modification of both the measure and the kinetic term: for mesoscopic (i.e., not too close to the UV) scales $\ell_*/\ell_{\rm Pl}\gtrsim O(10)$, the effect may actually become visible.\label{fig:gamma_L-distance}}
\end{figure}

In Fig.\ \ref{fig:gamma_L-distance}, we see two typical plots for $\Gamma$, which would arise if we change the measure and the graviton kinetic term. To get observable effects, we should approach the GR value $\Gamma=1$ from above. There are very few theories of QG that could satisfy this request and only three of them, among those that have been explored, could have a large-enough effect: loop quantum gravity, spin foams and GFT. In these scenarios, kinematical states of geometry give rise to a profile like the one in the figure, where the deviation from 1 can be, in principle, large enough. The problem is that we do not know whether the bump above 1 is physical or just a numerical artifact.

The general relation (\ref{eq.lumdist}) was tested with the only standard siren observed so far, the kilonova GW170817/GRB170817A caused by a binary neutron star merger \cite{Ab17b}, as well as with a simulated LISA super-massive black-hole standard siren \cite{Calcagni:2019ngc,Calcagni:2019kzo}. It was found that QG effects could get above the experimental uncertainty without violating any constraint when $\Gamma$ deviates from 1 by a few percent, depending on the value of $\ell_*$. For example, 
\begin{equation}
\Gamma\lesssim 1.02\qquad \textrm{when}\qquad \ell_*=O(\ell_{\rm Pl})\,,
\end{equation}
while for $\ell_*=O(10^{50}\ell_{\rm Pl})$ $\Gamma$ could be as large as $1.08$. Note also that, with the modified dispersion relation of section \ref{mdr}, in a spacetime with $d_\textsc{h}$ one has $\Gamma=1-n/2$ in the UV, so that the allowed range (\ref{n076}) translates into $\Gamma>0.72$, which does not place any further constraint for observable effects on the luminosity distance \cite{Calcagni:2019kzo}.

\subsection{SGWB}\label{sgwb}

We conclude by considering possible modifications to the SGWB. The amplitude of the stochastic background from inflation is relatively flat and well below the sensitivity in terms of present or future experiments. To have a detectable signal, we would need some mechanism blue tilting the primordial spectrum so that at higher frequencies, the amplitude of the spectrum would cross the sensitivity curves. 

In general, a blue tilted spectrum cannot be realized in standard slow-roll inflation. For instance, in Starobinsky gravity (\ref{starog}) the tensor spectrum is red tilted and with a very small amplitude. For these reasons, the predicted stochastic background is not detectable by present or future GW interferometers and neither is the SGWB of non-local Starobinsky inflation. The problem is that no matter which form factor we choose in (\ref{starnlP}), the blue tilt at CMB scales never gets a boost towards higher amplitudes at high frequencies. This happens because ${\rm H}_2\to 0$ at high frequencies and, eventually, the spectrum tends to the SGWB of the local Starobinsky model \cite{Calcagni:2020tvw}. In other words, a blue tilt at CMB scales does not guarantee a blue tilt at the scales of GW detectors.

\begin{figure}[ht]
    \centering
    \includegraphics[width=0.75\linewidth]{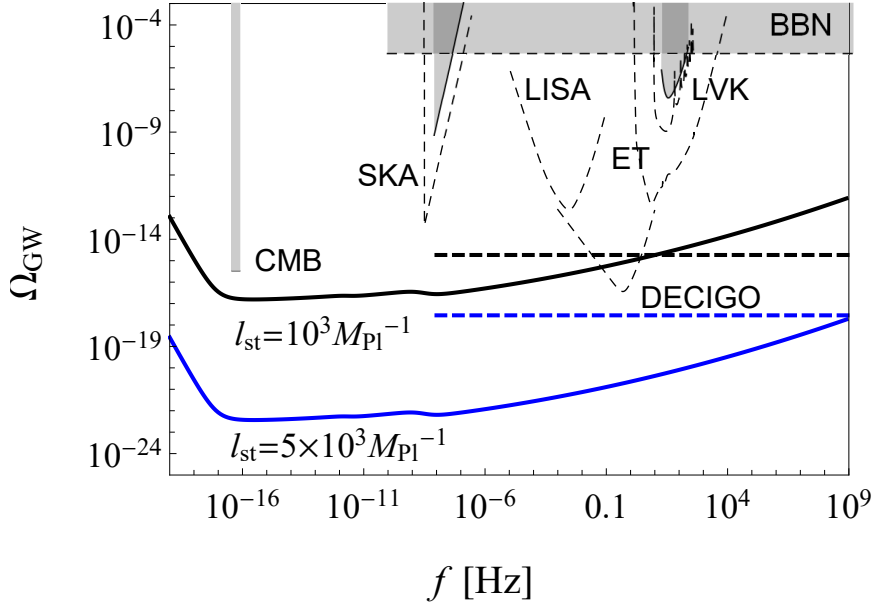}
    \caption{SGWB of string-gas cosmology compared with the sensitivity curves of LIGO-Virgo-KAGRA (LVK), SKA, LISA, Einstein Telescope (ET) and DECIGO. The black and blue solid curves corresponds to $l_{\rm st}=10^3 M_{\rm Pl}^{-1}$ and $5\times 10^3 M_{\rm Pl}^{-1}$, respectively, where $l_{\rm st}$ is the string length scale. The only known analytic expression of the tensor power spectrum is an approximation which fails for high-frequency GWs, since the corresponding curves (thick black and blue curves) go beyond the upper bound on the GW amplitude (dashed lines) obtained from the full spectrum. Note that we widened the single-frequency CMB constraint to a band in order to make it more visible. Figure adapted from \cite{Calcagni:2020tvw} (here we use the correct tensor-to-scalar ratio (\ref{newrsgg}) and adopt the updated upper bound $r=0.036$).}\label{fig:stringGas}
\end{figure}

\begin{figure}[ht]
    \centering
    \includegraphics[width=0.75\linewidth]{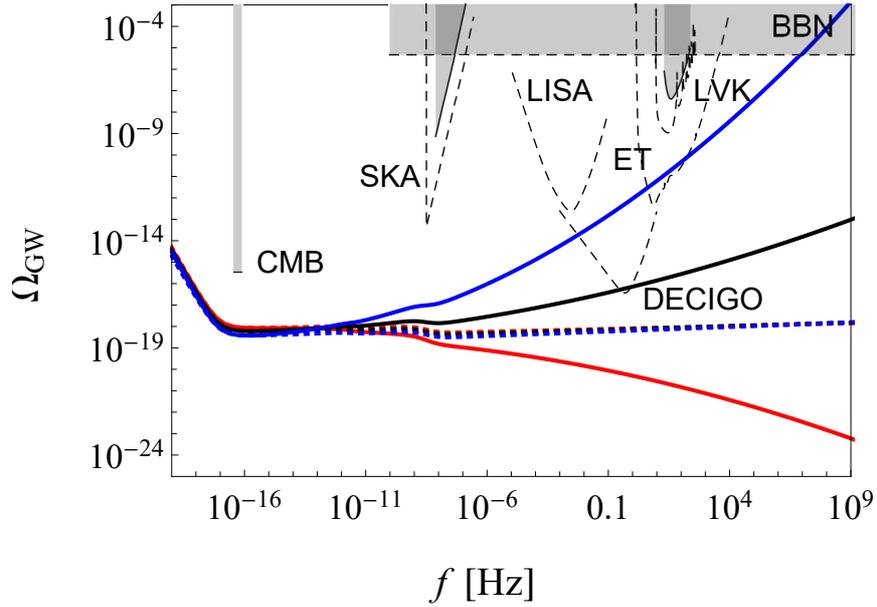}
    \caption{SGWB of the new ekpyrotic scenario compared with the sensitivity curves of LVK, SKA, LISA, ET and DECIGO. Denoting as $n_{\rm s}^{\rm obs}\pm\delta n_{\rm s}$ and $\alpha_{\rm s}^{\rm obs}\pm\delta \alpha_{\rm s}$ the preferred values obtained by \textsc{Planck}, we plot the worst case minimizing the tensor blue tilt and maximizing the negative tensor running at the $2\sigma$-level ($n_{\rm t}=1-(n_{\rm s}^{\rm obs}+2\delta n_{\rm s})\approx 0.026$, $\alpha_{\rm t}=-(\alpha_{\rm s}^{\rm obs}+2\delta \alpha_{\rm s})\approx -0.007$, red solid curve), the intermediate case taking the central values of the parameters (black solid curve) and the best case maximizing the tensor blue tilt and the positive tensor running at the $2\sigma$-level ($n_{\rm t}=1-(n_{\rm s}^{\rm obs}-2\delta n_{\rm s})\approx 0.042$, $\alpha_{\rm t}=-(\alpha_{\rm s}^{\rm obs}-2\delta \alpha_{\rm s})\approx 0.021$, blue solid curve). The dotted curves correspond to the above cases with no running. The high-frequency bend due to reheating is not shown. Note that we widened the single-frequency CMB constraint to a band in order to make it more visible. Figure adapted from \cite{Calcagni:2020tvw} (here we use the correct tensor-to-scalar ratio (\ref{rekpynew}) and adopt the updated upper bound $r=0.036$).}
    \label{fig:Escenario}
\end{figure}

\begin{figure}[ht]
   \centering
  \includegraphics[width=0.75\linewidth]{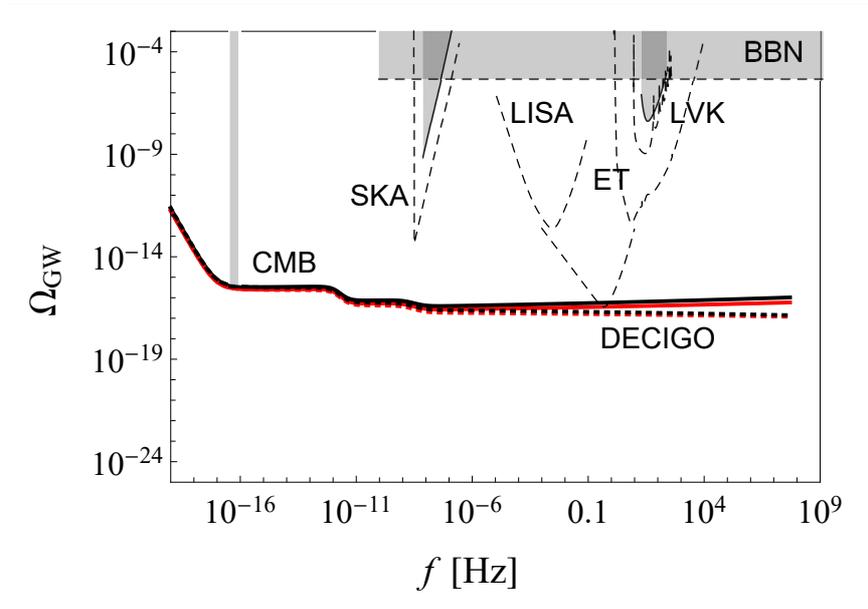}
 \caption{SGWB of the Brandenberger--Ho non-commutative model with ${\cal N}=50$ e-foldings compared with the sensitivity curves of LVK, SKA, LISA, ET and DECIGO. The red and black solid curves represent the cases with, respectively, $\phi_*^2=18M_{\rm Pl}^2$ and $\phi_*^2=21M_{\rm Pl}^2$, where $\phi_*$ is the field scale of natural inflation. Dotted curves correspond to the commutative cases with the above values of $\phi_*$. Note that we widened the single-frequency CMB constraint to a band in order to make it more visible and that the spectra in the plot satisfy the CMB bound at the pivot scale. Figure adapted from \cite{Calcagni:2020tvw} (here we report the correct LISA sensitivity curve and adopt the updated upper bound $r=0.036$).}
\label{fig:BH}
\end{figure}

\begin{figure}[ht]
   \centering
  \includegraphics[width=0.75\linewidth]{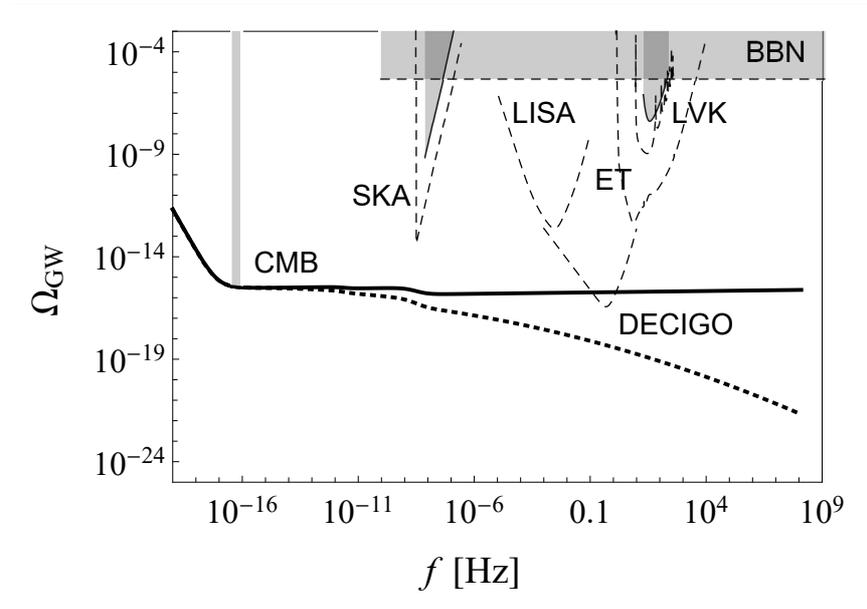}
 \caption{SGWB of multi-fractional inflation with no running ($\alpha_{\rm t}=0$) and fractional exponent $\alpha=-3$ (black solid curve), compared with the sensitivity curves of LVK, SKA, LISA, ET and DECIGO. Here $n_{\rm t}\approx -0.0045$ is given by the consistency relation $r=-8n_{\rm t}$ and we take $k_*=10^{-3}\,{\rm Mpc}^{-1}$. The dotted curve corresponds to the above case with non-zero running $\alpha_{\rm t}=-0.0001$ (typical order of magnitude of inflationary models). Note that we widened the single-frequency CMB constraint to a band in order to make it more visible and that the spectra in the plot satisfy the CMB bound at the pivot scale. Figure adapted from \cite{Calcagni:2020tvw} (here we report the correct LISA sensitivity curve and adopt the updated upper bound $r=0.036$).}
\label{fig:mf}
\end{figure}

It is natural to ask whether other models motivated by QG that give rise to a blue tilted primordial spectrum at inflationary scales exist. In \cite{Calcagni:2020tvw}, four such models were explored:
\begin{itemize}
    \item String-gas cosmology.
    \item New ekpyrotic scenario.
    \item Brandenberger--Ho noncommutative inflation.
		\item Multi-fractional inflation.
\end{itemize}
The first two models are partially embedded in string theory (where ``partially'' means with caveats, several assumptions and undeveloped aspects), the third and fourth are inspired by QG (non-commutative spacetimes and multi-scale spacetimes) but they are not fully embedded in it. However debatable they are, these models offer a prediction of the SGWB that can be compared with present or near-future observational data. Other models that could produce a detactable signal but that have not been studied yet in detail are the pre-big-bang scenario \cite{Gasperini:1992em,Gasperini:2016gre} and bouncing models with a slow contracting phase \cite{Ben-Dayan:2016iks,Ben-Dayan:2018ksd,Artymowski:2020pci}.
\nonBulletListing{String-gas cosmology}
In string-gas cosmology, the early universe is described as a gas of strings at thermal equilibrium \cite{BrVa1,Brandenberger:2015kga,Bernardo:2020bpa}. This model is an alternative to inflation and CMB fluctuations are explained as thermal fluctuations of the string gas. There is a maximum amplitude for the power spectrum derived from its full expression and thus we do not expect the signal to reach the bottom of the DECIGO curve (Fig.~\ref{fig:stringGas}). The tensor-to-scalar ratio is\footnote{Formul\ae\ (\ref{newrsgg}) and (\ref{rekpynew}) correct a typo in, respectively, equations (4.35) and (4.39) of \cite{Calcagni:2020tvw}, which were supposed to introduce a missing prefactor in the results of previous literature.}
\begin{equation}
r=\frac{4}{9}\left(1-\hat T\right)^2\ln^2\left[\frac{1-\hat T}{(l_{\rm st}k)^2}\right],
\label{newrsgg}
\end{equation}
where $\hat T:=T(k)/T_{\rm H}$ is the string-gas temperature divided by the Hagedorn temperature $T_{\rm H}$ and $l_{\rm st}$ is the string length scale.
\nonBulletListing{New ekpyrotic scenario}
In the new ekpyrotic scenario \cite{Brandenberger:2020tcr,Brandenberger:2020eyf,Brandenberger:2020wha}, inflation is replaced by the collision of two 3-branes. The tensor-to-scalar ratio at the pivot scale $k_0$ is
\begin{equation}\label{rekpynew}
r\simeq \frac{2^{n_{\rm s}+2}}{\Gamma^2\!\left(1-\frac{n_{\rm s}}{2}\right)}(k_0\tau_{\rm B})^{2(1-n_{\rm s})}\,(1-n_{\rm s})^2\,,
\end{equation}
where $\Gamma$ is Euler's function and $\tau_{\rm B}$ is the conformal time at which a cosmological bounce takes place. This model respects all the CMB bounds and predicts a blue-tilted tensor spectrum, which could be observed by the Einstein Telescope (Fig.~\ref{fig:Escenario}). However, if the spectrum goes up too sharply, it may hit the BBN bound. We have to take into account reheating, which in general bends the spectrum down \cite{Seto:2003kc,Kuroyanagi:2014nba}. Therefore, when we have reheating in our early universe model, the spectrum can respect the BBN constraints but it can also bend too early and go below the sensitivity curves. This bending depends on the reheating temperature. This model can only reach the Einstein Telescope sensitivity curve, a EU interferometer under proposal.
\nonBulletListing{Brandenberger--Ho non-commutative inflation}
This is a phenomenological model motivated by non-commutativity in string theory \cite{BH}. The inflation is realized by a scalar field on a non-commutative background and gives rise to a blue tilt in the SGWB, which barely crosses the DECIGO sensitivity curve (Fig.~\ref{fig:BH}), the model could give rise to an observable but weak signal, available only for one Japanese detector which is still in a pre-approval stage.
\nonBulletListing{Multi-fractional spacetimes}
In multi-fractional spacetimes, the scale-dependent Hausdorff and spectral dimensions of spacetime distort the primordial inflationary spectrum at post-CMB scales making it blue tilted. In particular, for the so-called theory with $q$-derivatives, the SGWB can be blue tilted even if the tensor spectral index $n_{\rm t}$ is negligible at CMB scales. The predicted SGWB crosses the DECIGO sensitivity curve (Fig.~\ref{fig:mf}).


\section*{Acknowledgments}

G.C.\ is supported by the I+D grant PID2020-118159GB-C41 of the Spanish Ministry of Science and Innovation and acknowledges networking support by the COST Action CA18108. He also thanks the organizers of the Training School for the invitation (E.\ Saridakis, N.\ Mavromatos, J.M.\ Carmona, G.\ Djordjevic, G.\ Gubitosi, A.\ di Matteo, C.\ P\'erez de los Heros, C.\ Pfeifer and T.\ Terzi\'c). T.F.\ thanks his supervisor J.R.\ H\"orandel for his support.



\begin{thebibliography}{99}
\bibitem{wei72} \au{S}{Weinberg}, \book{Gravitation and Cosmology}{Wiley}{New York}{NY}{1972}.
\bibitem{LiL}   \au{D.H}{Lyth} and \au{A.R}{Liddle}, \book{The Primordial Density Perturbation}{Cambridge University Press}{Cambridge}{UK}{2009}.
\bibitem{Muk}   \au{V}{Mukhanov}, \book{Physical Foundations of Cosmology}{Cambridge University Press}{Cambridge}{UK}{2005}.
\bibitem{Calcagni:2017sdq}   \au{G}{Calcagni}, \books{\href{10.1007/978-3-319-41127-9}{\cob Classical and Quantum Cosmology}}{Springer}{Switzerland}{2017}.
\bibitem{Hoffman:2017ako} \au{Y}{Hoffman}, \au{D}{Pomarede}, \au{R.B}{Tully} and \au{H}{Courtois}, \tia{The Dipole Repeller} \doinn{10.1038/s41550-016-0036}{Nature Astron.}{1}{0036}{2017} [\arX{1702.02483}].
\bibitem{Verde:2019ivm} \au{L}{Verde}, \au{T}{Treu} and \au{A.G}{Riess}, \tia{Tensions between the early and the late universe} \doinn{10.1038/s41550-019-0902-0}{Nature Astron.}{3}{891}{2019} [\arX{1907.10625}].
\bibitem{Perivolaropoulos:2021jda} \au{L}{Perivolaropoulos} and \au{F}{Skara}, \tia{Challenges for $\Lambda$CDM: an update} \arX{2105.05208}.
\bibitem{ParticleDataGroup:2020ssz} \au{P.A}{Zyla} {et al.} [Particle Data Group],
\tia{Review of particle physics} \doinn{10.1093/ptep/ptaa104}{Prog.\ Theor.\ Exp.\ Phys.}{2020}{083C01}{2020}.
\bibitem{Wong:2019kwg} \au{K.C}{Wong} {et al.}, \tia{H0LiCOW -- XIII. A 2.4 per cent measurement of $H_0$ from lensed quasars: $5.3\sigma$ tension between early- and late-Universe probes} \doinn{10.1093/mnras/stz3094}{Mon.\ Not.\ Roy.\ Astron.\ Soc.}{498}{1420}{2020} [\arX{1907.04869}].
\bibitem{Riess:2019cxk} \au{A.G}{Riess}, \au{S}{Casertano}, \au{W}{Yuan}, \au{L.M}{Macri} and \au{D}{Scolnic}, \tia{Large Magellanic Cloud Cepheid standards provide a 1\% foundation for the determination of the Hubble constant and stronger evidence for physics beyond $\Lambda$CDM} \doinn{10.3847/1538-4357/ab1422}{Astrophys.\ J.}{876}{85}{2019} [\arX{1903.07603}].
\bibitem{Freedman:2019jwv} \au{W.L}{Freedman} {et al.} \tia{The Carnegie--Chicago Hubble Program. VIII. An independent determination of the Hubble constant based on the tip of the red giant branch} \doinn{10.3847/1538-4357/ab2f73}{Astrophys.\ J.}{882}{34}{2019} [\arX{1907.05922}].
\bibitem{Pesce:2020xfe} \au{D.W}{Pesce} {et al.} \tia{The Megamaser Cosmology Project. XIII. Combined Hubble constant constraints} \doinn{10.3847/2041-8213/ab75f0}{Astrophys.\ J.\ Lett.}{891}{L1}{2020} [\arX{2001.09213}].
\bibitem{eBOSS:2020yzd} \au{S}{Alam} {et al.} [eBOSS Collaboration], \tia{Completed SDSS-IV extended Baryon Oscillation Spectroscopic Survey: Cosmological implications from two decades of spectroscopic surveys at the Apache Point Observatory} \doin{10.1103/PhysRevD.103.083533}{Phys.\ Rev.}{D}{103}{083533}{2021} [\arX{2007.08991}].
\bibitem{Planck:2018vyg} \au{N}{Aghanim} {et al.} [\textsc{Planck} Collaboration], \tia{Planck 2018 results. VI. Cosmological parameters} \doinn{10.1051/0004-6361/201833910}{Astron.\ Astrophys.}{641}{A6}{2020}; \doinn{10.1051/0004-6361/201833910e}{Erratum-Ibid.}{652}{C4}{2021} [\arX{1807.06209}].
\bibitem{ACT:2020gnv} \au{S}{Aiola} {et al.} [ACT Collaboration], \tia{The Atacama Cosmology Telescope: DR4 maps and cosmological parameters} \doij{10.1088/1475-7516/2020/12/047}{JCAP}{2012}{047}{2020} [\arX{2007.07288}].
\bibitem{Abbott:2016blz} \au{B.P}{Abbott} et al.\ [LIGO Scientific and \textsc{Virgo} Collaborations], \tia{Observation of gravitational waves from a binary black hole merger} \doinn{10.1103/PhysRevLett.116.061102}{Phys.\ Rev.\ Lett.}{116}{061102}{2016} [\arX{1602.03837}].
\bibitem{TheLIGOScientific:2016src} \au{B.P}{Abbott} et al.\ [LIGO Scientific and \textsc{Virgo} Collaborations], \tia{Tests of general relativity with GW150914} \doinn{10.1103/PhysRevLett.116.221101}{Phys.\ Rev.\ Lett.}{116}{221101}{2016}; \doinn{10.1103/PhysRevLett.121.129902}{Erratum-ibid.}{121}{129902}{2018} [\arX{1602.03841}].
\bibitem{CoLu}  \au{P}{Coles} and \au{F}{Lucchin}, \book{Cosmology {\rm (2nd ed.)}}{Wiley}{Chichester}{UK}{2002}.
\bibitem{Belgacem:2019pkk} \au{E}{Belgacem} {et al.} [LISA Cosmology Working Group], \tia{Testing modified gravity at cosmological distances with LISA standard sirens} \doij{10.1088/1475-7516/2019/07/024}{JCAP}{1907}{024}{2019} [\arX{1906.01593}].
\bibitem{Calcagni:2019ngc} \au{G}{Calcagni}, \au{S}{Kuroyanagi}, \au{S}{Marsat}, \au{M}{Sakellariadou}, \au{N}{Tamanini} and \au{G}{Tasinato}, \tia{Quantum gravity and gravitational-wave astronomy} \doij{10.1088/1475-7516/2019/10/012}{JCAP}{1910}{012}{2019} [\arX{1907.02489}].
\bibitem{Belgacem:2020pdz} \au{E}{Belgacem}, \au{Y}{Dirian}, \au{A}{Finke}, \au{S}{Foffa} and \au{M}{Maggiore}, \tia{Gravity in the infrared and effective nonlocal models} \doij{10.1088/1475-7516/2020/04/010}{JCAP}{2004}{010}{2020} [\arX{2001.07619}].
\bibitem{Fix96} \au{D.J}{Fixsen}, \au{E.S}{Cheng}, \au{J.M}{Gales}, \au{J.C}{Mather}, \au{R.A}{Shafer} and \au{E.L}{Wright}, \tia{The cosmic microwave background spectrum from the full COBE/FIRAS data set} \doinn{10.1086/178173}{Astrophys.\ J.}{473}{576}{1996} [\oarX{astro-ph/9605054}].
\bibitem{WMAP}  \url{http://map.gsfc.nasa.gov}.
\bibitem{Ben12} \au{C.L}{Bennett} {et al.} [WMAP Collaboration], \tia{Nine-year Wilkinson Microwave Anisotropy Probe (WMAP) observations: final maps and results} \doinn{10.1088/0067-0049/208/2/20}{Astrophys.\ J.\ Suppl.}{208}{20}{2013} [\arX{1212.5225}].
\bibitem{Hin12} \au{G}{Hinshaw} {et al.} [WMAP Collaboration], \tia{Nine-year Wilkinson Microwave Anisotropy Probe (WMAP) observations: cosmological parameter results} \doinn{10.1088/0067-0049/208/2/19}{Astrophys.\ J.\ Suppl.}{208}{19}{2013} [\arX{1212.5226}].
\bibitem{Planck:2018nkj} \au{N}{Aghanim} {et al.} [\textsc{Planck} Collaboration], \tia{Planck 2018 results. I. Overview and the cosmological legacy of Planck} \doinn{10.1051/0004-6361/201833880}{Astron.\ Astrophys.}{641}{A1}{2020} [\arX{1807.06205}].
\bibitem{WMAPc4}  \url{http://map.gsfc.nasa.gov}.
\bibitem{BICEP:2021xfz} \au{P.A.R}{Ade} {et al.} [BICEP/Keck Collaboration], \tia{Improved constraints on primordial gravitational waves using Planck, WMAP, and BICEP/Keck observations through the 2018 observing season} \doinn{10.1103/PhysRevLett.127.151301}{Phys.\ Rev.\ Lett.}{127}{151301}{2021} [\arX{2110.00483}].
\bibitem{WMAP:2003syu} \au{H.V}{Peiris} {et al.} [WMAP Collaboration], \tia{First year Wilkinson Microwave Anisotropy Probe (WMAP) observations: Implications for inflation} \doinn{10.1086/377228}{Astrophys.\ J.\ Suppl.}{148}{213}{2003} [\arX{astro-ph/0302225}].
\bibitem{Planck:2013jfk} \au{P.A.R}{Ade} {et al.} [\textsc{Planck} Collaboration],
\tia{Planck 2013 results. XXII. Constraints on inflation} \doinn{10.1051/0004-6361/201321569}{Astron.\ Astrophys.}{571}{A22}{2014} [\arX{1303.5082}].
\bibitem{BLT1}  \au{C.P}{Burgess}, \au{H.M}{Lee} and \au{M}{Trott}, \tia{Power-counting and the validity of the classical approximation during inflation} \doij{10.1088/1126-6708/2009/09/103}{JHEP}{0909}{103}{2009} [\arX{0902.4465}]. 
\bibitem{BaEs}  \au{J.L.F}{Barb\'on} and \au{J.R}{Espinosa}, \tia{Naturalness of Higgs inflation} \doin{10.1103/PhysRevD.79.081302}{Phys.\ Rev.}{D}{79}{081302}{2009} [\arX{0903.0355}].
\bibitem{BGS2}  \au{F}{Bezrukov}, \au{D}{Gorbunov} and \au{M}{Shaposhnikov}, \tia{Late and early time phenomenology of Higgs-dependent cutoff} \doij{10.1088/1475-7516/2011/10/001}{JCAP}{1110}{001}{2011} [\arX{1106.5019}].
\bibitem{BMSS}  \au{F}{Bezrukov}, \au{A}{Magnin}, \au{M}{Shaposhnikov} and \au{S}{Sibiryakov}, \tia{Higgs inflation: consistency and generalisations} \doij{10.1007/JHEP01(2011)016}{JHEP}{1101}{016}{2011} [\arX{1008.5157}].
\bibitem{AtC2}  \au{M}{Atkins} and \au{X}{Calmet}, \tia{Remarks on Higgs inflation} \doin{10.1016/j.physletb.2011.01.028}{Phys.\ Lett.}{B}{697}{37}{2011} [\arX{1011.4179}].
\bibitem{Yamaguchi:2011kg} \au{M}{Yamaguchi} \tia{Supergravity based inflation models: a review} \doinn{10.1088/0264-9381/28/10/103001}{Class.\ Quantum Grav.}{28}{103001}{2011} [\arX{1101.2488}].
\bibitem{Pen65} \au{R}{Penrose}, \tia{Gravitational collapse and space-time singularities} \doinn{10.1103/PhysRevLett.14.57}{Phys.\ Rev.\ Lett.}{14}{57}{1965}. 
\bibitem{Haw1}  \au{S.W}{Hawking}, \tia{Occurrence of singularities in open universes} \doinn{10.1103/PhysRevLett.15.689}{Phys.\ Rev.\ Lett.}{15}{689}{1965}.
\bibitem{Haw2}  \au{S.W}{Hawking}, \tia{The occurrence of singularities in cosmology} \doin{10.1098/rspa.1966.0221}{Proc.\ Roy.\ Soc.\ Lond.}{A}{294}{511}{1966}.
\bibitem{Haw3}  \au{S.W}{Hawking}, \tia{The occurrence of singularities in cosmology. II} \doin{10.1098/rspa.1966.0255}{Proc.\ Roy.\ Soc.\ Lond.}{A}{295}{490}{1966}.
\bibitem{Ger66} \au{R.P}{Geroch}, \tia{Singularities in closed universes} \doinn{10.1103/PhysRevLett.17.445}{Phys.\ Rev.\ Lett.}{17}{445}{1966}.
\bibitem{Haw4}  \au{S.W}{Hawking}, \tia{The occurrence of singularities in cosmology. III. Causality and singularities} \doin{10.1098/rspa.1967.0164}{Proc.\ Roy.\ Soc.\ Lond.}{A}{300}{187}{1967}.
\bibitem{HaP}   \au{S.W}{Hawking} and \au{R}{Penrose}, \tia{The singularities of gravitational collapse and cosmology} \doin{10.1098/rspa.1970.0021}{Proc.\ Roy.\ Soc.\ Lond.}{A}{314}{529}{1970}.
\bibitem{BGV}   \au{A}{Borde}, \au{A.H}{Guth} and \au{A}{Vilenkin}, \tia{Inflationary spacetimes are incomplete in past directions}
  \doinn{10.1103/PhysRevLett.90.151301}{Phys.\ Rev.\ Lett.}{90}{151301}{2003} [\oarX{gr-qc/0110012}].
\bibitem{Mis69} \au{C.W}{Misner},  \tia{Mixmaster universe} \doinn{10.1103/PhysRevLett.22.1071}{Phys.\ Rev.\ Lett.}{22}{1071}{1969}.
\bibitem{BKL70} \au{V.A}{Belinski\u{\i}}, \au{I.M}{Khalatnikov} and \au{E.M}{Lifshitz}, \tia{Oscillatory approach to a singular point in the relativistic cosmology} \doinn{10.1080/00018737000101171}{Adv.\ Phys.}{19}{525}{1970}.
\bibitem{BKL82} \au{V.A}{Belinski\u{\i}}, \au{I.M}{Khalatnikov} and \au{E.M}{Lifshitz}, \tia{A general solution of the Einstein equations with a time singularity} \doinn{10.1080/00018738200101428}{Adv.\ Phys.}{31}{639}{1982}.
\bibitem{Dad}   \au{T}{Damour} and \au{S}{de Buyl}, \tia{Describing general cosmological singularities in Iwasawa variables}
  \doin{10.1103/PhysRevD.77.043520}{Phys.\ Rev.}{D}{77}{043520}{2008} [\arX{0710.5692}].
\bibitem{Martin:2012bt} \au{J}{Martin}, \tia{Everything you always wanted to know about the cosmological constant problem (but were afraid to ask)} \doinn{10.1016/j.crhy.2012.04.008}{Comptes Rendus Phys.}{13}{566}{2012} [\arX{1205.3365}].
\bibitem{Zel68} \au{Ya.B}{Zel'dovich}, \tia{The cosmological constant and the theory of elementary particles} \doinn{10.1070/PU1968v011n03ABEH003927}{Sov.\ Phys.\ Usp.}{11}{381}{1968}.
\bibitem{Akhmedov:2002ts} \au{E.K}{Akhmedov}, \tia{Vacuum energy and relativistic invariance} \oarX{hep-th/0204048}.
\bibitem{KoP}   \au{J.F}{Koksma} and \au{T}{Prokopec}, \tia{The cosmological constant and Lorentz invariance of the vacuum state} \arX{1105.6296}.
\bibitem{Ezk14} \au{I}{Ezkurdia}, \au{D}{Juan}, \au{J.M}{Rodr\'iguez}, \au{A}{Frankish}, \au{M}{Diekhans}, \au{J}{Harrow}, \au{J}{V\'azquez}, \au{A}{Valencia} and \au{M.L}{Tress}, \tia{Multiple evidence strands suggest that there may be as few as $19000$ human protein-coding genes} \doinn{10.1093/hmg/ddu309}{Hum.\ Mol.\ Genetics}{23}{5866}{2014}.
\bibitem{ensembl} \url{http://www.ensembl.org/Homo_sapiens/Info/Annotation}.
\bibitem{Lew67} \au{R.C}{Lewontin}, \tia{An estimate of average heterozygosity in man} \ndoinn{https://www.ncbi.nlm.nih.gov/pmc/articles/PMC1706234}{Am.\ J.\ Hum.\ Genet.}{19}{681}{1967}.
\bibitem{HaHo} \au{H}{Harris} and \au{D.A}{Hopkinson}, \tia{Average heterozygosity per locus in man: an estimate based on the incidence of enzyme polymorphisms} \doinn{10.1111/j.1469-1809.1972.tb00578.x}{Ann.\ Hum.\ Genet.}{36}{9}{1972}.
\bibitem{Wil13} \au{H.F}{Willard}, \tia{\href{https://doi.org/10.1016/B978-0-12-382227-7.00001-X}{\cob The human genome: a window on human genetics, biology and medicine}} \procsinm{Genomic and Personalized Medicine (2nd ed.)}{\au{G.S}{Ginsburg} and \au{H.F}{Willard}}{Academic Press}{San Diego}{2013}.
\bibitem{Wei89} \au{S}{Weinberg}, \tia{The cosmological constant problem} \doinn{10.1103/RevModPhys.61.1}{Rev.\ Mod.\ Phys.}{61}{1}{1989}.
\bibitem{Pad14} \au{T}{Padmanabhan} and \au{H}{Padmanabhan}, \tia{Cosmological constant from the emergent gravity perspective} \doin{10.1142/S0218271814300110}{Int.\ J.\ Mod.\ Phys.}{D}{23}{1430011}{2014} [\arX{1404.2284}].
\bibitem{Tsu13} \au{S}{Tsujikawa}, \tia{Quintessence: a review} \doinn{10.1088/0264-9381/30/21/214003}{Class.\ Quantum Grav.}{30}{214003}{2013} [\arX{1304.1961}].
\bibitem{Wetterich:1994bg} \au{C}{Wetterich}, \tia{The Cosmon model for an asymptotically vanishing time dependent cosmological `constant'} \ndoinn{https://ui.adsabs.harvard.edu/abs/1995A\%26A...301..321W/abstract}{Astron.\ Astrophys.}{301}{321}{1995} [\arX{hep-th/9408025}].
\bibitem{Amendola:1999er} \au{L}{Amendola}, \tia{Coupled quintessence} \doin{10.1103/PhysRevD.62.043511}{Phys.\ Rev.}{D}{62}{043511}{2000} [\arX{astro-ph/9908023}].
\bibitem{Peri05} \au{L}{Perivolaropoulos}, \tia{Crossing the phantom divide barrier with scalar tensor theories} \doij{10.1088/1475-7516/2005/10/001}{JCAP}{0510}{001}{2005} [\oarX{astro-ph/0504582}].
\bibitem{Davari:2019tni} \au{Z}{Davari}, \au{V}{Marra} and \au{M}{Malekjani}, \tia{Cosmological constrains on minimally and non-minimally coupled scalar field models} \doinn{10.1093/mnras/stz3096}{Mon.\ Not.\ Roy.\ Astron.\ Soc.}{491}{1920}{2020} [\arX{1911.00209}].
\bibitem{Banerjee:2020xcn} \au{A}{Banerjee}, \au{H}{Cai}, \au{L}{Heisenberg}, \au{E}{\'O Colg\'ain}, \au{M.M}{Sheikh-Jabbari} and \au{T}{Yang}, \tia{Hubble sinks in the low-redshift swampland} \doin{10.1103/PhysRevD.103.L081305}{Phys.\ Rev.}{D}{103}{L081305}{2021} [\arX{2006.00244}].
\bibitem{Lee:2022cyh} \au{B.-H}{Lee}, \au{W}{Lee}, \au{E}{\'O Colg\'ain}, \au{M.M}{Sheikh-Jabbari} and \au{S}{Thakur}, \tia{Is local $H_0$ at odds with dark energy EFT?} \arX{2202.03906}.
\bibitem{Tsu10} \au{S}{Tsujikawa}, \tia{Modified gravity models of dark energy} \doinn{10.1007/978-3-642-10598-2_3}{Lect.\ Notes Phys.}{800}{99}{2010} [\arX{1101.0191}].
\bibitem{ClFPS} \au{T}{Clifton}, \au{P.G}{Ferreira}, \au{A}{Padilla} and \au{C}{Skordis}, \tia{Modified gravity and cosmology} \doinn{10.1016/j.physrep.2012.01.001}{Phys.\ Rept.}{513}{1}{2012} [\arX{1106.2476}].
\bibitem{GoSa2}  \au{M.H}{Goroff} and \au{A}{Sagnotti}, \tia{The ultraviolet behavior of Einstein gravity} \doin{10.1016/0550-3213(86)90193-8}{Nucl.\ Phys.}{B}{266}{709}{1986}.
\bibitem{Bojowald:2011hd} \au{M}{Bojowald}, \au{G}{Calcagni} and \au{S}{Tsujikawa}, \tia{Observational constraints on loop quantum cosmology} \doinn{10.1103/PhysRevLett.107.211302}{Phys.\ Rev.\ Lett.}{107}{211302}{2011} [\arX{1101.5391}].
\bibitem{Modesto:2017sdr} \au{L}{Modesto} and \au{L}{Rachwa\l}, \tia{Nonlocal quantum gravity: a review} \doin{10.1142/S0218271817300208}{Int.\ J.\ Mod.\ Phys.}{D}{26}{1730020}{2017}.
\bibitem{Calcagni:2021hve} \au{G}{Calcagni}, \tia{\href{https://doi.org/10.1007/978-3-030-83715-0_9}{\cob Non-local gravity}} \procsinm{Modified Gravity and Cosmology}{\au{E.N}{Saridakis} {et al.}}{Springer}{Switzerland}{2021}. 
\bibitem{Barrau:2014maa} \au{A}{Barrau}, \au{M}{Bojowald}, \au{G}{Calcagni}, \au{J}{Grain} and \au{M}{Kagan}, \tia{Anomaly-free cosmological perturbations in effective canonical quantum gravity} \doij{10.1088/1475-7516/2015/05/051}{JCAP}{1505}{051}{2015} [\arX{1404.1018}].
\bibitem{Bojowald:2015iga} \au{M}{Bojowald}, \tia{Quantum cosmology: a review} \doinn{10.1088/0034-4885/78/2/023901}{Rept.\ Prog.\ Phys.}{78}{023901}{2015} [\arX{1501.04899}].
\bibitem{ElizagaNavascues:2020uyf} \au{B}{Elizaga Navascu\'es} and \au{G.A}{Mena Marug\'an}, \tia{Hybrid loop quantum cosmology: an overview} \doinn{10.3389/fspas.2021.624824}{Front.\ Astron.\ Space Sci.}{8}{81}{2021} [\arX{2011.04559}].
\bibitem{Li:2021mop} \au{B.-F}{Li}, \au{P}{Singh} and \au{A}{Wang}, \tia{Phenomenological implications of modified loop cosmologies: an overview} \doinn{10.3389/fspas.2021.701417}{Front.\ Astron.\ Space Sci.}{8}{701417}{2021} [\arX{2105.14067}].
\bibitem{BaMcA} \au{D}{Baumann} and \au{L}{McAllister}, \book{\href{http://dx.doi.org/10.1017/CBO9781316105733}{\cob Inflation and String Theory}}{Cambridge University Press}{Cambridge}{UK}{2015} [\arX{1404.2601}].
\bibitem{GaVe4} \au{M}{Gasperini} and \au{G}{Veneziano}, \tia{The pre-big bang scenario in string cosmology} \doinn{10.1016/S0370-1573(02)00389-7}{Phys.\ Rept.}{373}{1}{2003} [\oarX{hep-th/0207130}].
\bibitem{LMS3}  \au{H}{Liu}, \au{G.W}{Moore} and \au{N}{Seiberg}, \tia{The challenging cosmic singularity} \oarX{gr-qc/0301001}.
\bibitem{CoCo2} \au{L}{Cornalba} and \au{M.S}{Costa}, \tia{Time-dependent orbifolds and string cosmology} \doinn{10.1002/prop.200310123}{Fortsch.\ Phys.}{52}{145}{2004} [\oarX{hep-th/0310099}].
\bibitem{Cra06} \au{B}{Craps}, \tia{Big bang models in string theory} \doinn{10.1088/0264-9381/23/21/S01}{Class.\ Quantum Grav.}{23}{S849}{2006} [\oarX{hep-th/0605199}].
\bibitem{BeRe}  \au{M}{Berkooz} and \au{D}{Reichmann}, \tia{A short review of time dependent solutions and space-like singularities in string theory} \doinn{10.1016/j.nuclphysbps.2007.06.008}{Nucl.\ Phys.\ Proc.\ Suppl.}{171}{69}{2007} [\arX{0705.2146}].
\bibitem{DaNi3} \au{T}{Damour} and \au{H}{Nicolai}, \tia{Symmetries, singularities and the de-emergence of space} \doin{10.1142/S0218271808012206}{Int.\ J.\ Mod.\ Phys.}{D}{17}{525}{2008} [\arX{0705.2643}].
\bibitem{Biswas:2005qr}   \au{T}{Biswas}, \au{A}{Mazumdar} and \au{W}{Siegel}, \tia{Bouncing universes in string-inspired gravity} \doij{10.1088/1475-7516/2006/03/009}{JCAP}{0603}{009}{2006} [\oarX{hep-th/0508194}].
\bibitem{Koshelev:2013lfm} \au{A.S}{Koshelev}, \tia{Stable analytic bounce in non-local Einstein--Gauss--Bonnet cosmology} \doinn{10.1088/0264-9381/30/15/155001}{Class.\ Quantum Gravity}{30}{155001}{2013} [\arX{1302.2140}].
\bibitem{Calcagni:2013vra}  \au{G}{Calcagni}, \au{L}{Modesto} and \au{P}{Nicolini}, \tia{Super-accelerating bouncing cosmology in asymptotically-free non-local gravity} \doin{10.1140/epjc/s10052-014-2999-8}{Eur.\ Phys.\ J.}{C}{74}{2999}{2014} [\arX{1306.5332}].
\bibitem{De671} \au{B.S}{DeWitt}, \tia{Quantum theory of gravity. I. The canonical theory} \doinn{10.1103/PhysRev.160.1113}{Phys.\ Rev.}{160}{1113}{1967}.
\bibitem{Kim5}  \au{S.P}{Kim}, \tia{Quantum potential and cosmological singularities}
  \doin{10.1016/S0375-9601(97)00744-5}{Phys.\ Lett.}{A}{236}{11}{1997} [\oarX{gr-qc/9703065}].
\bibitem{Bojowald:2020wuc} \au{M}{Bojowald}, \tia{Critical evaluation of common claims in loop quantum cosmology} \doinn{10.3390/universe6030036}{Universe}{6}{36}{2020} [\arX{2002.05703}].
\bibitem{Per13} \au{A}{Perez}, \tia{The spin-foam approach to quantum gravity} \doinn{10.12942/lrr-2013-3}{Living Rev.\ Rel.}{16}{3}{2013}.
\bibitem{Ren13} \au{J}{Rennert} and \au{D}{Sloan}, \tia{A homogeneous model of spinfoam cosmology} \doinn{10.1088/0264-9381/30/23/235019}{Class.\ Quantum Grav.}{30}{235019}{2013} [\arX{1304.6688}].
\bibitem{Fousp} \au{G.F.R}{Ellis}, \au{J}{Murugan} and \au{A}{Weltman} (eds.), \book{Foundations of Space and Time}{Cambridge University Press}{Cambridge}{UK}{2012}.
\bibitem{BaO11} \au{A}{Baratin} and \au{D}{Oriti}, \tia{Ten questions on group field theory (and their tentative answers)} \doinn{10.1088/1742-6596/360/1/012002}{J.\ Phys.\ Conf.\ Ser.}{360}{012002}{2012} [\arX{1112.3270}].
\bibitem{GiSi}   \au{S}{Gielen} and \a{L}{Sindoni}, \tia{Quantum cosmology from group field theory condensates: a review} \doinn{10.3842/SIGMA.2016.082}{SIGMA}{12}{082}{2016} [\arX{1602.08104}].
\bibitem{revmu} \au{G}{Calcagni}, \tia{Multifractional theories: an unconventional review} \doij{10.1007/JHEP03(2017)138}{JHEP}{1703}{138}{2017} [\arX{1612.05632}].
\bibitem{Calcagni:2021ipd} \au{G}{Calcagni}, \tia{Multifractional theories: an updated review} \doin{10.1142/S021773232140006X}{Mod.\ Phys.\ Lett.}{A}{36}{2140006}{2021} [\arX{2103.06557}].
\bibitem{NiR}   \au{M}{Niedermaier} and \au{M}{Reuter}, \tia{The asymptotic safety scenario in quantum gravity} \doinn{10.12942/lrr-2006-5}{Living Rev.\ Rel.}{9}{5}{2006}. 
\bibitem{Eichhorn:2018yfc} \au{A}{Eichhorn}, \tia{An asymptotically safe guide to quantum gravity and matter} \doinn{10.3389/fspas.2018.00047}{Front.\ Astron.\ Space Sci.}{5}{47}{2019} [\arX{1810.07615}].
\bibitem{Bonanno:2020bil} \au{A}{Bonanno}, \au{A}{Eichhorn}, \au{H}{Gies}, \au{J.M}{Pawlowski}, \au{R}{Percacci}, \au{M}{Reuter}, \au{F}{Saueressig} and \au{G.P}{Vacca}, \tia{Critical reflections on asymptotically safe gravity} \doinn{10.3389/fphy.2020.00269}{Front.\ Phys.}{8}{269}{2020} [\arX{2004.06810}].
\bibitem{Kofinas:2016lcz} \au{G}{Kofinas} and \au{V}{Zarikas}, \tia{Asymptotically safe gravity and non-singular inflationary big bang with vacuum birth} \doin{10.1103/PhysRevD.94.103514}{Phys.\ Rev.}{D}{94}{103514}{2016} [\arX{1605.02241}].
\bibitem{BOTu}  \au{R}{Bufalo}, \au{M}{Oksanen} and \au{A}{Tureanu}, \tia{How unimodular gravity theories differ from general relativity at quantum level} \doin{10.1140/epjc/s10052-015-3683-3}{Eur.\ Phys.\ J.}{C}{75}{477}{2015} [\arX{1505.04978}].
\bibitem{Dow13} \au{F}{Dowker}, \tia{Introduction to causal sets and their phenomenology} \doinn{10.1007/s10714-013-1569-y}{Gen.\ Rel.\ Grav.}{45}{1651}{2013}.
\bibitem{Surya:2019ndm} \au{S}{Surya}, \tia{The causal set approach to quantum gravity} \doinn{10.1007/s41114-019-0023-1}{Living Rev.\ Rel.}{22}{5}{2019}. 
\bibitem{AGJL4} \au{J}{Ambj{\o}rn}, \au{A}{G\"orlich}, \au{J}{Jurkiewicz} and \au{R}{Loll}, \tia{Nonperturbative quantum gravity} \doinn{10.1016/j.physrep.2012.03.007}{Phys.\ Rept.}{519}{127}{2012} [\arX{1203.3591}].
\bibitem{Loll:2019rdj} \au{R}{Loll}, \tia{Quantum gravity from causal dynamical triangulations: a review} \doinn{10.1088/1361-6382/ab57c7}{Class.\ Quantum Grav.}{37}{013002}{2020} [\arX{1905.08669}].
\bibitem{Calcagni:2020ads} \au{G}{Calcagni} and \au{A}{De Felice}, \tia{Dark energy in multifractional spacetimes} \doin{10.1103/PhysRevD.102.103529}{Phys.\ Rev.}{D}{102}{103529}{2020} [\arX{2004.02896}].
\bibitem{PaPa2} \au{H}{Padmanabhan} and \au{T}{Padmanabhan}, \tia{CosMIn: the solution to the cosmological constant problem} \doin{10.1142/S0218271813420017}{Int.\ J.\ Mod.\ Phys.}{D}{22}{1342001}{2013} [\arX{1302.3226}].
\bibitem{Dou03} \au{M.R}{Douglas}, \tia{The statistics of string/M theory vacua} \doij{10.1088/1126-6708/2003/05/046}{JHEP}{0305}{046}{2003} [\oarX{hep-th/0303194}].
\bibitem{Do04b} \au{M.R}{Douglas}, \tia{Basic results in vacuum statistics} \doinn{10.1016/j.crhy.2004.09.008}{Comptes Rendus Phys.}{5}{965}{2004} [\oarX{hep-th/0409207}].
\bibitem{DDKa}  \au{F}{Denef}, \au{M.R}{Douglas} and \au{S}{Kachru}, \tia{Physics of string flux compactifications} \doinn{10.1146/annurev.nucl.57.090506.123042}{Ann.\ Rev.\ Nucl.\ Part.\ Sci.}{57}{119}{2007} [\oarX{hep-th/0701050}].
\bibitem{BoPo}  \au{R}{Bousso} and \au{J}{Polchinski}, \tia{Quantization of four form fluxes and dynamical neutralization of the cosmological constant} \doij{10.1088/1126-6708/2000/06/006}{JHEP}{0006}{006}{2000} [\oarX{hep-th/0004134}].
\bibitem{Sus03} \au{L}{Susskind}, \tia{The anthropic landscape of string theory} \procm{Universe or Multiverse?}{\au{B}{Carr}}{Cambridge University Press}{Cambridge}{UK}{2007} [\oarX{hep-th/0302219}]. 
\bibitem{Wei87} \au{S}{Weinberg}, \tia{Anthropic bound on the cosmological constant} \doinn{10.1103/PhysRevLett.59.2607}{Phys.\ Rev.\ Lett.}{59}{2607}{1987}.
\bibitem{Palti:2019pca} \au{E}{Palti}, \tia{The swampland: introduction and review} \doinn{10.1002/prop.201900037}{Fortsch.\ Phys.}{67}{1900037}{2019} [\arX{1903.06239}].
\bibitem{Bau83} \au{E}{Baum}, \tia{Zero cosmological constant from minimum action} \doin{10.1016/0370-2693(83)90556-7}{Phys.\ Lett.}{B}{133}{185}{1983}.
\bibitem{Ha84b} \au{S.W}{Hawking}, \tia{The cosmological constant is probably zero} \doin{10.1016/0370-2693(84)91370-4}{Phys.\ Lett.}{B}{134}{403}{1984}.
\bibitem{Wu07}  \au{Z.C}{Wu}, \tia{The cosmological constant is probably zero, and a proof is possibly right} \doin{10.1016/j.physletb.2007.12.019}{Phys.\ Lett.}{B}{659}{891}{2008}
 [\arX{0709.3314}].
\bibitem{Vil82} \au{A}{Vilenkin}, \tia{Creation of universes from nothing} \doin{10.1016/0370-2693(82)90866-8}{Phys.\ Lett.}{B}{117}{25}{1982}.
\bibitem{HaH83} \au{J.B}{Hartle} and \au{S.W}{Hawking}, \tia{Wave function of the Universe} \doin{10.1103/PhysRevD.28.2960}{Phys.\ Rev.}{D}{28}{2960}{1983}.
\bibitem{Alexander:2008yg} \au{S.H.S}{Alexander} and \au{G}{Calcagni}, \tia{Quantum gravity as a Fermi liquid} \doinn{10.1007/s10701-008-9257-6}{Found.\ Phys.}{38}{1148}{2008} [\arX{0807.0225}].
\bibitem{Fal14} \au{K}{Falls}, \tia{Asymptotic safety and the cosmological constant} \doij{10.1007/JHEP01(2016)069}{JHEP}{1601}{069}{2016} [\arX{1408.0276}].
\bibitem{Anagnostopoulos:2018jdq} \au{F.K}{Anagnostopoulos}, \au{S}{Basilakos}, \au{G}{Kofinas} and \au{V}{Zarikas}, \tia{Constraining the asymptotically safe cosmology: cosmic acceleration without dark energy} \doij{10.1088/1475-7516/2019/02/053}{JCAP}{1902}{053}{2019} [\arX{1806.10580}].
\bibitem{Anagnostopoulos:2019mrc} \au{F.K}{Anagnostopoulos}, \au{G}{Kofinas} and \au{V}{Zarikas}, \tia{IR quantum gravity solves naturally cosmic acceleration and its coincidence problem} \doin{10.1142/S0218271819440139}{Int.\ J.\ Mod.\ Phys.}{D}{28}{14}{2019} [\arX{2102.07578}].
\bibitem{War13} \au{B.F.L}{Ward}, \tia{An estimate of $\Lambda$ in resummed quantum gravity in the context of asymptotic safety} \doinn{10.1016/j.dark.2013.06.002}{Phys.\ Dark Univ.}{2}{97}{2013}.
\bibitem{War14} \au{B.F.L}{Ward}, \tia{Running of the cosmological constant and estimate of its value in quantum general relativity} \doin{10.1142/S0217732315400301}{Mod.\ Phys.\ Lett.}{A}{30}{1540030}{2015} [\arX{1412.7417}].
\bibitem{War15} \au{B.F.L}{Ward}, \tia{Einstein--Heisenberg consistency condition interplay with cosmological constant prediction in resummed quantum gravity} \doin{10.1142/S0217732315502065}{Mod.\ Phys.\ Lett.}{A}{30}{1550206}{2015} [\arX{1507.00661}].
\bibitem{Sor90}	\au{R.D}{Sorkin}, \tia{\href{http://www.physics.syr.edu/~sorkin/some.papers/66.cocoyoc.pdf}{\cob Spacetime and causal sets}} \procsinm{Relativity and Gravitation: Classical and Quantum}{\au{J.C}{D'Olivo}, \au{E}{Nahmad-Achar}, \au{M}{Rosenbaum}, \au{M.P}{Ryan}, \au{L.F}{Urrutia} and \au{F}{Zertuche}}{World Scientific}{Singapore}{1991}. 
\bibitem{ADGS}  \au{M}{Ahmed}, \au{S}{Dodelson}, \au{P.B}{Greene} and \au{R}{Sorkin}, \tia{Everpresent $\Lambda$} \doin{10.1103/PhysRevD.69.103523}{Phys.\ Rev.}{D}{69}{103523}{2004} [\oarX{astro-ph/0209274}].
\bibitem{AhSo}  \au{M}{Ahmed} and \au{R}{Sorkin}, \tia{Everpresent $\Lambda$. II. Structural stability} \doin{10.1103/PhysRevD.87.063515}{Phys.\ Rev.}{D}{87}{063515}{2013} [\arX{1210.2589}].
\bibitem{Bar06} \au{J.D}{Barrow}, \tia{Strong constraint on ever-present $\Lambda$} \doin{10.1103/PhysRevD.75.067301}{Phys.\ Rev.}{D}{75}{067301}{2007} [\oarX{gr-qc/0612128}].
\bibitem{Zun07} \au{J.A}{Zuntz}, \tia{The cosmic microwave background in a causal set universe} \doin{10.1103/PhysRevD.77.043002}{Phys.\ Rev.}{D}{77}{043002}{2008} [\arX{0711.2904}].
\bibitem{Zwane:2017xbg} \au{N}{Zwane}, \au{N}{Afshordi} and \au{R.D}{Sorkin}, \tia{Cosmological tests of Everpresent $\Lambda$} \doinn{10.1088/1361-6382/aadc36}{Class.\ Quantum Grav.}{35}{194002}{2018} [\arX{1703.06265}].
\bibitem{Calcagni:2020ume} \au{G}{Calcagni}, \tia{\href{https://www.doi.org/10.1007/978-981-15-4702-7_30-1}{\cob Quantum gravity and gravitational-wave astronomy}} \procsinm{Handbook of Gravitational Wave Astronomy}{\au{C}{Bambi}, \au{S}{Katsanevas} and \au{K.D}{Kokkotas}}{Springer}{Singapore}{2021} [\arX{2012.08251}]. 
\bibitem{Canizares:2012is} \au{P}{Canizares}, \au{J.R}{Gair} and \au{C.F}{Sopuerta}, \tia{Testing Chern--Simons modified gravity with gravitational-wave detections of extreme-mass-ratio binaries} \doin{10.1103/PhysRevD.86.044010}{Phys.\ Rev.}{D}{86}{044010}{2012} [\arX{1205.1253}].
\bibitem{Yunes:2016jcc}   \au{N}{Yunes}, \au{K}{Yagi} and \au{F}{Pretorius}, \tia{Theoretical physics implications of the binary black-hole merger GW150914} \doin{10.1103/PhysRevD.94.084002}{Phys.\ Rev.}{D}{94}{084002}{2016} [\arX{1603.08955}].
\bibitem{Barausse:2020rsu} \au{E}{Barausse} {et al.} [LISA Fundamental Physics Working Group], \tia{Prospects for fundamental physics with LISA} \doinn{10.1007/s10714-020-02691-1}{Gen.\ Rel.\ Grav.}{52}{81}{2020} [\arX{2001.09793}].
\bibitem{Calcagni:2019kzo} \au{G}{Calcagni}, \au{S}{Kuroyanagi}, \au{S}{Marsat}, \au{M}{Sakellariadou}, \au{N}{Tamanini} and \au{G}{Tasinato}, \tia{Gravitational-wave luminosity distance in quantum gravity} \doin{10.1016/j.physletb.2019.135000}{Phys.\ Lett.}{B}{798}{135000}{2019} [\arX{1904.00384}].
\bibitem{EMNan} \au{J}{Ellis}, \au{N.E}{Mavromatos} and \au{D.V}{Nanopoulos}, \tia{Comments on graviton propagation in light of GW150914} \doin{10.1142/S0217732316750018}{Mod.\ Phys.\ Lett.}{A}{31}{1650155}{2016} [\arX{1602.04764}].
\bibitem{ArCa2} \au{M}{Arzano} and \au{G}{Calcagni}, \tia{What gravity waves are telling about quantum spacetime} \doin{10.1103/PhysRevD.93.124065}{Phys.\ Rev.}{D}{93}{124065}{2016} [\arX{1604.00541}].
\bibitem{Mirshekari:2011yq} \au{S}{Mirshekari}, \au{N}{Yunes} and \au{C.M}{Will}, \tia{Constraining generic Lorentz violation and the speed of the graviton with gravitational waves} \doin{10.1103/PhysRevD.85.024041}{Phys.\ Rev.}{D}{85}{024041}{2012} [\arX{1110.2720}].
\bibitem{Addazi:2021xuf} \au{A}{Addazi} {et al.}, \tia{Quantum gravity phenomenology at the dawn of the multi-messenger era -- A review} to appear in Prog.\ Part.\ Nucl.\ Phys.\ [\arX{2111.05659}].
\bibitem{CDL} \au{V}{Cardoso}, \au{\'O.J.C}{Dias} and \au{J.P.S}{Lemos}, \tia{Gravitational radiation in $D$-dimensional spacetimes} \doin{10.1103/PhysRevD.67.064026}{Phys.\ Rev.}{D}{67}{064026}{2003} [\oarX{hep-th/0212168}].
\bibitem{Ab17b} \au{B.P}{Abbott} {et al.} [LIGO Scientific and Virgo and Fermi-GBM and INTEGRAL Collaborations], \tia{Gravitational waves and gamma-rays from a binary neutron star merger: GW170817 and GRB 170817A} \doinn{10.3847/2041-8213/aa920c}{Astrophys.\ J.}{848}{L13}{2017} [\arX{1710.05834}].
\bibitem{Starobinsky:1980te} \au{A.A}{Starobinsky}, \tia{A new type of isotropic cosmological models without singularity} \doin{10.1016/0370-2693(80)90670-X}{Phys.\ Lett.}{B}{91}{99}{1980}.
\bibitem{Koshelev:2016xqb} \au{A.S}{Koshelev}, \au{L}{Modesto}, \au{L}{Rachwa\l} and \au{A.A}{Starobinsky}, \tia{Occurrence of exact $R^2$ inflation in non-local UV-complete gravity} \doij{10.1007/JHEP11(2016)067}{JHEP}{1611}{067}{2016} [\arX{1604.03127}].
\bibitem{Koshelev:2017tvv}  \au{A.S}{Koshelev}, \au{K.S}{Kumar} and \au{A.A}{Starobinsky}, \tia{$R^2$ inflation to probe non-perturbative quantum gravity} \doij{10.1007/JHEP03(2018)071}{JHEP}{1803}{071}{2018} [\arX{1711.08864}].
\bibitem{Calcagni:2020tvw} \au{G}{Calcagni} and \au{S}{Kuroyanagi}, \tia{Stochastic gravitational-wave background in quantum gravity}, \doij{10.1088/1475-7516/2021/03/019}{JCAP}{2103}{019}{2021} [\arX{2012.00170}].
\bibitem{Gasperini:1992em} \au{M}{Gasperini} and \au{G}{Veneziano}, \tia{Pre-big bang in string cosmology} \doinn{10.1016/0927-6505(93)90017-8}{Astropart.\ Phys.}{1}{317}{1993} [\arX{hep-th/9211021}].
\bibitem{Gasperini:2016gre} \au{M}{Gasperini}, \tia{Observable gravitational waves in pre-big bang cosmology: an update} \doij{10.1088/1475-7516/2016/12/010}{JCAP}{1612}{010}{2016} [\arX{1606.07889}].
\bibitem{Ben-Dayan:2016iks} \au{I}{Ben-Dayan}, \tia{Gravitational waves in bouncing cosmologies from gauge field production} \doij{10.1088/1475-7516/2016/09/017}{JCAP}{1609}{017}{2016} [\arX{1604.07899}].
\bibitem{Ben-Dayan:2018ksd} \au{I}{Ben-Dayan} and \au{J}{Kupferman}, \tia{Sourced scalar fluctuations in bouncing cosmology} \doij{10.1088/1475-7516/2019/07/050}{JCAP}{1907}{050}{2019}; \doij{10.1088/1475-7516/2020/12/E01}{Erratum-ibid.}{2012}{E01}{2020} [\arX{1812.06970}].
\bibitem{Artymowski:2020pci} \au{M}{Artymowski}, \au{I}{Ben-Dayan} and \au{U}{Thattarampilly}, \tia{Sourced fluctuations in generic slow contraction} \doij{10.1088/1475-7516/2021/06/010}{JCAP}{2106}{010}{2021} [\arX{2011.00626}].
\bibitem{Brandenberger:2015kga} \au{R.H}{Brandenberger}, \tia{String gas cosmology after Planck} \doinn{10.1088/0264-9381/32/23/234002}{Class.\ Quantum Grav.}{32}{234002}{2015} [\arX{1505.02381}].
\bibitem{BrVa1} \au{R}{Brandenberger} and \au{C}{Vafa}, \tia{Superstrings in the early universe} \doin{10.1016/0550-3213(89)90037-0}{Nucl.\ Phys.}{B}{316}{391}{1989}.
\bibitem{Bernardo:2020bpa} \au{H}{Bernardo}, \au{R}{Brandenberger} and \au{G}{Franzmann}, \tia{Solution of the size and horizon problems from classical string geometry} \doij{10.1007/JHEP10(2020)155}{JHEP}{2010}{155}{2020} [\arX{2007.14096}].
\bibitem{Brandenberger:2020tcr} \au{R}{Brandenberger} and \au{Z}{Wang}, \tia{Nonsingular ekpyrotic cosmology with a nearly scale-invariant spectrum of cosmological perturbations and gravitational waves} \doin{10.1103/PhysRevD.101.063522}{Phys.\ Rev.}{D}{101}{063522}{2020} [\arX{2001.00638}].
\bibitem{Brandenberger:2020eyf} \au{R}{Brandenberger} and \au{Z}{Wang}, \tia{Ekpyrotic cosmology with a zero-shear S-brane} \doin{10.1103/PhysRevD.102.023516}{Phys.\ Rev.}{D}{102}{023516}{2020} [\arX{2004.06437}].
\bibitem{Brandenberger:2020wha} \au{R}{Brandenberger}, \au{K}{Dasgupta} and \au{Z}{Wang}, \tia{Reheating after S-brane ekpyrosis} \doin{10.1103/PhysRevD.102.063514}{Phys.\ Rev.}{D}{102}{063514}{2020} [\arX{2007.01203}].
\bibitem{Seto:2003kc} \au{N}{Seto} and \au{J}{Yokoyama}, \tia{Probing the equation of state of the early universe with a space laser interferometer} \doinn{10.1143/JPSJ.72.3082}{J.\ Phys.\ Soc.\ Jap.}{72}{3082}{2003} [\arX{gr-qc/0305096}].
\bibitem{Kuroyanagi:2014nba} \au{S}{Kuroyanagi}, \au{T}{Takahashi} and \au{S}{Yokoyama}, \tia{Blue-tilted tensor spectrum and thermal history of the universe} \doij{10.1088/1475-7516/2015/02/003}{JCAP}{02}{003}{2015} [\arX{1407.4785}].
\bibitem{BH}    \au{R}{Brandenberger} and \au{P.-M}{Ho}, \tia{Noncommutative spacetime, stringy spacetime uncertainty principle, and density fluctuations} \doin{10.1103/PhysRevD.66.023517}{Phys.\ Rev.}{D}{66}{023517}{2002} [\oarX{hep-th/0203119}].
\end{thebibliography}
\end{document}